\documentclass[a4paper,11pt]{article}
\pdfoutput=1 

\usepackage{jheppub} 

\usepackage[T1]{fontenc} 

\usepackage{amsfonts}
\usepackage{amsmath}
\usepackage{amsthm}
\usepackage{amssymb}
\usepackage{array}
\usepackage{dcolumn}
\usepackage{placeins}
\usepackage[svgnames]{xcolor}

\usepackage{graphics}

\usepackage{graphicx}
\usepackage{caption}
\usepackage{subcaption}
\usepackage{adjustbox}
\usepackage{gensymb}
\usepackage{breqn}
\usepackage{braket}
\usepackage{enumitem}
\usepackage{makecell}
\usepackage{ragged2e}
\usepackage{comment}
\usepackage{url}
\usepackage{xcolor}

\preprint{MI-HET-774, KIAS-P22016}

\newcommand\be{\begin{equation}}
\newcommand\ee{\end{equation}}
\newcommand\bea{\begin{eqnarray}}
\newcommand\eea{\end{eqnarray}}

\newcommand\mev{\rm MeV}
\newcommand\gev{\rm GeV}
\newcommand\tev{\rm TeV}

\title{\boldmath Sensitivity to Dark Sector Scales from Gravitational Wave Signatures }

\author[a]{James B.~Dent,}
\author[b]{Bhaskar Dutta,}
\author[b,c]{Sumit Ghosh,}
\author[d]{Jason Kumar,}
\author[d]{and Jack Runburg}

\affiliation[a]{Department of Physics, Sam Houston State University, Huntsville, TX 77341, USA}

\affiliation[b]{Mitchell Institute for Fundamental Physics and Astronomy, Department of Physics  and Astronomy, Texas A$\&$M University, College Station, Texas 77843,  USA}

\affiliation[c]{School of Physics, Korea Institute for Advanced Study, Seoul 02455, Korea}

\affiliation[d]{Department of Physics and Astronomy, University of Hawaii,  Honolulu, Hawaii 96822, USA}

\emailAdd{jbdent@shsu.edu}
\emailAdd{dutta@physics.tamu.edu}
\emailAdd{ghosh@kias.re.kr}
\emailAdd{jkumar@hawaii.edu}
\emailAdd{runburg@hawaii.edu}

\abstract{We consider gravitational sound wave signals produced by a first-order phase transition in a theory with a generic renormalizable thermal effective potential of power law form.  We find the frequency and amplitude of the gravitational wave signal can be related in a straightforward manner to the parameters of the thermal effective potential.  This leads to a general conclusion; if the mass of the dark Higgs is less than 1\% of the dark Higgs vacuum expectation value, then the gravitational wave signal will be unobservable at all upcoming and planned gravitational wave observatories.  Although the understanding of gravitational wave production at cosmological phase transitions is still evolving, we expect this result to be robust. }

\keywords{Gravitational Wave, Cosmological phase transition, Dark Higgs}

\arxivnumber{2203.11736}

\begin{document} 
\maketitle
\flushbottom

\section{Introduction}
\label{sec:intro}

With the advent of gravitational wave (GW) detection~\cite{LIGOScientific:2016aoc}, and the development of new instruments (see Table~\ref{tab:Experiments}, and references therein), gravitational wave signals have become among the most promising signatures for new physics beyond the Standard Model (SM) in the early Universe.  In particular, cosmological first order phase transitions in the early Universe can generate gravitational wave signatures \cite{Kosowsky:1992rz,Kosowsky:1991ua,Apreda:2001us,Grojean:2006bp,Huber:2008hg} which can be observed at current and upcoming experiments (for reviews, see~\cite{Cai:2017cbj,Weir:2017wfa,Caprini:2015zlo,Mazumdar:2018dfl,Caprini:2018mtu,Caprini:2019egz,Caldwell:2022qsj}).  Considerable work has focused on determining if particular models can produce GW signals which are observable at upcoming experiments. There has been a recent flurry of works examining extensions of the Standard Model which could provide gravitational wave signatures \cite{Ashoorioon:2009nf,Alves:2018jsw,Balazs:2016tbi,Kang:2017mkl,Matsui:2017ggm,Vaskonen:2016yiu,Lewicki:2021pgr,Croon:2018new,Beniwal:2018hyi,Hashino:2018wee,Ahriche:2018rao,Shajiee:2018jdq,Vieu:2018nfq,Alves:2018oct,Chala:2018opy,Fujikura:2018duw,Graf:2021xku,Li:2020eun,Ghorbani:2017lyk,Marzo:2018nov,Marzola:2017jzl,Verweij:2019xch,Prokopec:2018tnq,Zhou:2018zli,Archer-Smith:2019gzq,Cai:2022bcf,Paul:2020wbz,Chao:2017ilw,Pimentel:2020rkw,Blinov:2015sna,Inoue:2015pza,Greljo:2019xan,Croon:2018kqn,Demidov:2017lzf,Bian:2017wfv,Eichhorn:2020upj,Miura:2018dsy,Azatov:2019png,Aoki:2017aws,Chen:2017cyc,Croon:2019ugf,Croon:2019iuh, Costa:2022oaa, Chao:2017vrq, Ghosh:2020ipy, Chao:2021xqv, Romero:2021kby, Alves:2020bpi, Hall:2019ank, Hall:2019rld}, including models with a focus on dark sectors and dark matter \cite{Madge:2018gfl,Croon:2018erz,Hashino:2018zsi,Schwaller:2015tja,Alanne:2014bra,Addazi:2017gpt,Fairbairn:2019xog,Bertone:2019irm,Huang:2021rrk,Halverson:2020xpg,Nakai:2020oit,Ratzinger:2020koh,Bhoonah:2020oov,Helmboldt:2019pan,Beniwal:2017eik,Baldes:2017ygu,Baldes:2018emh, Huang:2017kzu,Tsumura:2017knk,Croon:2019rqu,Addazi:2017nmg,Imtiaz:2018dfn,Addazi:2017oge,Bai:2018dxf,Bian:2018bxr,Pandey:2020hoq,Davoudiasl:2021ijv,Huang:2017rzf,Paul:2019pgt,Hektor:2018esx,Kannike:2019mzk,Kannike:2019wsn,Mohamadnejad:2019vzg,Breitbach:2018ddu,Arakawa:2021wgz, Azatov:2021ifm, Azatov:2020ufh}.

In this work, we approach this problem from a different angle.  We scan the parameter space of a simple and well-motivated thermal effective potential with a single dark Higgs field, which can arise from a wide class of UV-complete models.  We correlate the properties of the GW signal arising from a first-order phase transition with the parameters of the effective potential (for other work examining general models, as well as model and signal features and classes, see for example, \cite{Chung:2012vg,Dev:2016feu,Jinno:2017ixd,Chala:2019rfk,Croon:2019kpe,Ellis:2019flb,Ellis:2020awk,Megevand:2021llq,Schmitz:2020syl,Alanne:2019bsm}, including cosmological contraints \cite{Bai:2021ibt}, non-gaussianities \cite{Kumar:2021ffi}, and circular polarization \cite{Ellis:2020uid}).

The result of this analysis is a characterization of 
the  
sound wave
GW signal which can arise from any UV-complete 
model whose thermal effective potential is polynomial 
and renormalizable.  Interestingly, we find that 
dimensional analysis and scaling relations control much 
of the analysis, allowing us to relate specific features 
of the GW signal to specific parameters of the effective 
potential.

As a specific example, we find that amplitude of the 
gravitational wave signal is strongly related to the 
ratio of the dark Higgs mass to the dark Higgs 
vacuum expectation value (vev) 
at zero temperature.  If the dark Higgs mass is less 
than ${\cal O}(1\%)$ of the dark Higgs vev, then 
the GW signal will be too small to be observed at 
any upcoming experiments.  Because the amplitude of the signal 
scales as several powers of ratio of dark Higgs mass to 
vev, our result is robust.  Indeed, theoretical work connecting 
the amplitude of a gravitational wave signal to the underlying 
parameters of the phase transition is constantly evolving (for example, more refined calculations for determining the kinetic energy fraction~\cite{Espinosa:2010hh,Giese:2020rtr,Giese:2020znk}, the wall velocity~\cite{Bodeker:2009qy,Bodeker:2017cim,Hoche:2020ysm}, and adoption of more general parameterized forms for the frequency spectra~\cite{Schmitz:2020rag} which can account for the variations from different studies, such as~\cite{Hindmarsh:2013xza,Hindmarsh:2015qta,Hindmarsh:2017gnf,Jinno:2017fby,Cutting:2018tjt,Cutting:2020nla,Jinno:2020eqg,Dahl:2021wyk,Auclair:2022jod}), leading 
one to wonder how well one can trust any even state-of-the-art 
result.  But in order to change our bound by an order-of-magnitude, 
one would need a correction to the gravitational wave signal of 
many orders of magnitude, giving our result some robustness to 
future developments.

We also find a general relationship between the 
symmetry-breaking vev and peak frequency of the 
GW signal.  We then characterize the optimal 
instruments for probing generic first-order 
phase transitions, in terms of the symmetry-breaking 
scale at zero temperature.

But it is also important to point out what we do not 
do --  we do not provide a complete analysis of any 
particular UV-complete model.  We assume a thermal 
effective potential for a single scalar field  
which is renormalizable and of 
the polynomial form.  Although well-motivated, for 
any particular model, this 
may or may not be a good approximation to the thermal 
effective potential.  Our results apply directly to 
models for which this thermal effective potential is 
a good approximation, while for other models, our 
results are at best indicative.

The plan of this paper is as follows.  In Section~\ref{sec:Transition}, we parameterize 
the thermal effective potential, and characterize 
the allowed phase transitions and resulting 
gravitational wave signals.  
In Section~\ref{sec:Results}, we describe the results 
of our scans over the parameter space of the 
thermal effective potential.  We conclude in 
Section~\ref{sec:Conclusion} with a discussion of 
our results.

\section{First-Order Phase Transitions with a  
Renormalizable Scalar Potential}
\label{sec:Transition}

We consider a thermal effective potential for a 
single real scalar field $\phi$.  We consider a 
renormalizable 
polynomial potential~\cite{Croon:2018erz} of the form
\bea
V(T,\phi) &=& \Lambda^4 \left[\left( -\frac{1}{2} +c 
\left( \frac{T}{v} \right)^2\right) \left(\frac{\phi}{v} \right)^2
+ b \frac{T}{v}  \left(\frac{\phi}{v} \right)^3
+ \frac{1}{4} \left(\frac{\phi}{v} \right)^4  \right] ,
\eea
where $V(T=0, \phi)$ is minimized at $\phi = \pm v$, and 
where the mass of the dark Higgs (the excitation of 
$\phi$ about this minimum) is given by 
$m^2 / v^2 = 2(\Lambda / v)^4$.  
This form of the potential arises, for example, from 
thermal corrections in the high temperature limit, where 
the dimensionless coefficients $b$ and $c$ depend on the 
details of the UV model.  We assume $\Lambda / v 
\lesssim 1$, 
ensuring that effective potential has a perturbative 
quartic coupling.  We also take $b<0$, as this coefficient 
arises in the high temperature limit from 
fermion loops which contribute with a negative sign to 
the effective potential.

It is convenient to express the potential 
in terms of the 
scale-free quantities $\tilde \phi = \phi /v$ and 
$\tilde T = T /v$.  
Essentially, the symmetry-breaking scale $v$ is used as 
the scale against which all other energies are measured.
Defining 
$\tilde V (\tilde T, \tilde \phi) = \Lambda^{-4} 
V(T, \phi)$, we have 
\bea
\tilde V (\tilde T, \tilde \phi) &=& 
\left( -\frac{1}{2} +c  \tilde T^2 \right) 
\tilde \phi^2 
+ b \tilde T \tilde \phi^3 
+ \frac{1}{4} \tilde \phi^4   .
\eea

We can now consider the generic behavior of this 
thermal effective potential.  
At finite temperature, the potential can have up to 
three extrema, at $\phi = 0$ and at factor of 2 under the radical is fixed
\bea
\tilde \phi  &=& \frac{-3b \tilde T \pm 
\sqrt{9b^2 \tilde T^2 + 4(1-2c\tilde T^2)}}{2} .
\eea
If $c \tilde T^2 < (1/2) + (9/8) b^2 \tilde T^2$, then 
the discriminant is positive, and three distinct 
extrema exist.
In that case, one extremum is a local maximum, while the 
other two are local minima, and at least one is a 
symmetry-breaking minimum.

$V(T, \phi=0)=0$, where this extremum at $\phi=0$ is a 
local minimum for $c \tilde T^2 > 1/2$ and a local 
maximum for $c \tilde T^2 <1/2$.  The condition for 
$V(T, \phi)=0$ for $\phi>0$ is
\bea
c \tilde T^2 \leq 
\frac{1}{2} + b^2 \tilde T^2 .
\eea

We can thus classify the behavior of the potential 
as the Universe cools from high temperature.  At 
sufficiently high temperature, the symmetry-preserving 
extremum is always a local minimum at $V=0$.  But for 
$c / b^2 <1$, the potential has two other 
zeros, with a global minimum between them, even at 
high temperature.  In this case, there is no phase 
transition, since the Universe is always in the 
symmetry-breaking phase.  We thus restrict ourselves 
to the case $c / b^2 >1$.

For $c / b^2 >1$, then at high temperature there 
is a global minimum at $\phi=0$ and a local 
minimum at $\phi_+ >0$, with a local maximum 
at $\phi_-$ between them.  But at low enough 
temperature ($\tilde T_C = 1/\sqrt{2(c-b^2)}$), 
we find $V(\phi_+) =0$.  

For $T < T_C$, $\phi_+$ is now the global 
minimum, and a first-order phase transition is 
possible.
The transition will not occur 
until the nucleation temperature, $T_N$, at which 
time the bubble nucleation rate is larger than the 
Hubble expansion rate.  But at sufficiently low 
temperature ($\tilde T< 1/\sqrt{2c}$), the potential 
barrier goes to zero; the symmetry-preserving 
extremum is now a local maximum, and if the phase 
transition has not yet completed, then $\phi$ can  
simply roll to the global minimum, yielding a 
second-order phase transition.

Note that our analysis of the minima of the potential  
depended only on the parameters $b$ and $c$, as $v$ appears 
only as a scale parameter, and  $\Lambda$ has 
factored out.  
But the dependence on $\Lambda$ 
returns when we  determine 
the nucleation temperature.  The bubble nucleation rate 
per unit volume ($p(T)$) is given by 
$p(T) = T^4 \exp [-S_E/T]$, where $S_E$ is the 
Euclidean action for the radial bounce solution 
between the symmetry-preserving and symmetry-breaking 
vacua.  The nucleation temperature $T_N$ is the temperature 
at the nucleation time $t_N$, at which a fraction $e^{-1}$ of 
the Universe has tunnelled to the global minimum of the potential.  
The nucleation time is given approximately by the relation 
$p(t_N ) t_N^4 = 1$.  Assuming a radiation-dominated universe with 
$g_* = {\cal O}(100)$ relativistic degrees of freedom, we 
then find~\cite{Caprini:2019egz,Quiros:1999jp,wainwright2012:cosmotransitions,Ramsey-Musolf:2019lsf}
\bea
\frac{S_E}{T_N} = 140 + {\cal O}(\log T_N / \tev) .
\eea
$S_E$ is the Euclidean action evaluated on the bounce 
solution to the radial equation~\cite{Akula:2016gpl,wainwright2012:cosmotransitions}
\bea
\frac{d^2 \phi}{dr^2} + \frac{2}{r} \frac{\partial \phi}
{\partial r} &=& \frac{\partial V_E (T,\phi)}{\partial \phi} .
\label{eq:Bounce}
\eea
where $V_{E}(T,\phi)$ is the temperature and field dependent potential in the Euclidean action. This solution interpolates between 
$\phi=0$ and $\phi = \phi_+$.  
We will utilize an analytic approximation to the bounce solution for this 
form of the effective potential~\cite{Dine:1992wr}, given by 
\bea
\frac{S_E}{T} = \frac{4.85M^3}{E^2T^3}\left[1 +\frac{\alpha}{4}\left(1 + \frac{2.4}{1-\alpha}+ \frac{0.26}{(1-\alpha)^2}\right)\right] ,
\eea
where 
\bea
M^2 \equiv 2\frac{\Lambda^4}{v^2}\left(\frac{cT^2}{v^2}-\frac{1}{2}\right),
\hspace{1cm} E \equiv -\frac{b\Lambda^4}{v^4},\hspace{1cm}
\alpha\equiv \frac{M^2 \Lambda^4}{2 E^2 T^2 v^4} .
\eea
For certain 
models, this analytic solution was compared to the Euclidean 
action computed using \texttt{CosmoTransitions} \cite{wainwright2012:cosmotransitions}, a Python package that numerically computes the properties of phase transitions in scalar field theories, and good agreement was found.  
A more detailed comparison was performed in~\cite{Guo:2020grp}, 
similarly finding good agreement.

To put eqn.~\ref{eq:Bounce} in 
scale-free form, we define $\tilde r \equiv r (\Lambda^2/v)$,
yielding
\bea
\frac{d^2 \tilde \phi}{d\tilde r^2} 
+ \frac{2}{\tilde r} \frac{\partial \tilde \phi}
{\partial \tilde r} &=& \frac{\partial \tilde V_E 
(\tilde T,\tilde \phi)}{\partial \tilde \phi} .
\eea
The scale-free Euclidean action,  $\tilde S_E$, 
is obtained by integrating the scale-free Lagrangian, 
evaluated 
on the bounce solution of the above equation, with respect 
to $d^3 \tilde r ~d(1/\tilde T)$, and 
is related to $S_E$ by 
\bea
\frac{S_E}{T} &=& \frac{v^2}{\Lambda^2} 
\frac{\tilde S_E}{\tilde T}.
\eea
We thus find
\bea
\frac{\tilde S_E}{\tilde T_N} &\sim& 
140 \left(\frac{\Lambda}{v} \right)^2 ,
\label{eqn:SE}
\eea
where the left-hand side of eqn.~\ref{eqn:SE} 
depends only on $b$, $c$ and $\tilde T_N$.

\subsection{Thermal Parameters of the Phase Transition }  

We are now able to determine determine $S_E (\tilde T)$ 
and 
$\tilde T_N$ for any parameters $b$, $c$ and $\Lambda / v$.  
From these, we can determine the thermal parameters of 
the phase transition, which in turn feed into the 
gravitational wave signature.  These are speed parameter 
of the phase transition ($\beta / H$), and 
the latent heat parameter ($\xi$).

We can write the thermal parameters as 
\bea
\left(\frac{\beta}{H} \right) &=& 
\left(\frac{\tilde  \beta}{H} (b,c,\Lambda/v) \right) 
\times \left(\frac{\Lambda}{v} \right)^{-2} ,
\nonumber\\
\xi &=& \tilde \xi (b,c,\Lambda /v) \times 
\left(\frac{\Lambda}{v} \right)^4 
\left(\frac{g_*}{100} \right)^{-1} ,
\label{eq:ThermalParamScaling}
\eea
where 
\bea
\frac{\tilde  \beta}{H} (b,c,\Lambda/v) &=& 
\left(\tilde T \frac{d (\tilde S_E / \tilde T)}
{d \tilde T} \right)_{\tilde T = \tilde T_N} ,
\nonumber\\
\tilde \xi (b,c,\Lambda /v) &=& 
\left[\left( \frac{3}{10\pi^2 \tilde T_N^4} \right)
\left(\Delta \tilde V - \tilde T 
\Delta \frac{d \tilde V}{d \tilde T} \right)
\right]_{\tilde T = \tilde T_N} ,
\eea
and $g_*$ is the number of relativistic degrees of freedom.
Essentially, the parameter $\Lambda / v$ controls an overall rescaling 
of the potential, and the power-law dependence of the 
thermal parameters on $\Lambda / v$ in eq.~\ref{eq:ThermalParamScaling} 
encodes the dependence of these parameters on that scaling.

Note that the scale-free thermal parameters 
$(\tilde \beta  / H)$ and $\tilde \xi$ depend on 
$\Lambda / v$ only through the determination of 
the scale-free nucleation temperature, $\tilde T_N$.  
Thus, the actual thermal parameters $(\beta /H)$ and 
$\xi$ have a dependence on $\Lambda / v$ which is nearly 
power-law.  This dependence would be exactly power law if 
the nucleation temperature were exactly the same as the 
critical temperature.  In fact, the nucleation temperature 
will be less than the critical temperature, leading to a 
deviation from power-law behavior.  But we will see that 
this power-law behavior largely determines the maximum 
allowed gravitational wave signal.
Moreover, this 
form of the thermal effective potential is a good 
approximation in the high temperature limit; if the 
nucleation temperature is significantly smaller than the 
critical temperature, this may not be a good 
approximation. A typical scale where the high-temperature approximation remains valid is given by $T \gtrsim m$ (for example, in~\cite{Croon:2018erz} the criteria $T^2 > 2m^2$ was employed). This can be translated to the constraint $\tilde{T}\gtrsim \sqrt{2}(\Lambda/v)^2$. However, the size of the effects of non-power law terms that would be generated if this criteria is not met would be model dependent, thus we do not pursue that line of inquiry further in this work.

\subsection{The Gravitational Wave Signal}

First order phase transitions can produce gravitational waves via three different mechanisms: i) bubble collisions, ii) sound waves from bubble expansion in the fluid, and iii) turbulence in the fluid. In a runaway bubble expansion, where bubble wall velocities are not sufficiently slowed by friction from the fluid, bubble collisions can provide the dominant gravitational wave signal. This situation can arise, for example, for extremely supercooled systems~\cite{Kosowsky:1992vn,Cutting:2018tjt,Lewicki:2019gmv}. The study of turbulence is ongoing, with the efficiency of converting kinetic energy into turbulent energy being one source of some uncertainty~\cite{Hindmarsh:2015qta,Schmitz:2020syl,Auclair:2022jod}. Thus, as we are not examining extremely supercooled situations, we will assume that the dominant gravitational wave signature arises 
from sound waves~\cite{Hindmarsh:2016lnk}.
We can express the amplitude ($h^2 \Omega_{sw}$) and 
frequency ($f_{sw}$) of this gravitational wave signal 
as~\cite{Weir:2017wfa,Guo:2020grp,Guo:2021qcq}
\bea
h^2 \Omega_{sw} (f) &=& h^2 \Omega_{sw}^{max}
\times \left(\frac{f}{f_{sw}} \right)^3 
\left(\frac{7}{4+3(f/f_{sw})^2 } \right)^{7/2} , 
\eea
where 
\bea
f_{sw} &=& \tilde f_{sw} (b, c, \Lambda/v) 
\times \left(\frac{\Lambda}{v} \right)^{-2} 
\left(\frac{T_N}{100~\gev} \right) 
\left( \frac{g_*}{100} \right)^{1/6} ,
\nonumber\\
h^2 \Omega_{sw}^{max} &=&  
h^2 \tilde \Omega_{sw}^{max} (b, c, \Lambda/v)
\times \left(\frac{\Lambda}{v} \right)^{10+8n} 
\left( \frac{g_*}{100}\right)^{-5/3-2n} 
\nonumber\\
&\,& 
\times  \left[ 1- \frac{1}{\sqrt{1+2 \tau_{sh} H_s}} \right],
\eea
and~\cite{Espinosa:2010hh}
\bea
\tilde f_{sw} (b,c, \Lambda / v) &=& 8.9 \times 10^{-3} {\rm mHz} \left(\frac{1}{v_w} \right) \left(\frac{\tilde \beta}{H} 
\right) 
\left[\times \left(\frac{z_p}{10} \right) \frac{1}{\sqrt{1+2 \tau_{sh} H_s}} \right] ,
\nonumber\\
h^2 \tilde \Omega_{sw}^{max} (b,c, \Lambda / v) &=& 
8.5 \times 10^{-6} 
\left(\frac{\Gamma}{4/3}\right)^2 
(\tilde \kappa_f \tilde \xi)^2 
\left(\frac {\tilde \beta}{H} \right)^{-1} 
v_w ,
\nonumber\\
\tilde \kappa_f &\sim& \tilde \xi^n .
\label{eqn:tilde_gravwave_param}
\eea
Here, $v_w$ is the sound speed, which determines the exponent $n$; 
for $v_w \sim 1$, we have 
$n = 1$, while for $v_w \sim 0.5$, we have $n \sim 2/5$~\cite{Espinosa:2010hh}.
$\Gamma$ is the adiabatic index, which we will take to 
be $\Gamma =4/3$~\cite{Croon:2018erz}, and $z_p \sim 10$~\cite{Hindmarsh:2017gnf,Guo:2020grp}.  

There are additional factors listed in brackets~\cite{Guo:2020grp}, 
which 
depend on the details of the model, encoded in quantities 
such as the time scale for turbulence ($\tau_{sh}$) and 
the Hubble scale at which the source becomes active 
($H_S$).  Indeed, as pointed out in~\cite{Guo:2021qcq}, 
there are a variety of suppression factors which can 
arise from a more diligent treatment of gravitational 
wave production, which can suppress the signal amplitude 
by more than an order of magnitude.  
As we begin with 
an effective potential for the scalar, but no detailed 
microphysics, it is not possible to attain this level of 
precision.  But for our purpose, it is not necessary.  Since 
the gravitational wave amplitude scales as more than 
10 powers of $(\Lambda / v)$, this analysis will be 
sufficient to determine classes of models which will be 
easily inaccessible to all upcoming experiments.  
Moreover, although future theoretical 
developments may uncover additional correction factors, 
unless their collective size is several orders of magnitude, 
they will not change our results appreciably.
For similar 
reasons, it is sufficient for us define the thermal parameters 
at the nucleation temperature, rather than the percolation 
temperature.

In a similar vein, there are a variety of theoretical uncertainties 
with both the first-order phase transition and the associated 
gravitational wave signature, and there has been much recent 
theoretical progress, including studies of gauge dependence, renormalization group equations, and the one-loop effective potential \cite{Chiang:2017zbz,Niemi:2021qvp}, discussions within an effective field theory framework \cite{Cai:2017tmh,Chala:2018ari,Gould:2021ccf}, issues related to the perturbative expansion including its scale dependence \cite{Gould:2021oba,Schicho:2021gca,Gould:2021dzl,Postma:2020toi}, and non-perturbative models and resummation  and use of numerical simulations (for example, see~\cite{Hindmarsh:2013xza,Hindmarsh:2015qta,Hindmarsh:2017gnf,Cutting:2018tjt,Cutting:2019zws,RoperPol:2019wvy,Giese:2020rtr,Cutting:2020nla,LatticeStrongDynamics:2020jwi,Lewicki:2020azd,Cutting:2021tqt,Jinno:2021ury,Schicho:2022wty}). 
It would be interesting to incorporate these effects in a 
more detailed analysis, but it is not necessary 
for our purpose 
here.

The main quantities we are interested in are the 
peak amplitude of the GW signal ($h^2 \Omega_{sw}^{max}$), 
and the frequency at which that peak is obtained 
($f_{sw}$).  If the nucleation temperature were equal 
to the critical temperature, then 
$h^2 \tilde \Omega_{sw}^{max}$ and 
$\tilde f_{sw}$ would be independent of $\Lambda /v$, 
the dependence of the GW signal parameters on $\Lambda / v$ 
would be entirely determined by the scaling relations in 
eq.~\ref{eq:ThermalParamScaling}.
We would then find
\bea
\left(h^2 \Omega_{sw}^{max} \right)_{T_N = T_C} 
&\propto& \xi^{2(n+2)} \left(\frac{\beta}{H} \right)^{-1} 
\propto\left(\frac{\Lambda}{v} \right)^{10+8n} ,
\nonumber\\
\left( f_{sw} \right)_{T_N = T_C} &\propto&
\frac{\beta}{H}  v 
\propto \left(\frac{\Lambda}{v} \right)^{-2} v ,
\label{eq:BasicRelations}
\eea
where $(\Lambda / v )^4 = m^2 / 2v^2$, and the power-law dependence on $\Lambda / v$ is 
induced by the dependence of the thermal parameters on the overall scale of the potential.  

Reducing $m/v$ by an order of magnitude would reduce the 
amplitude of the GW signal by a factor of at least 
$10^5$ (taking $n=0$).  We thus expect to find a sharp 
minimum for $m /v$; for dark Higgs masses which are 
sufficiently smaller than the symmetry-breaking scale, the 
amplitude of the GW signal will be unobservable.  

Note, however that although the amplitude of the GW 
signal depends on $m/v$, it has no actual dependence on  
$v$ itself, a result which could be anticipated from 
dimensional analysis.  But the symmetry-breaking scale 
does enter into $f_{sw}$, to which it is directly 
proportional.  Thus far, our analysis has been 
completely scale-free.  We now see that the energy 
scale of the phase transition only enters in the last 
step, setting the frequency scale of the GW signal.

Some of these considerations will be modified once we 
include the dependence of the nucleation temperature on 
$(\Lambda / v)$.  But we expect that this modification 
should not change the result substantially; if 
$T_N \ll T_C$, then the phase transition is very 
supercooled, and this form of the thermal effective 
potential may in any case not be a good approximation 
to the thermal potential of a UV complete model.  We 
will see from more detailed numerical calculation that 
this intuition is correct.

\section{Results}
\label{sec:Results}

In this section, we describe our strategy for scanning over the parameter space of the thermal effective potential and then present  the results of the scan. First, we briefly specify  the conditions for a parameter space point to 
be of interest.
\begin{itemize}
\item{$\Lambda / v < 1$: This condition ensures that the zero-temperature 
quartic coupling is perturbative.}
\item{$c/b^2 >1$: This condition is required for a phase transition 
to occur at all.  If it is not satisfied, then the symmetry-breaking minimum 
is the global minimum even at high temperature.}
\item{$c \tilde T_N^2 > 1/2$: This is necessary in order for the  
transition to be first-order.  If this is not 
satisfied, then the barrier disappears  before the nucleation temperature is reached, 
leading to a smooth phase 
transition.}
\item{$\beta / H > 1$: This condition is needed for the bubble growth rate to 
exceed the Hubble expansion, so that bubbles grow.  This is equivalent to 
$\tilde \beta / H > (\Lambda / v)^2$.}
\item{$\xi > 0$: a first order transition will only proceed if the latent heat 
parameter is positive at the nucleation temperature.  This is equivalent to 
$\tilde \xi >0$.}
\end{itemize}
It may seem unusual that the latent heat parameter can be 
negative.  It is easy to demonstrate that, if the phase  
transition occurs at the critical temperature, then the 
latent heat parameter must be positive.  But it may 
become negative if the phase transition is sufficiently 
supercooled.  As we have already mentioned, such a transition 
would potentially stretch the effective potential beyond 
its regime of validity.  But in any case, if the latent 
heat parameter is negative, no GW signal will be produced.


\begin{figure}[tbp]
\centering 
\includegraphics[width=.485\textwidth]{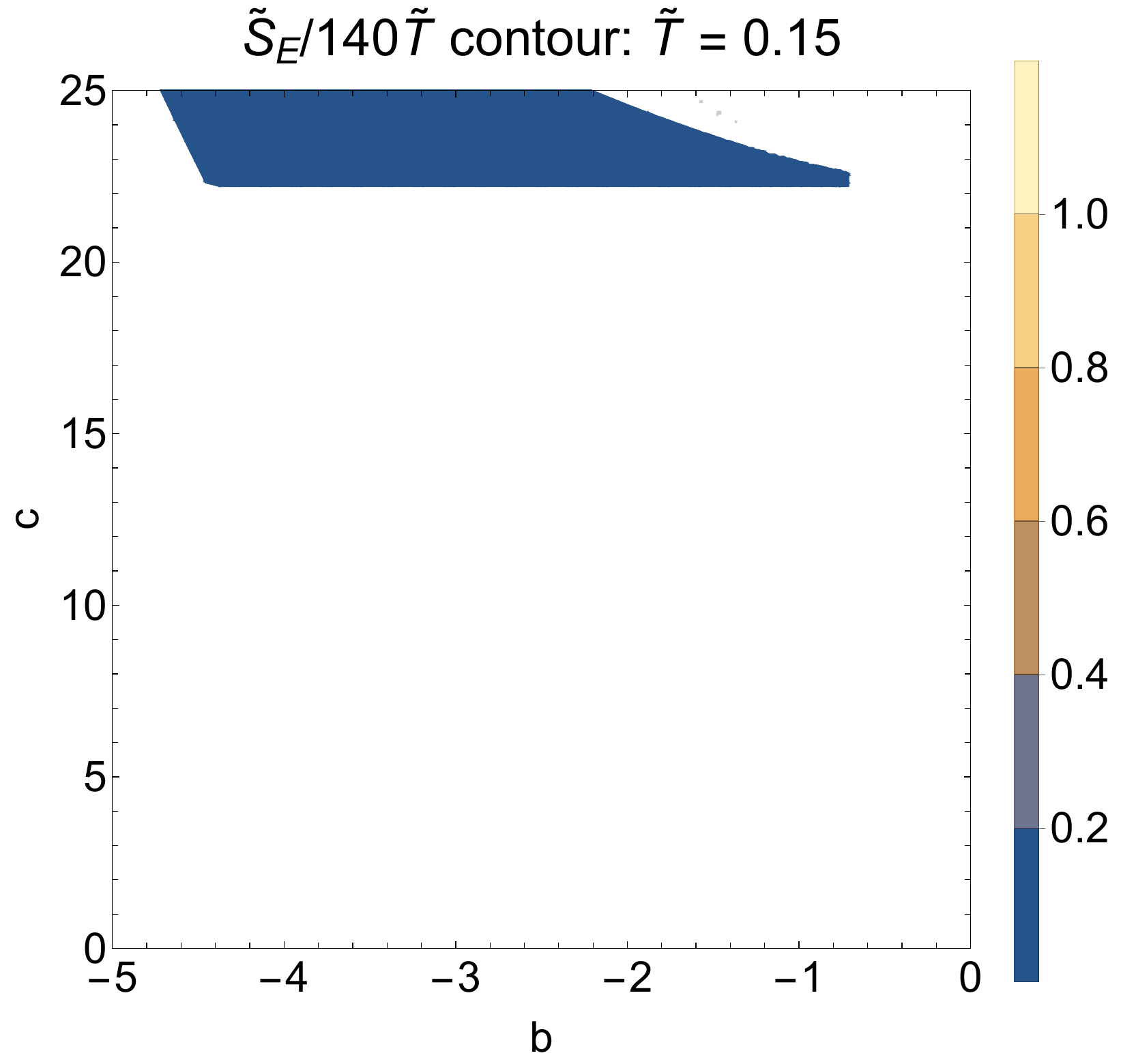}
\hfill
\includegraphics[width=.485\textwidth]{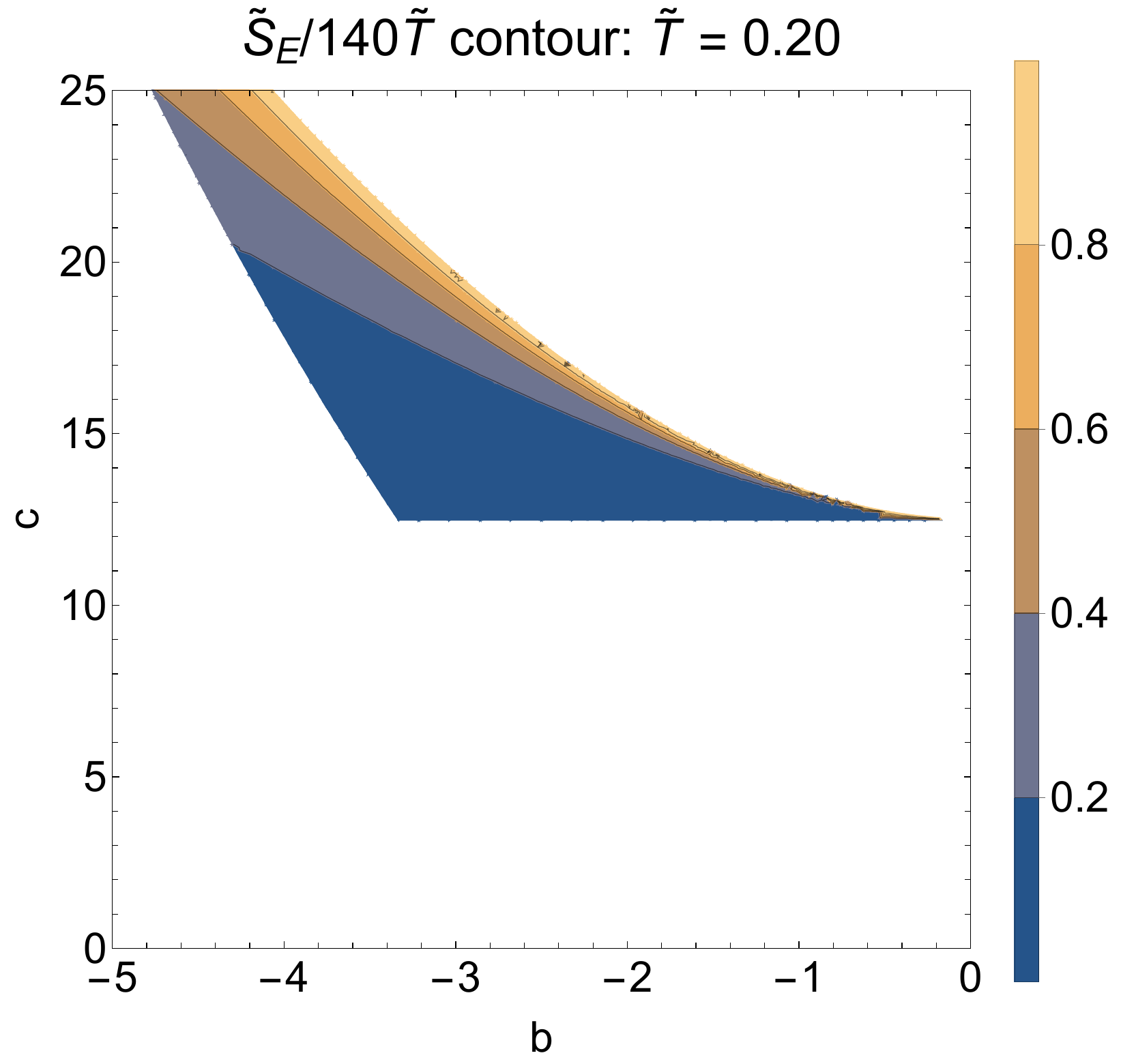}
\hfill
\includegraphics[width=.485\textwidth]{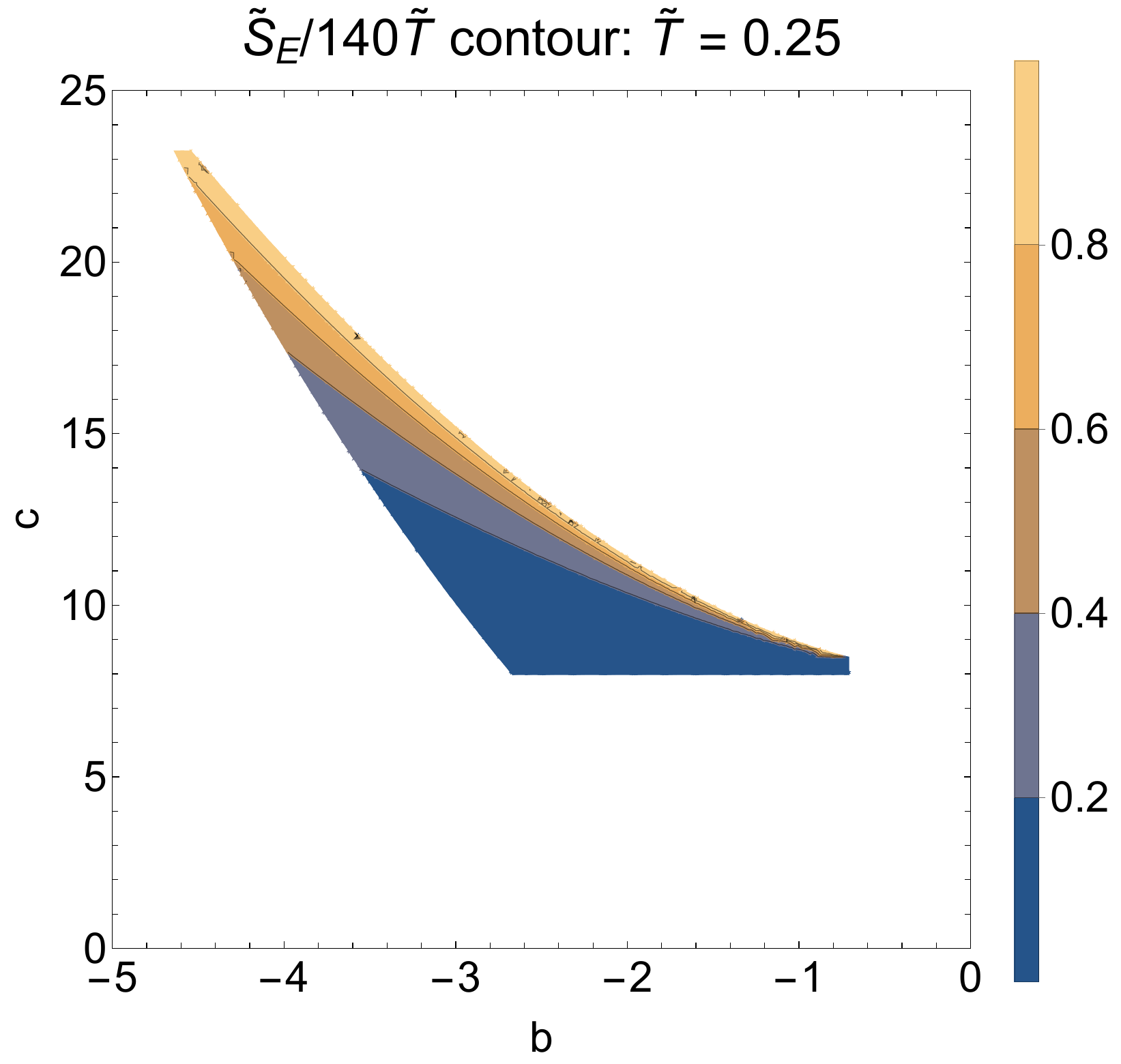}
\hfill
\includegraphics[width=.485\textwidth]{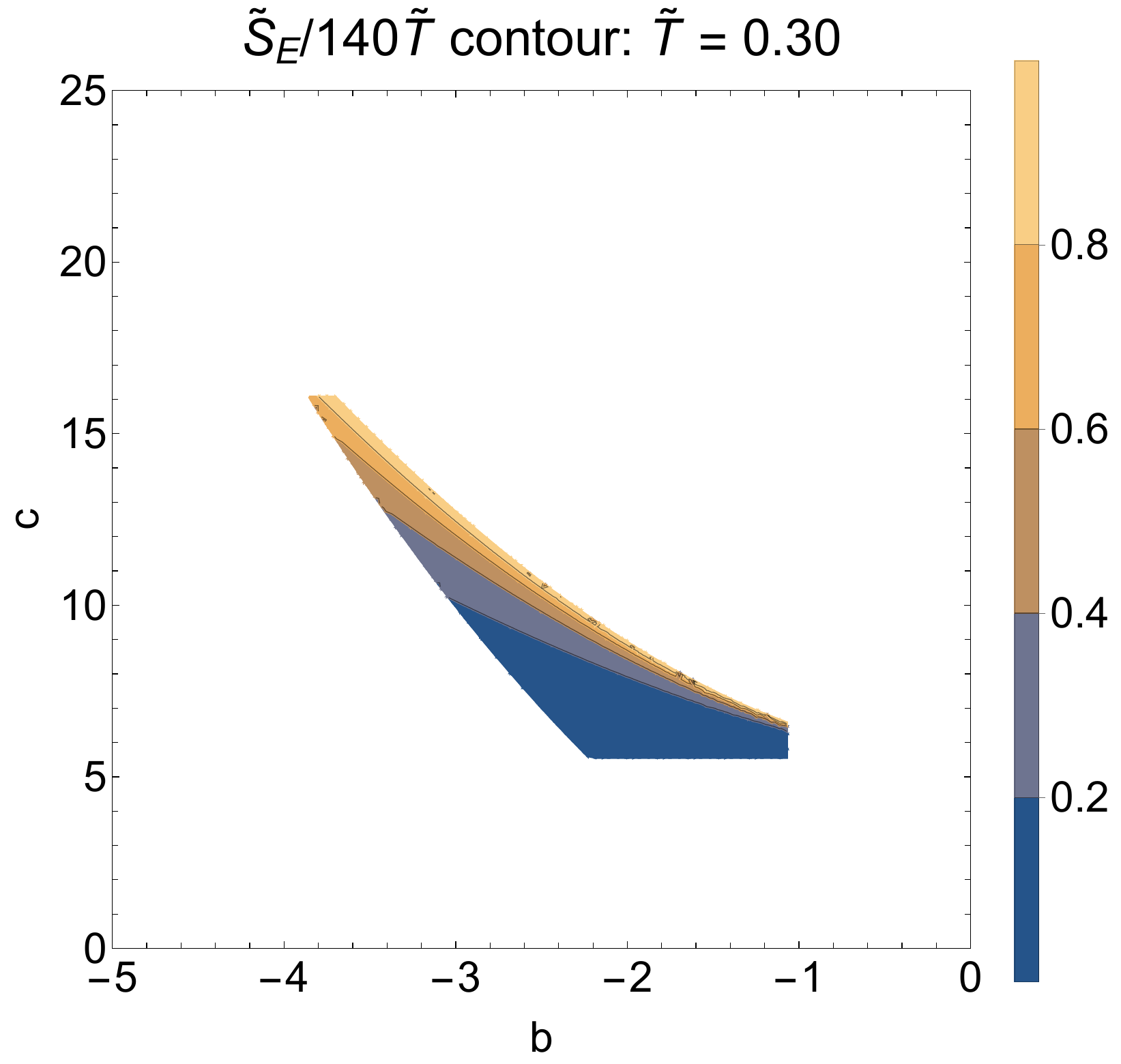}

\captionsetup{justification   = RaggedRight,
             labelfont = bf}
\caption{Contours of $\tilde{S}_E/140 \tilde{T}$ are shown in the $(b,c)$ plane for smaller values of $\tilde{T}$: $\tilde T = 0.15$ (upper left), $0.20$ (upper right), $0.25$ (lower left), and $0.30$ 
(lower right). The contours also satisfy the conditions: $c\ge b^2$, $c \tilde{T} \ge 1/2$, $\alpha \le 1 $, $0 \le \tilde{S}_E/140 \tilde{T} \le 1 $, $\tilde{\beta}/H\ge \tilde{S}_E/140 \tilde{T}$ and $\tilde{\xi} \ge 0 $. \label{fig:parameterspacelowT} } 
\end{figure}

\begin{figure}[tbp]
\centering 
\includegraphics[width=.485\textwidth]{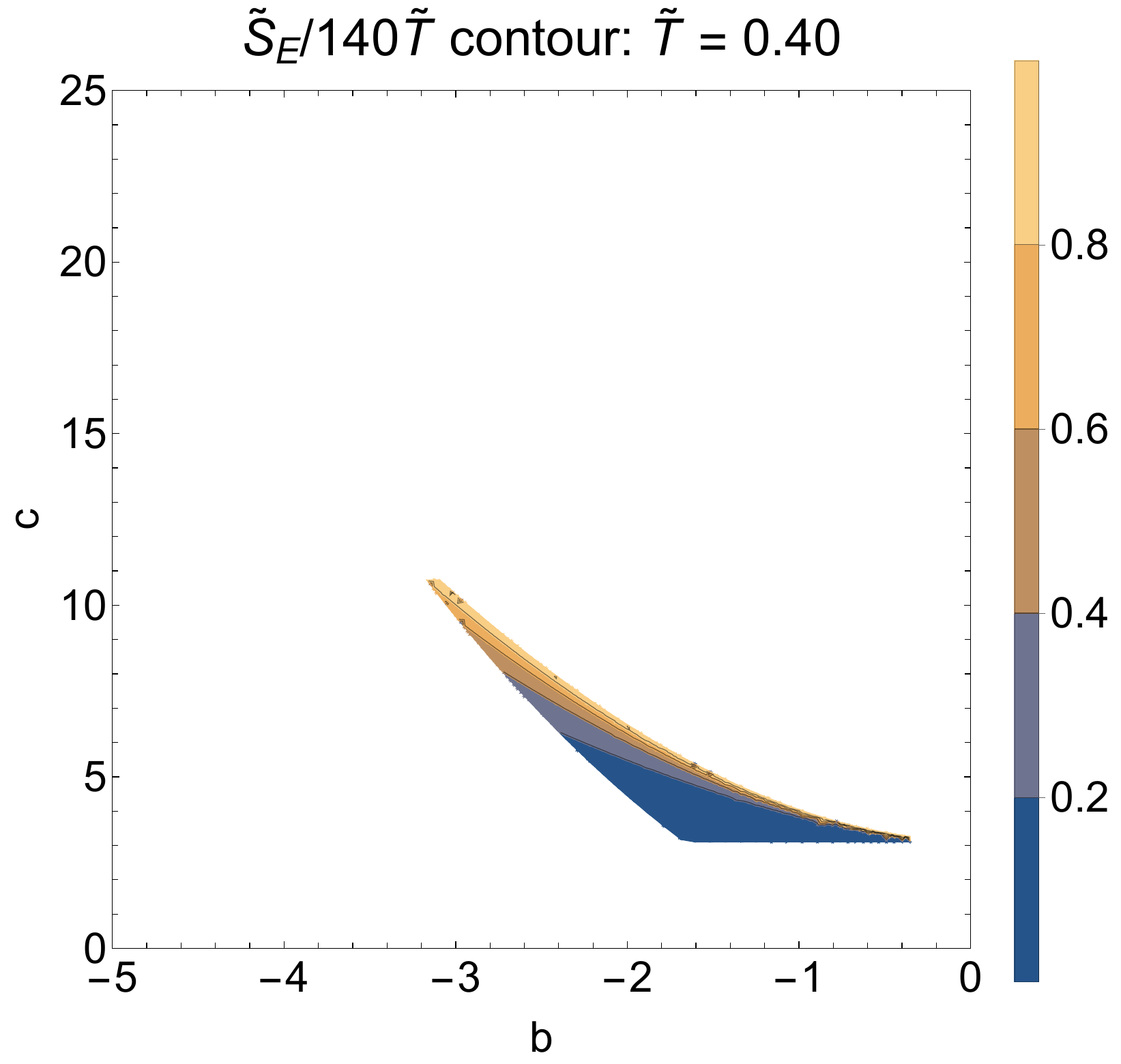}
\hfill
\includegraphics[width=.485\textwidth]{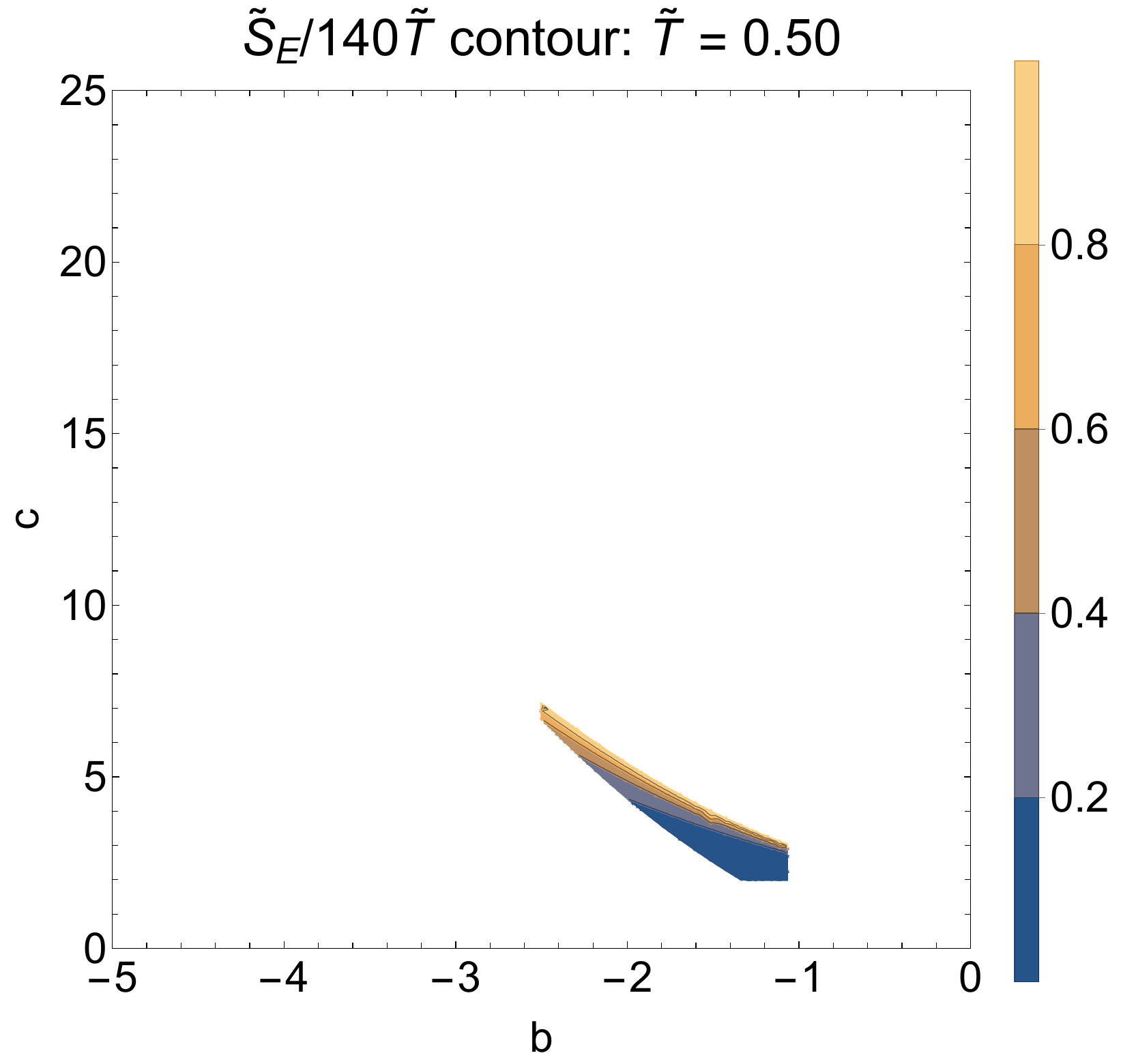}
\hfill
\includegraphics[width=.485\textwidth]{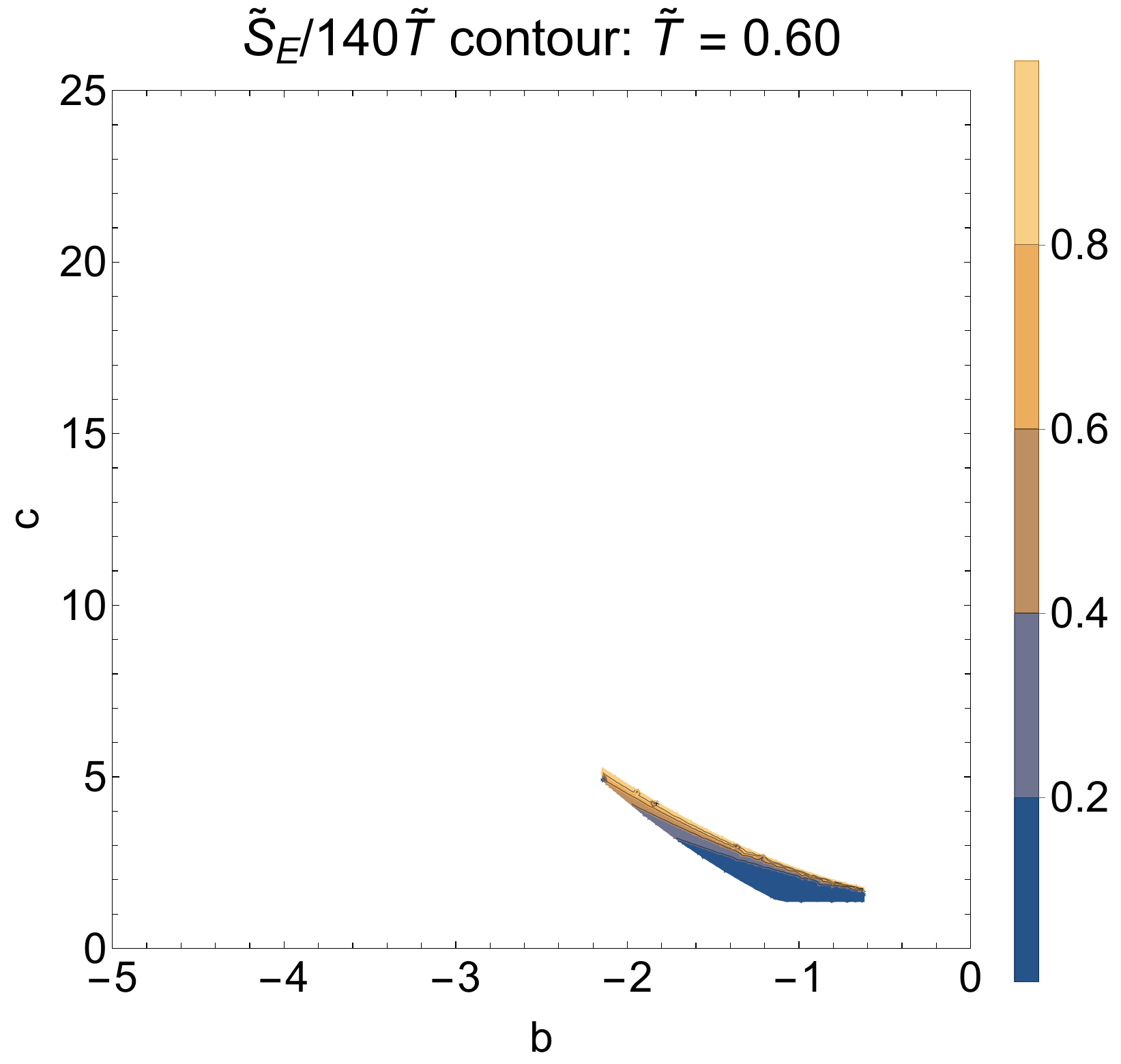}
\hfill
\includegraphics[width=.485\textwidth]{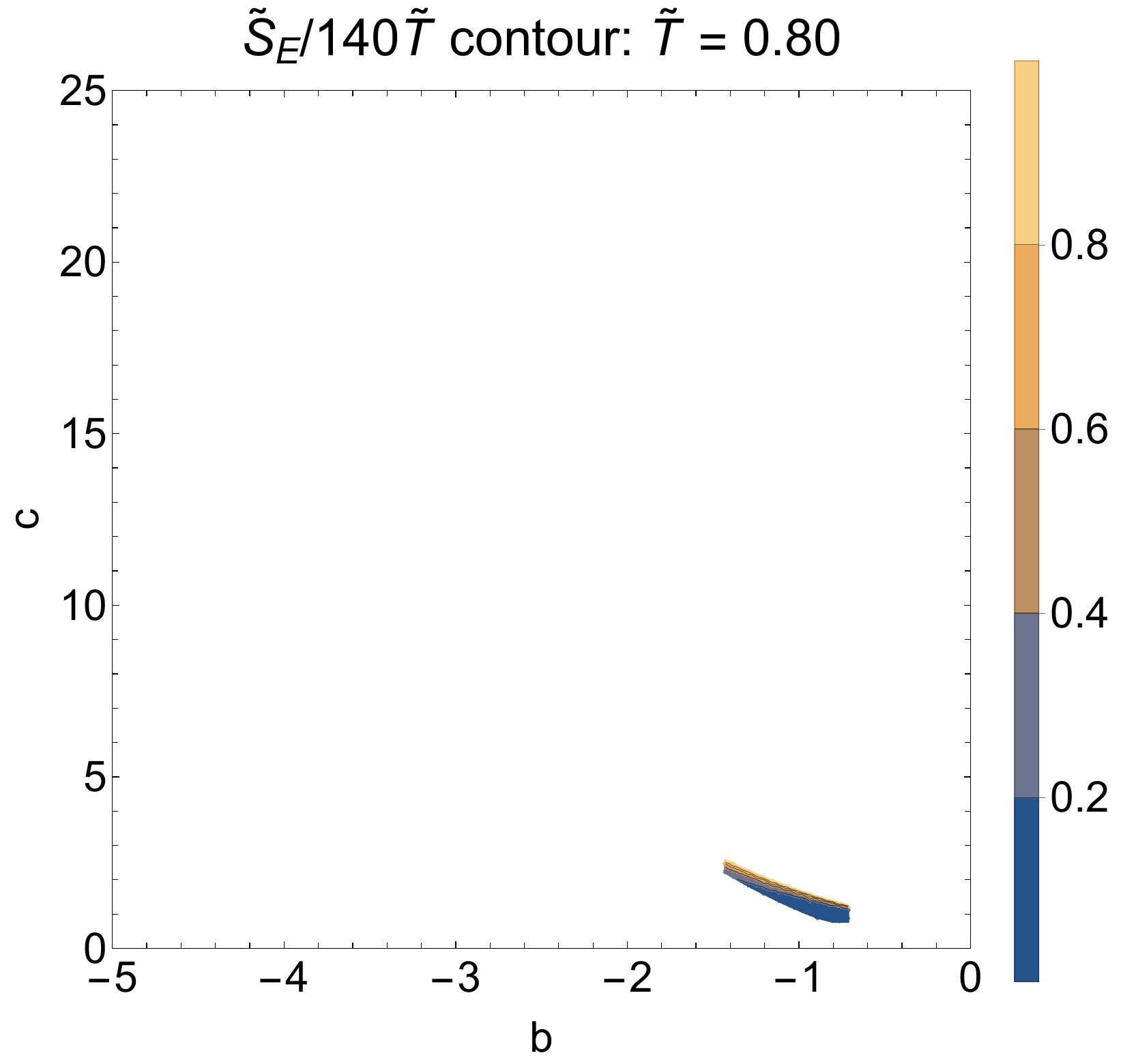}

\captionsetup{justification   = RaggedRight,
             labelfont = bf}
\caption{Contours of $\tilde{S}_E/140 \tilde{T}$ are shown in the $(b,c)$ plane for larger values of $\tilde{T}$: $\tilde T = 0.40$ (upper left), $0.50$ (upper right), $0.60$ (lower left), and $0.80$ 
(lower right).
The contours also satisfy the conditions: $c\ge b^2$, $c \tilde{T} \ge 1/2$, $\alpha \le 1 $, $0 \le \tilde{S}_E/140 \tilde{T} \le 1 $, $\tilde{\beta}/H\ge \tilde{S}_E/140 \tilde{T}$ and $\tilde{\xi} \ge 0 $. \label{fig:parameterspacehighT} } 
\end{figure}

The scale-free Euclidean action, $\tilde S_E$, is determined 
by $\tilde T$ and the parameters $b$ and $c$.
For any choice of parameters $b$, $c$ and $\Lambda / v$, 
the nucleation temperature $\tilde T_N$ is 
approximately determined by 
the solution of the equation 
\bea
\frac{\tilde S_E (b,c,\tilde T_N)}{140 \tilde T_N} 
&=& \left(\frac{\Lambda}{v} \right)^2.
\label{eqn:TN}
\eea
In Figures \ref{fig:parameterspacelowT} and 
\ref{fig:parameterspacehighT}, we plot contours 
of $\tilde S_E (b, c, \tilde T) / 140 \tilde T$  in the $(b,c)$-plane, 
for eight different choices of $\tilde T$.  
We can then solve eqn.~\ref{eqn:TN} by setting $(\Lambda / v)^2$ equal to the 
$\tilde S_E / 140 \tilde T$, for a value of $\tilde T$ which is taken as the 
nucleation temperature. 
These figures thus 
encode all choices of the parameters 
$(b,c,\Lambda / v)$ for which the scale-free 
nucleation temperature 
is given by a particular choice $\tilde T_N$.
Note, we only plot parameters 
points which are of interest, as defined by the criteria 
described at the beginning of this section.


Note that the range of parameter space 
which is of interest is very small for 
$\tilde T_N \sim 1$, near $(b,c) \sim (-\sqrt{2},2)$.  
More parameter space is available for $\tilde T_N < 1$, 
and for relatively large $\tilde T_N$, the available 
parameter space is focused near the curve $c = b^2$.
But as $\tilde T_N$ becomes small, the high temperature 
approximation becomes less valid; in that case, it is not 
clear if this effective potential can be realized from a 
UV complete model.  In particular, note that, although we 
do not consider the range $c > 25$ for reasons of computational 
ease, that range in any case corresponds to small $\tilde T_N$.

\begin{figure}[tbp]
\centering 
\includegraphics[width=.492\textwidth]{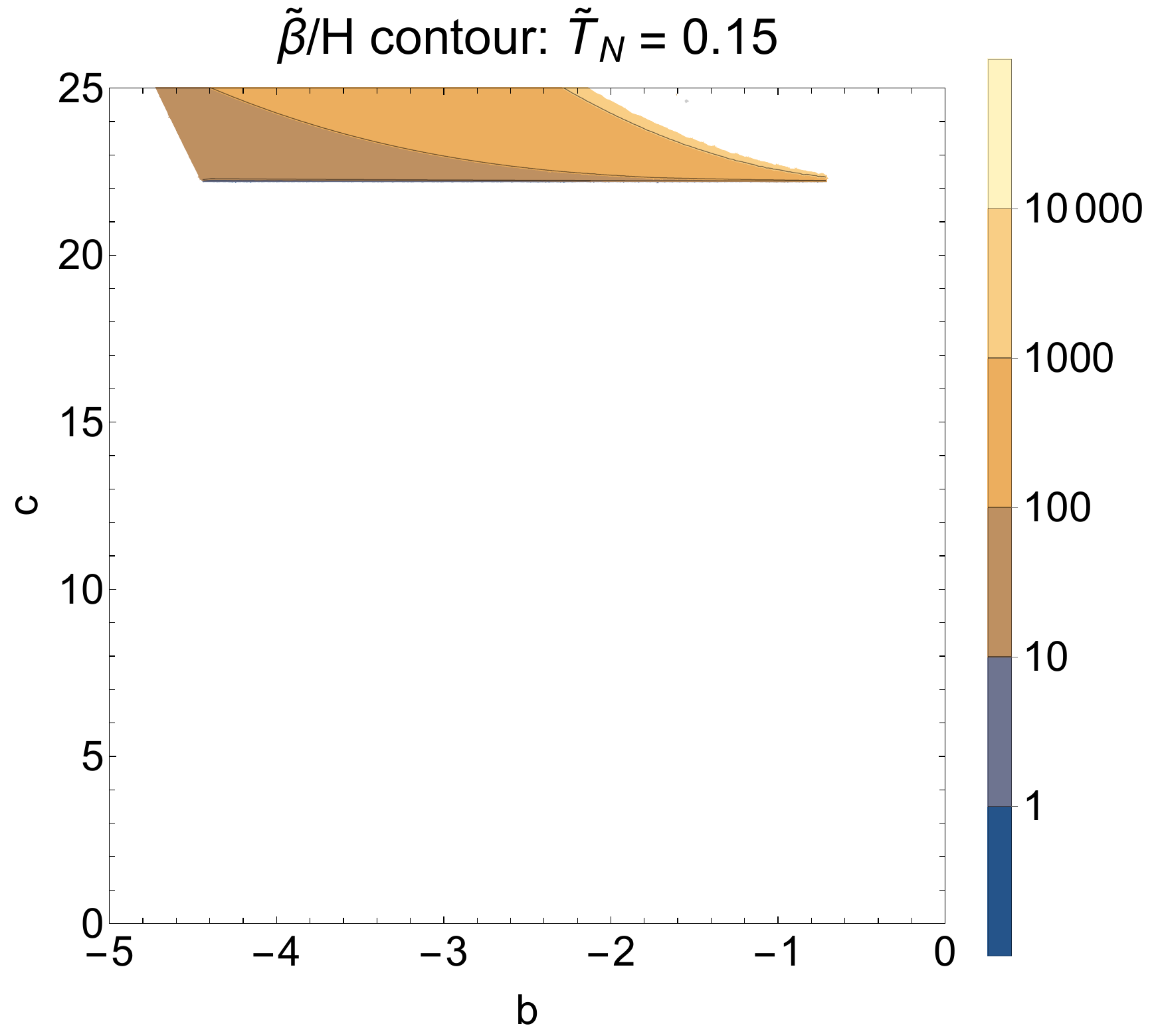}
\hfill
\includegraphics[width=.485\textwidth]{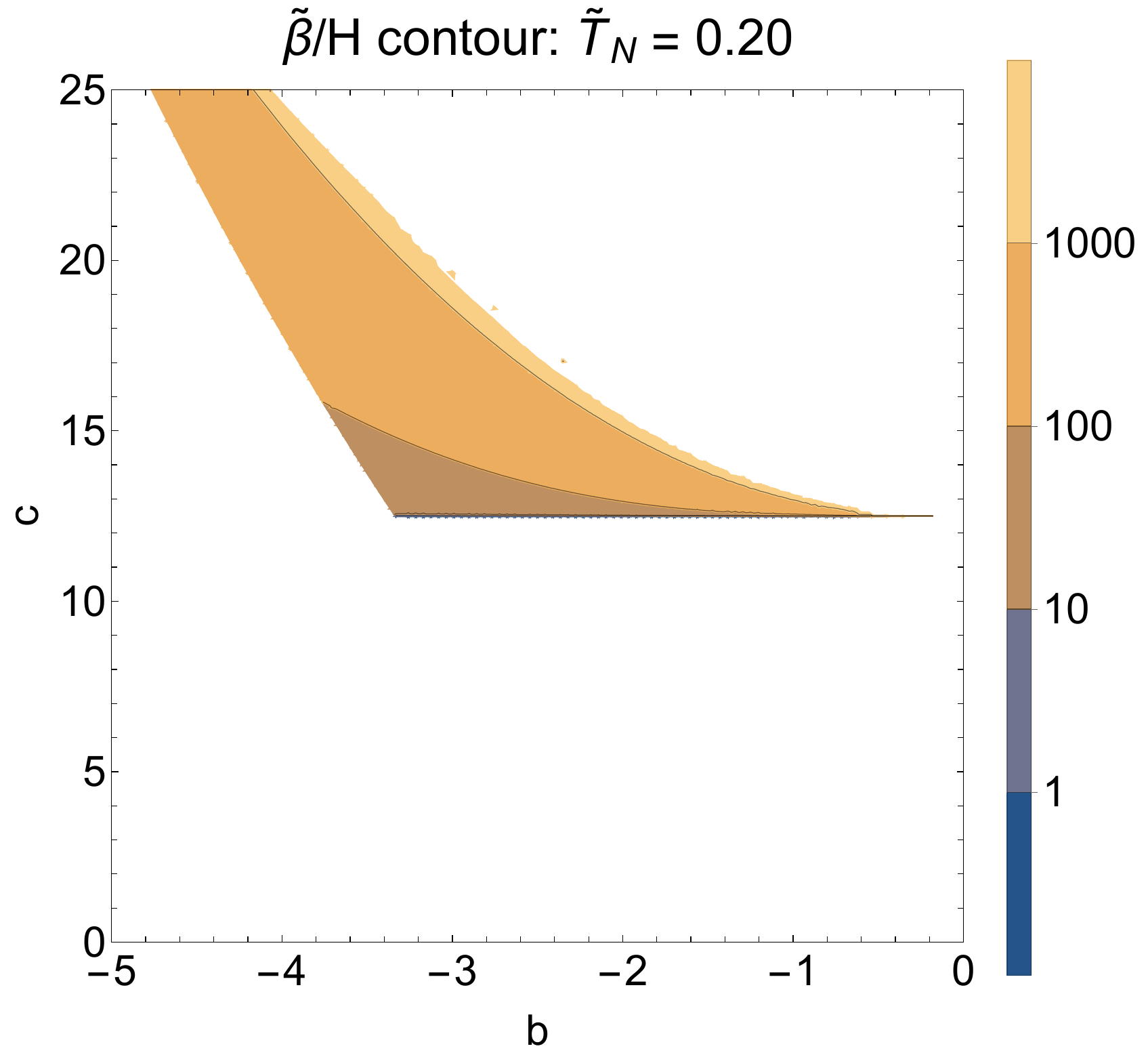}
\hfill
\includegraphics[width=.49\textwidth]{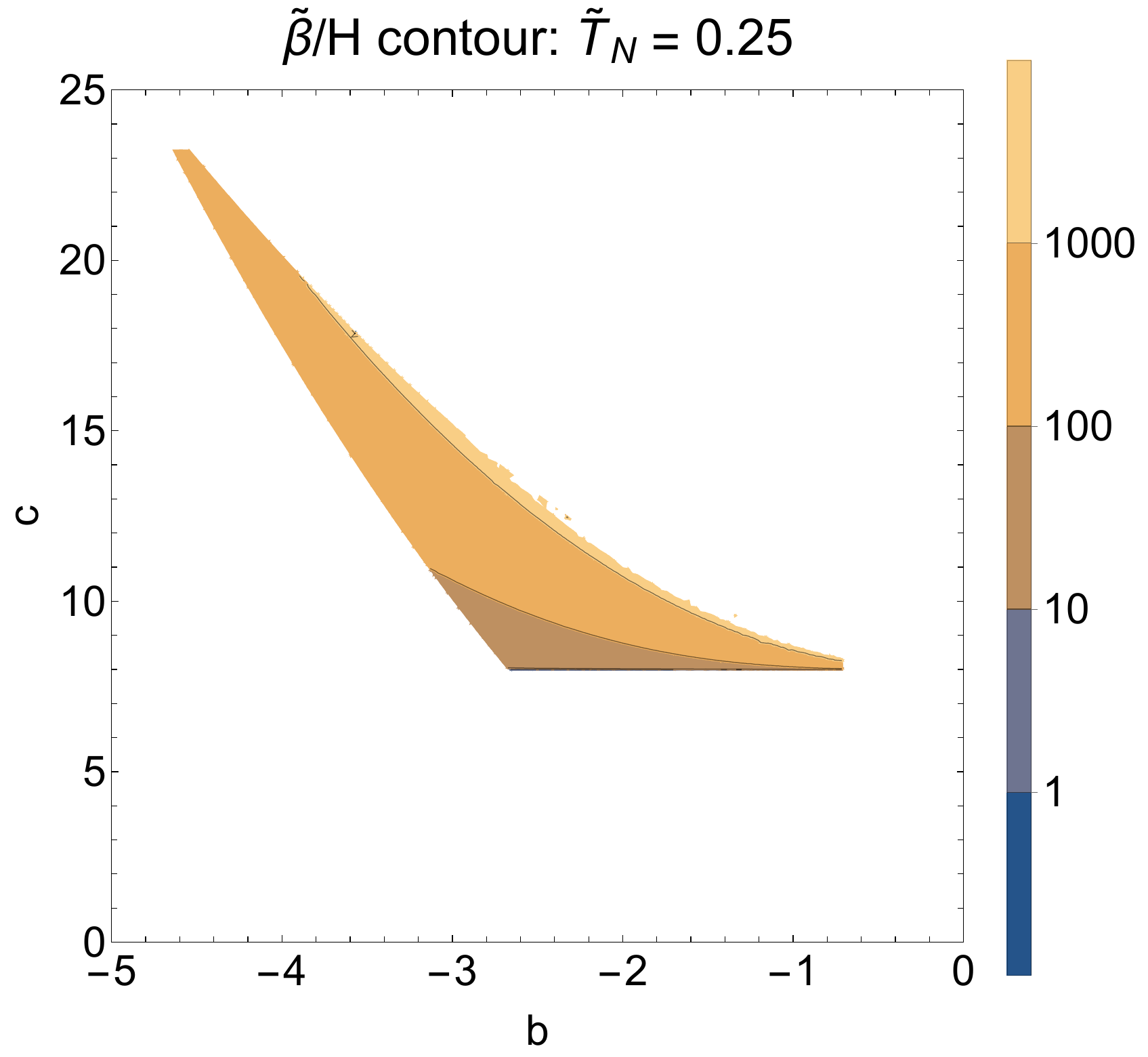}
\hfill
\includegraphics[width=.49\textwidth]{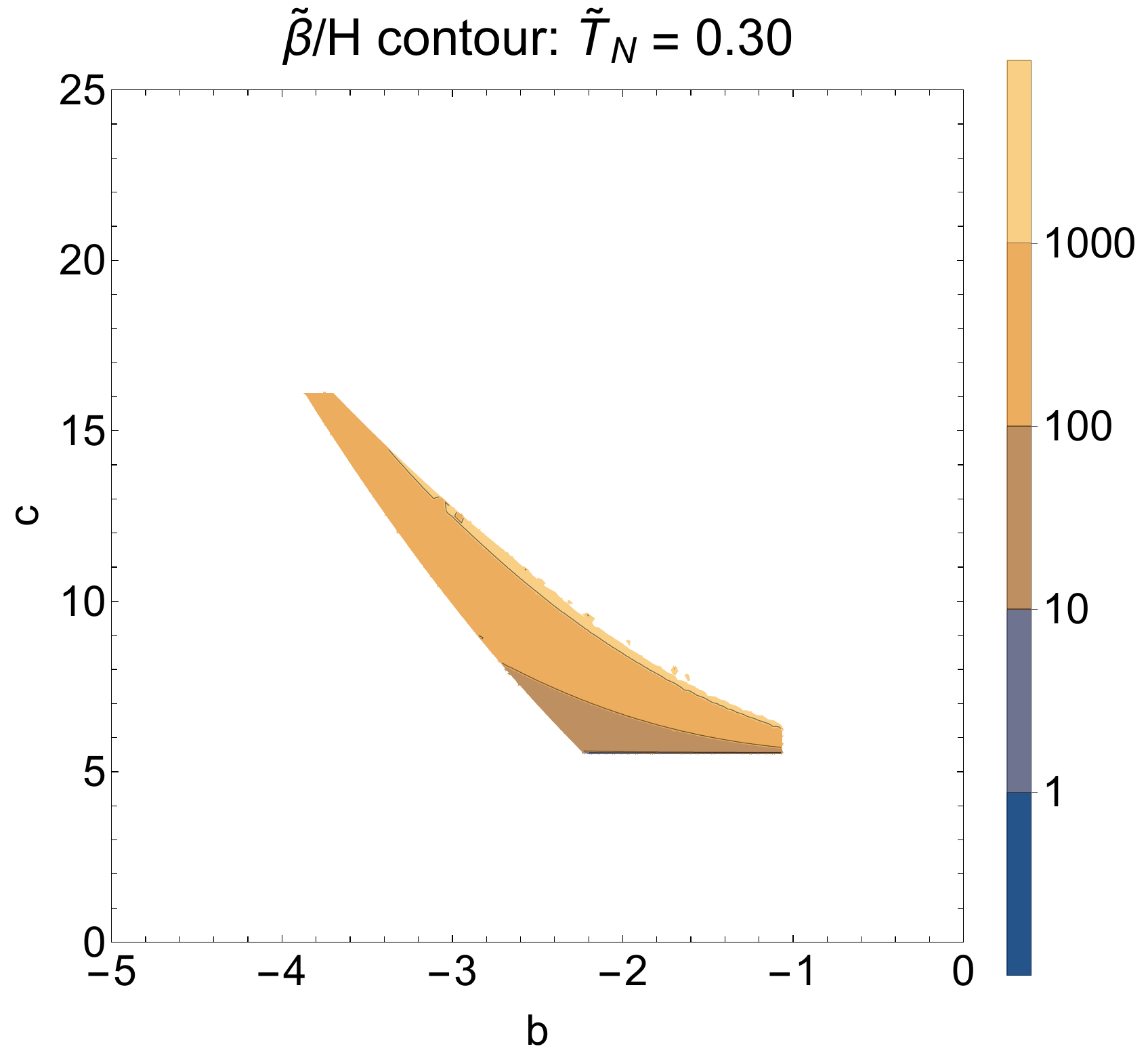}

\captionsetup{justification   = RaggedRight,
             labelfont = bf}
\caption{Contours of $\tilde{\beta}/H$ 
are shown in the $(b,c)$ plane for the allowed parameter space  
for lower values of $\tilde{T}_N$: $\tilde T_N = 0.15$ (upper left), $0.20$ (upper right), $0.25$ (lower left), and $0.30$ 
(lower right).
The contours also satisfy the conditions: $c\ge b^2$, $c \tilde{T}_N \ge 1/2$, $\alpha \le 1 $, $0 \le \tilde{S}_E/140 \tilde{T}_N \le 1 $, $\tilde{\beta}/H\ge \tilde{S}_E/140 \tilde{T}_N$ and $\tilde{\xi} \ge 0 $. 
\label{fig:betaoverHplotlowT} }  
\end{figure}

\begin{figure}[tbp]
\centering 
\includegraphics[width=.49\textwidth]{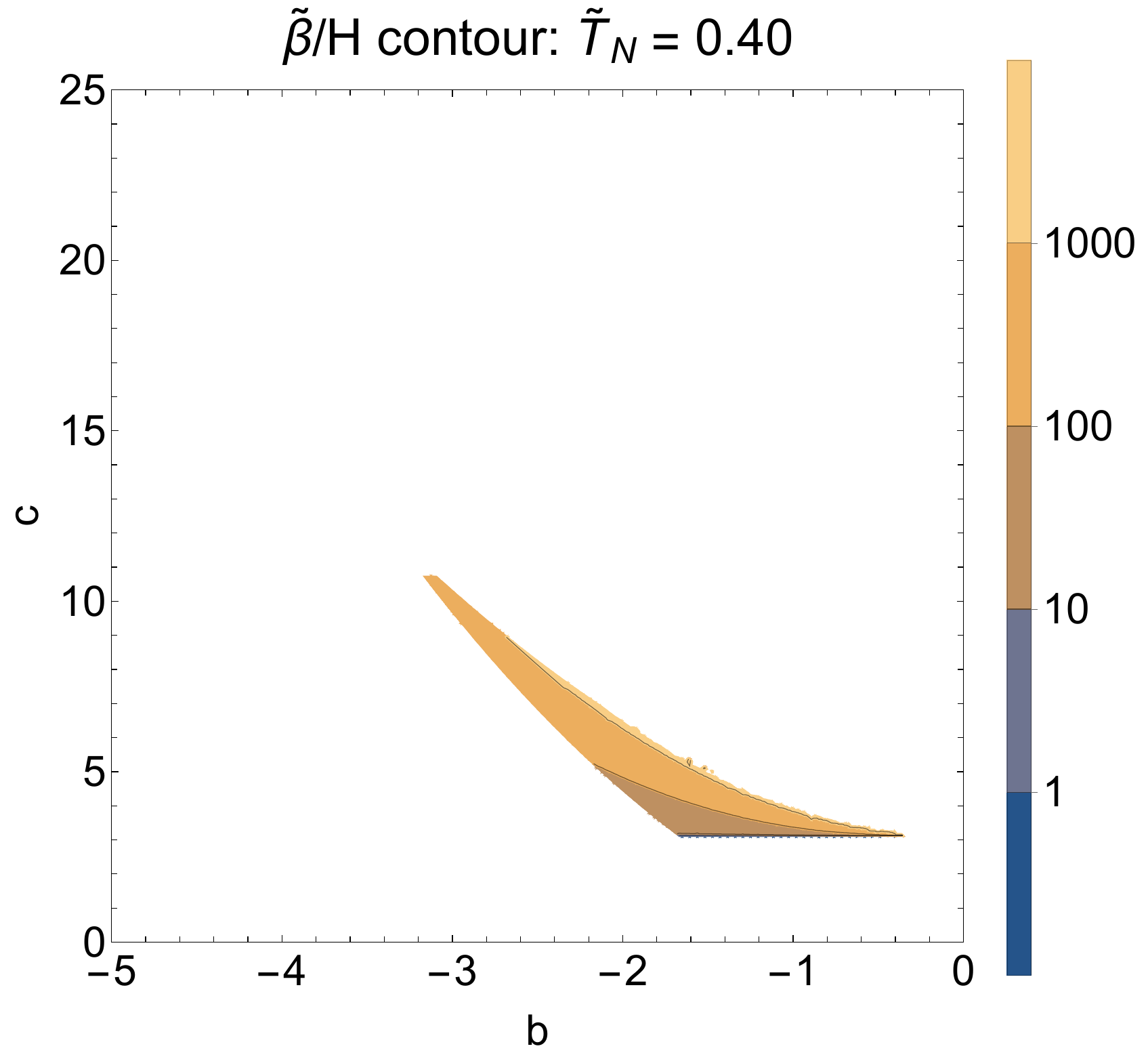}
\hfill
\includegraphics[width=.49\textwidth]{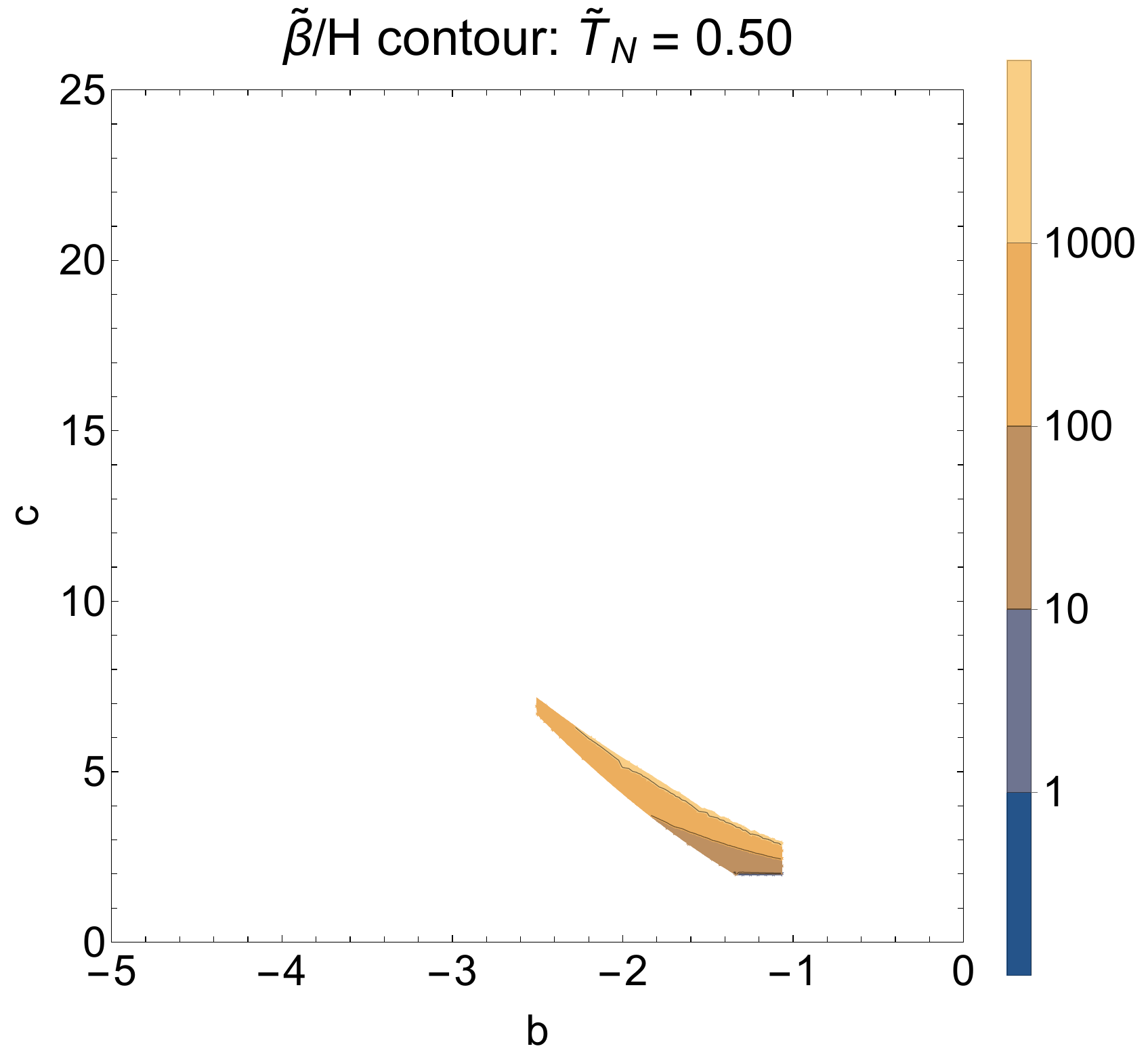}
\hfill
\includegraphics[width=.49\textwidth]{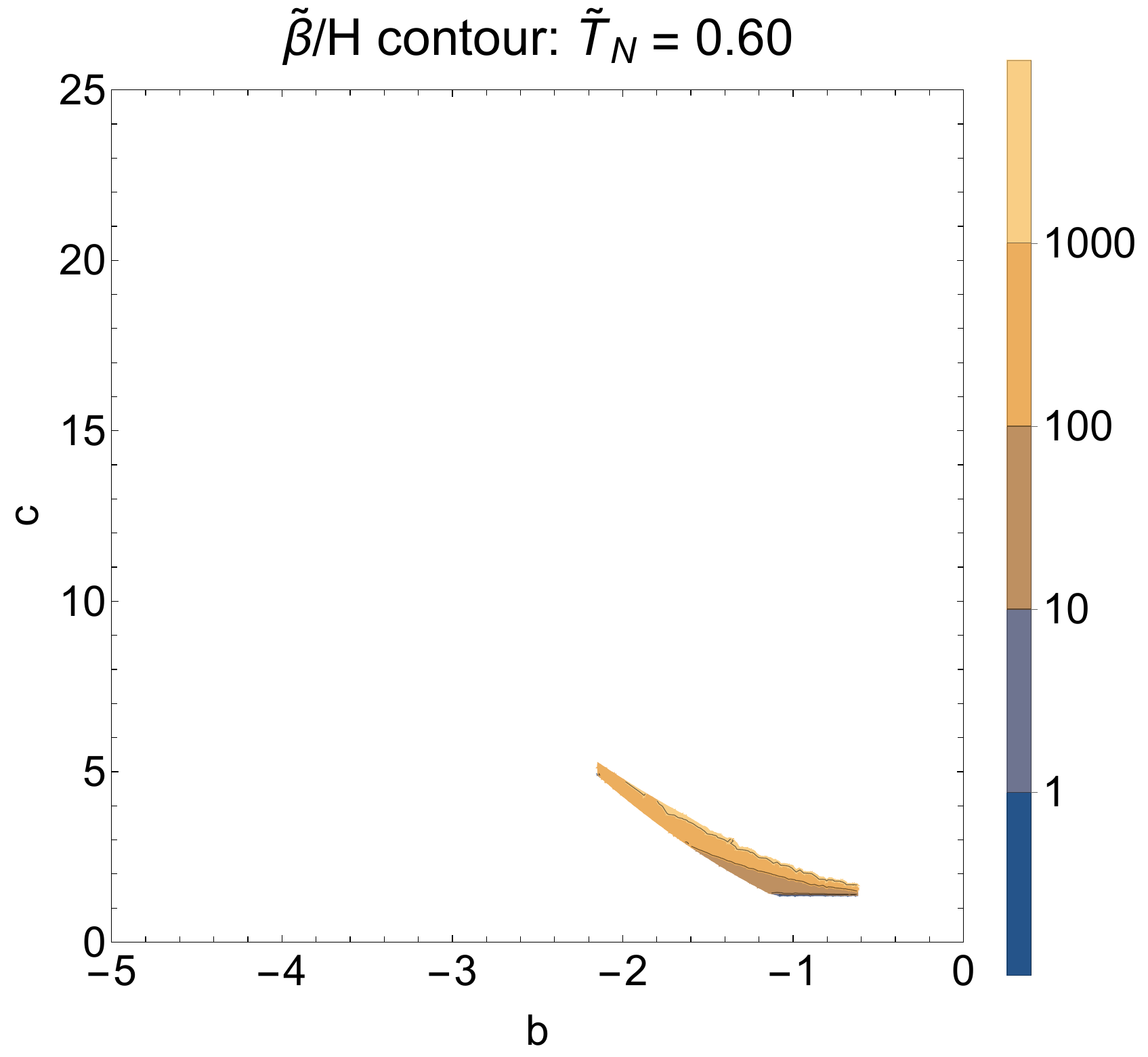}
\hfill
\includegraphics[width=.49\textwidth]{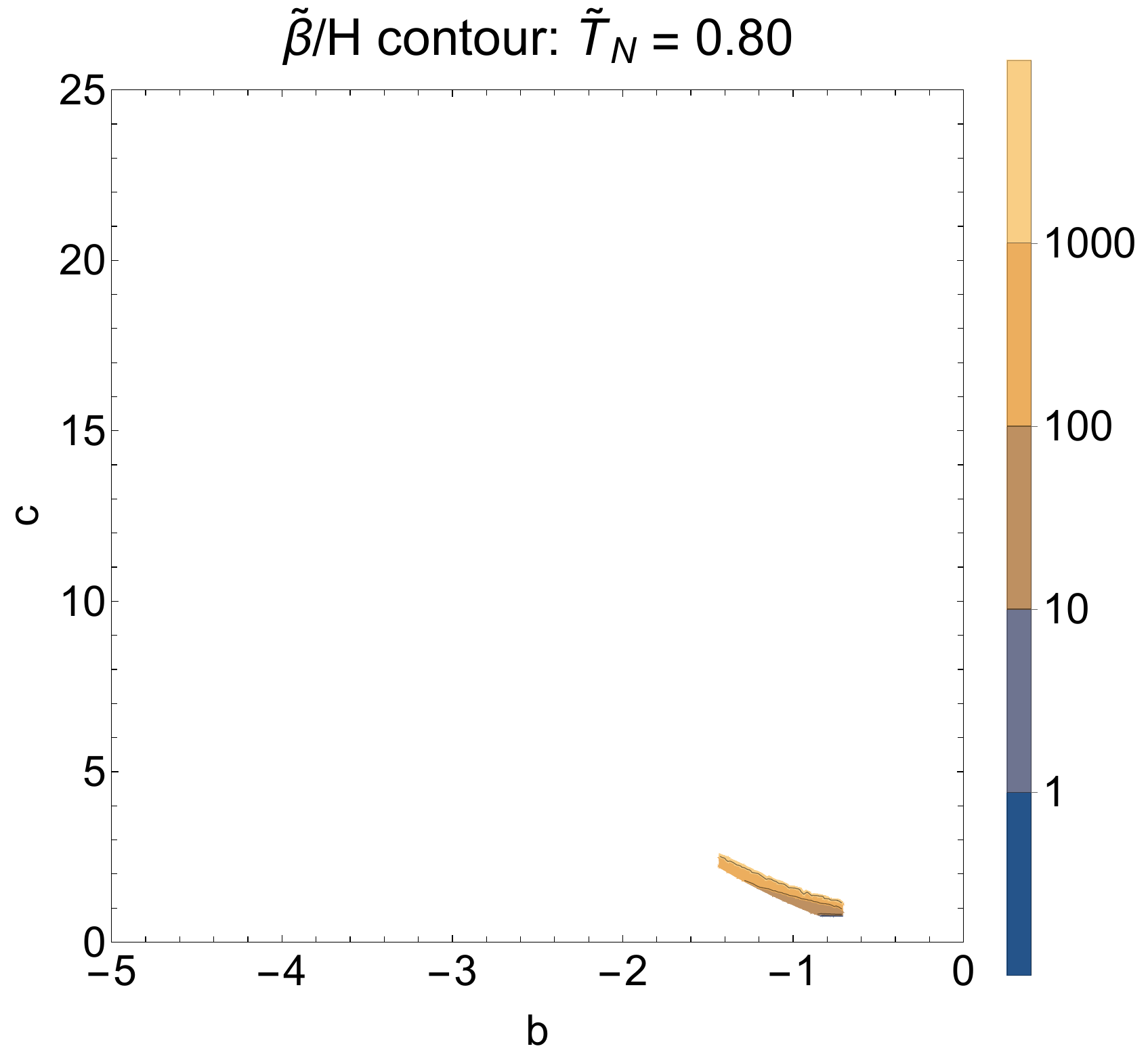}

\captionsetup{justification   = RaggedRight,
             labelfont = bf}
\caption{ Contours of $\tilde{\beta}/H$ 
are shown in the $(b,c)$ plane for the allowed parameter space  
for higher values of $\tilde{T}_N$: $\tilde T_N = 0.40$ (upper left), $0.50$ (upper right), $0.60$ (lower left), and $0.80$ 
(lower right).
The contours also satisfy the conditions: $c\ge b^2$, $c \tilde{T}_N \ge 1/2$, $\alpha \le 1 $, $0 \le \tilde{S}_E/140 \tilde{T}_N \le 1 $, $\tilde{\beta}/H\ge \tilde{S}_E/140 \tilde{T}_N$ and $\tilde{\xi} \ge 0 $. 
\label{fig:betaoverHplothighT} } 
\end{figure}

In Figures~\ref{fig:betaoverHplotlowT} 
and~\ref{fig:betaoverHplothighT}, we plot contours 
of $\tilde \beta / H$ in the $(b,c)$-plane, for 
different choices of $\tilde T_N$.  Only parameter points 
satisfying the criteria described in the beginning of this 
section are considered.  In particular, for any choice 
of $\tilde T_N$, only points in the $(b,c)$-plane are 
shown for which there exists some choice of $\Lambda / v$ 
consistent with the given $\tilde T_N$.


\begin{figure}[tbp]
\centering 
\includegraphics[width=.455\textwidth]{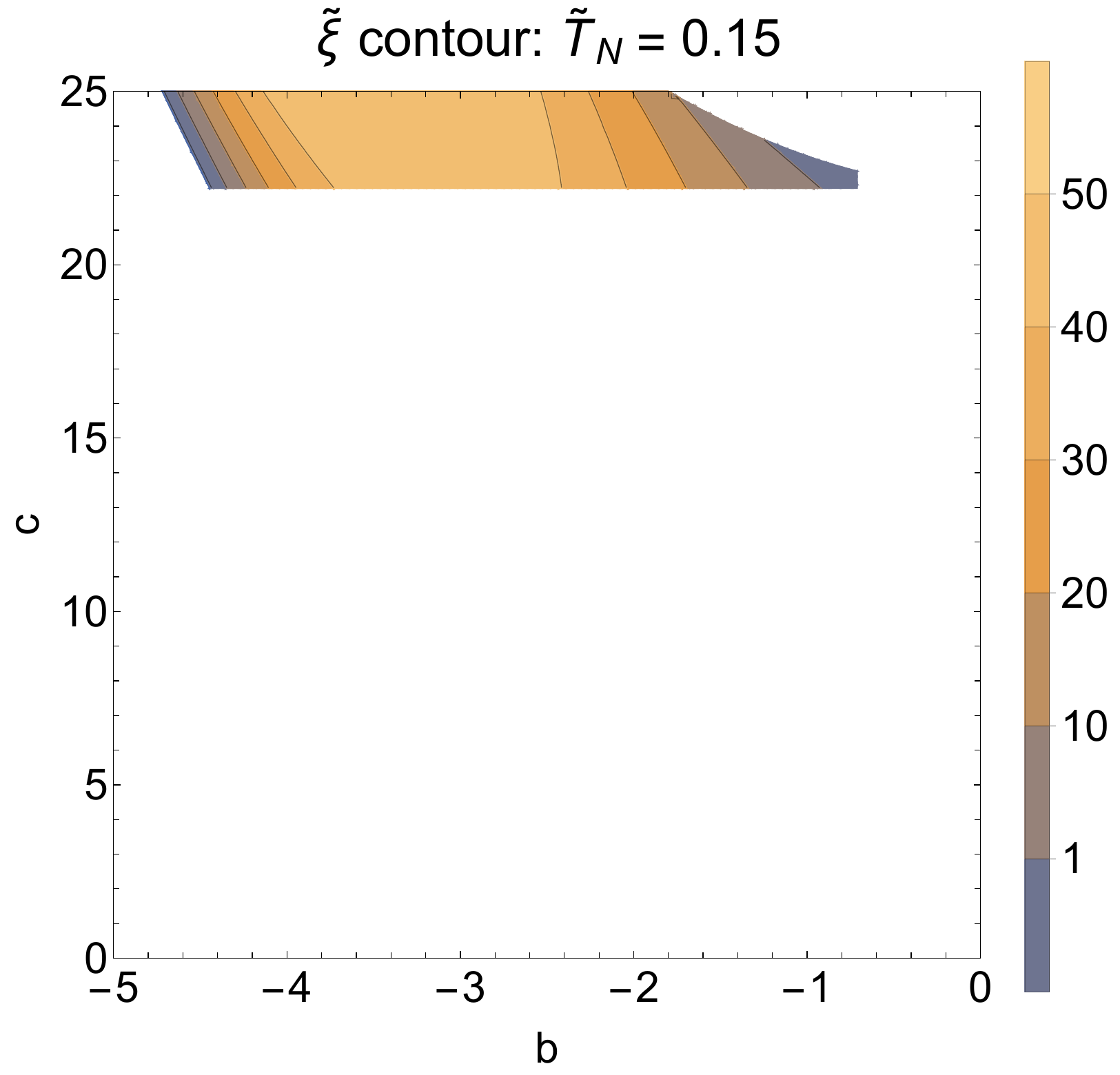}
\hfill
\includegraphics[width=.49\textwidth]{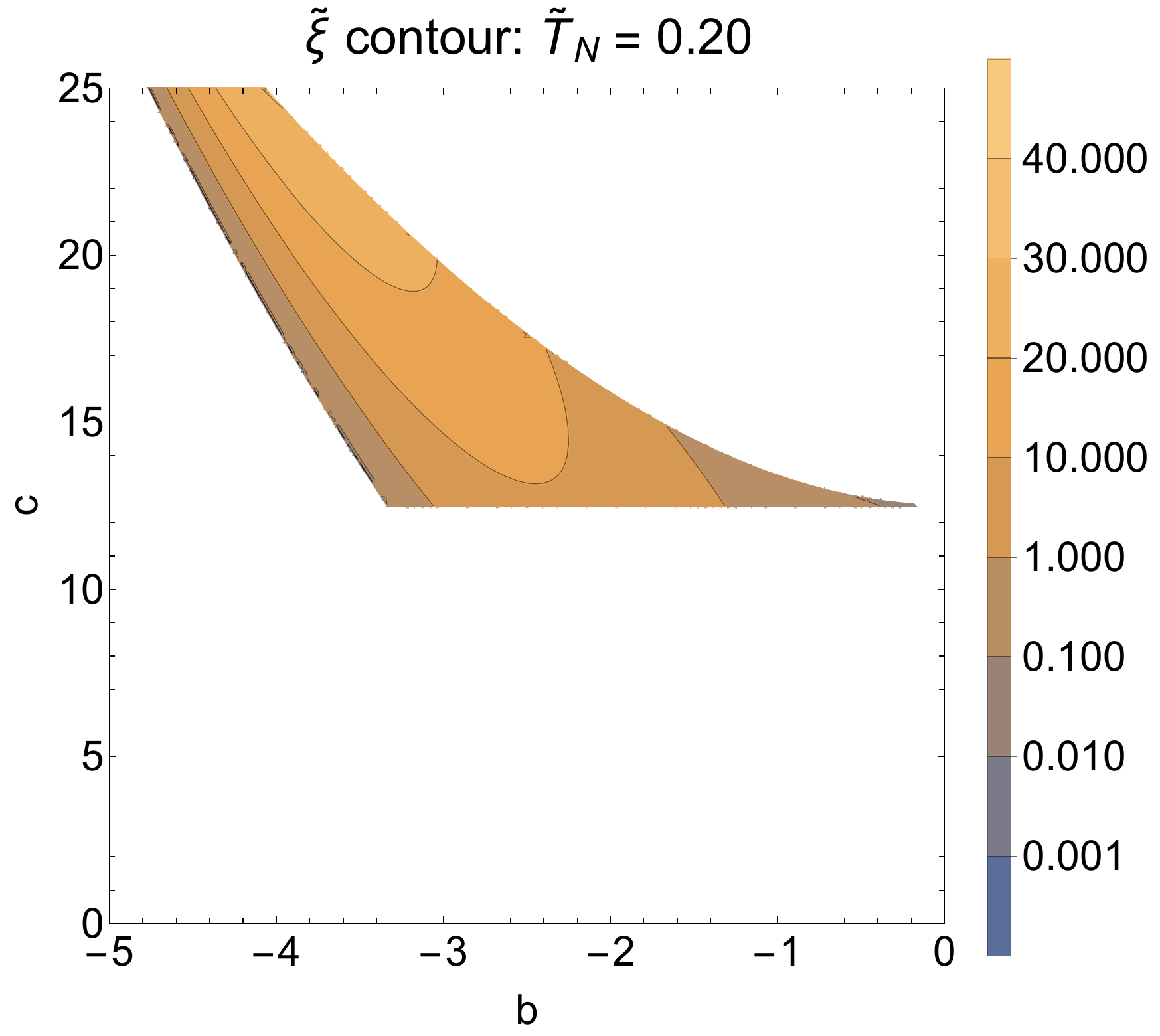}
\hfill
\includegraphics[width=.49\textwidth]{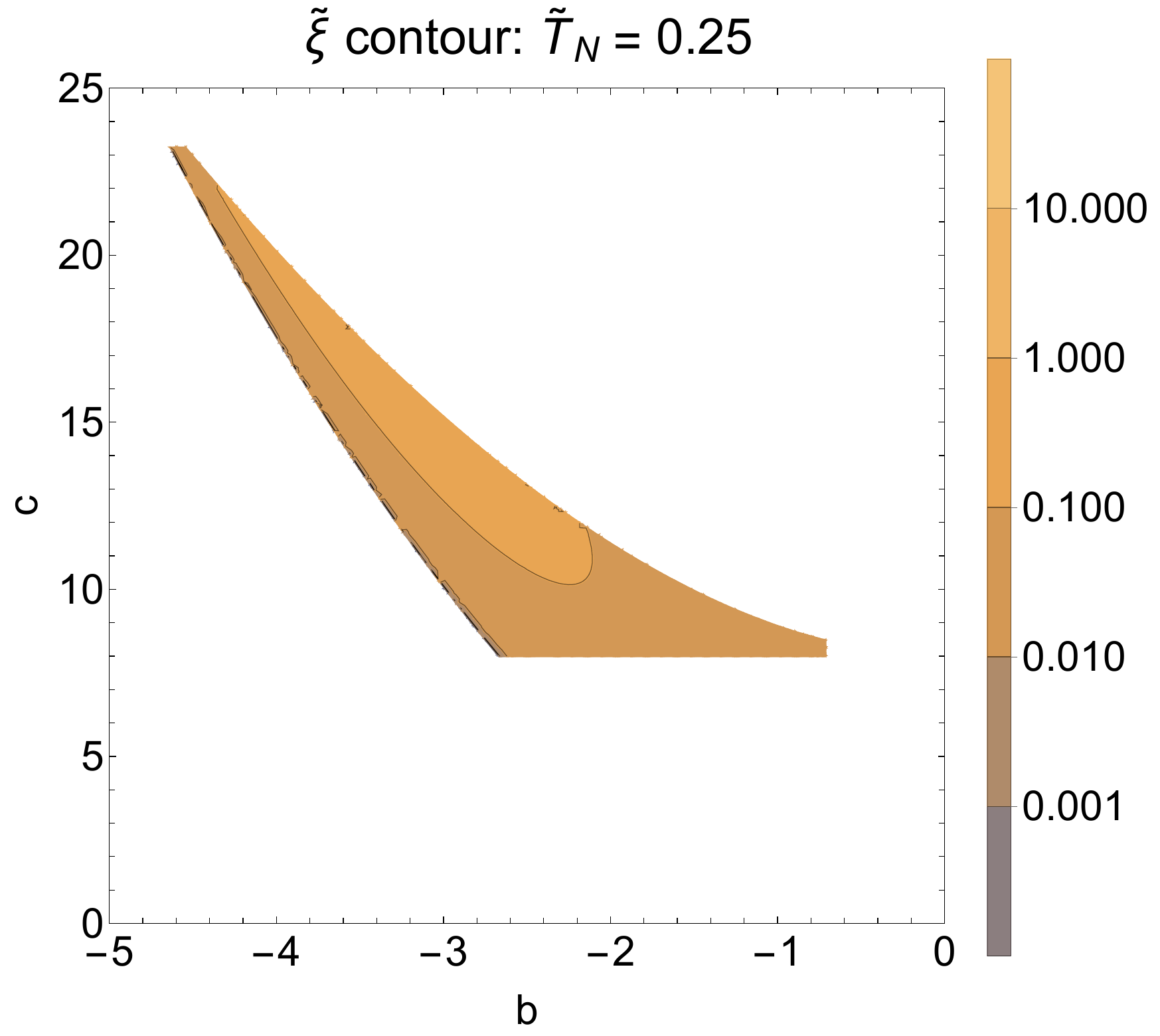}
\hfill
\includegraphics[width=.49\textwidth]{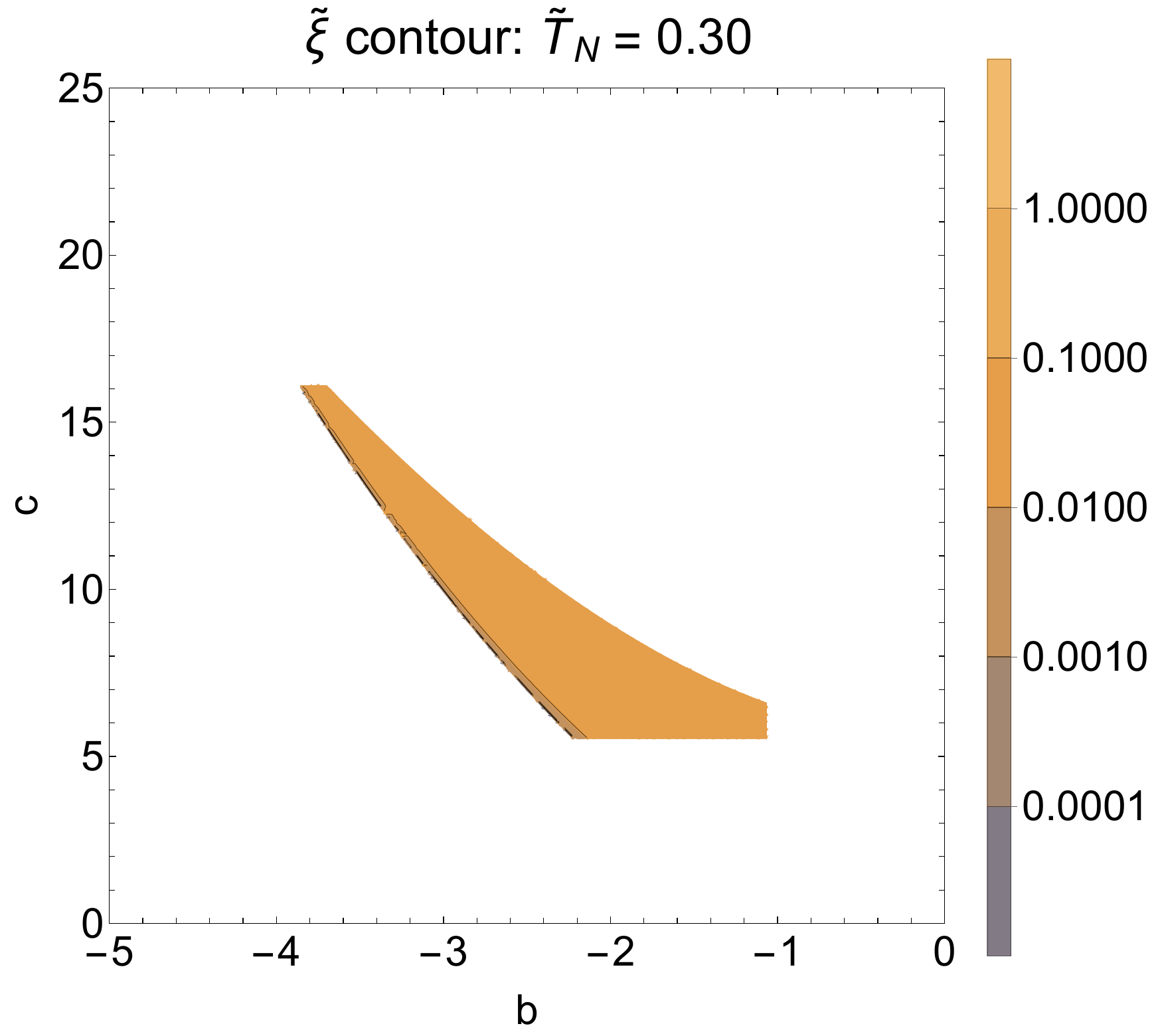}

\captionsetup{justification   = RaggedRight,
             labelfont = bf}
\caption{Contours of $\tilde \xi$ 
are shown in the $(b,c)$ plane for the allowed parameter space  
for lower values of $\tilde{T}_N$: $\tilde T_N = 0.15$ (upper left), $0.20$ (upper right), $0.25$ (lower left), and $0.30$ 
(lower right).
The contours also satisfy the conditions: $c\ge b^2$, $c \tilde{T}_N \ge 1/2$, $\alpha \le 1 $, $0 \le \tilde{S}_E/140 \tilde{T}_N \le 1 $, $\tilde{\beta}/H\ge \tilde{S}_E/140 \tilde{T}_N$ and $\tilde{\xi} \ge 0 $. 
\label{fig:xicontourplotlowT} }  
\end{figure}

\begin{figure}[tbp]
\centering 
\includegraphics[width=.485\textwidth]{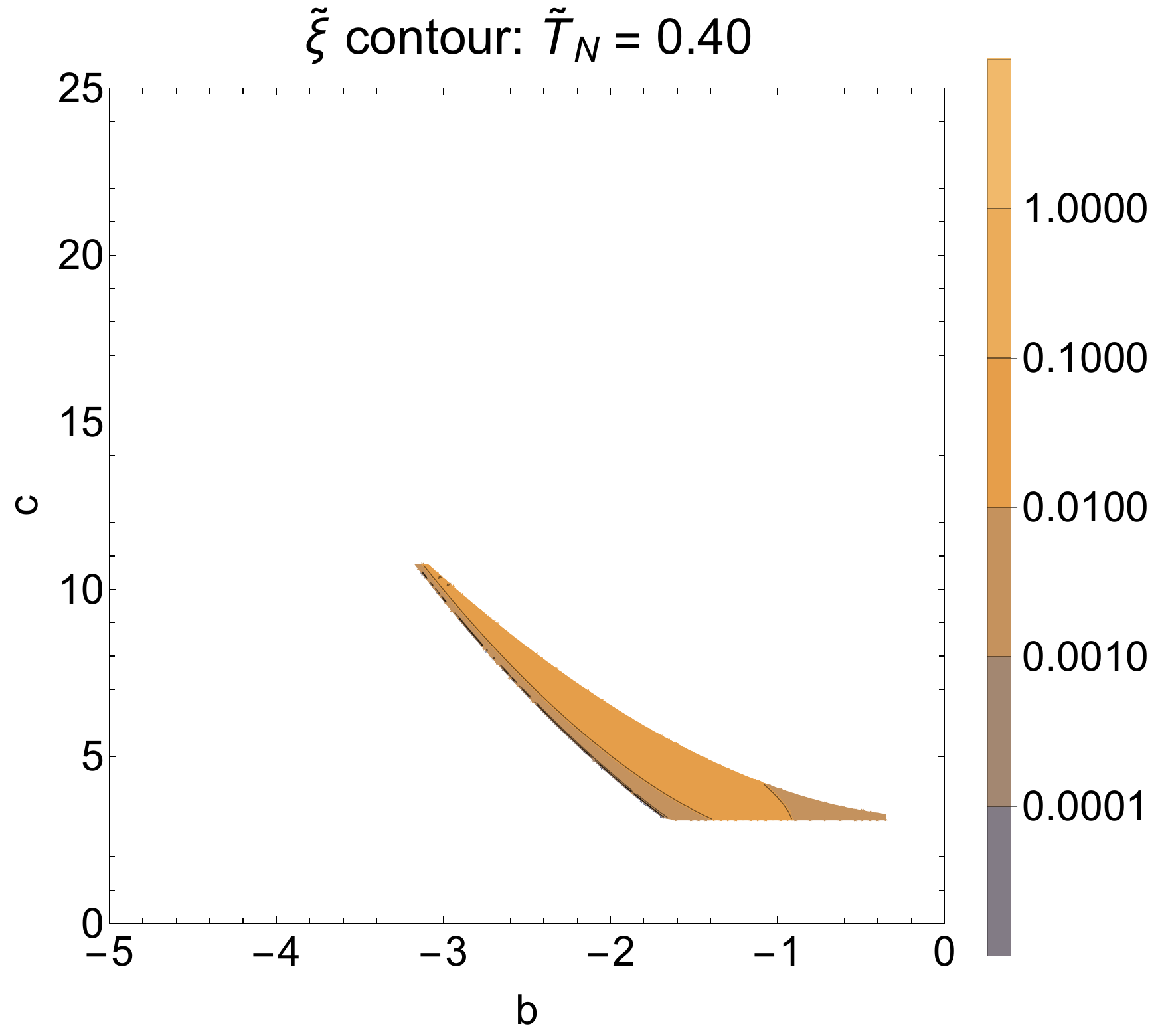}
\hfill
\includegraphics[width=.485\textwidth]{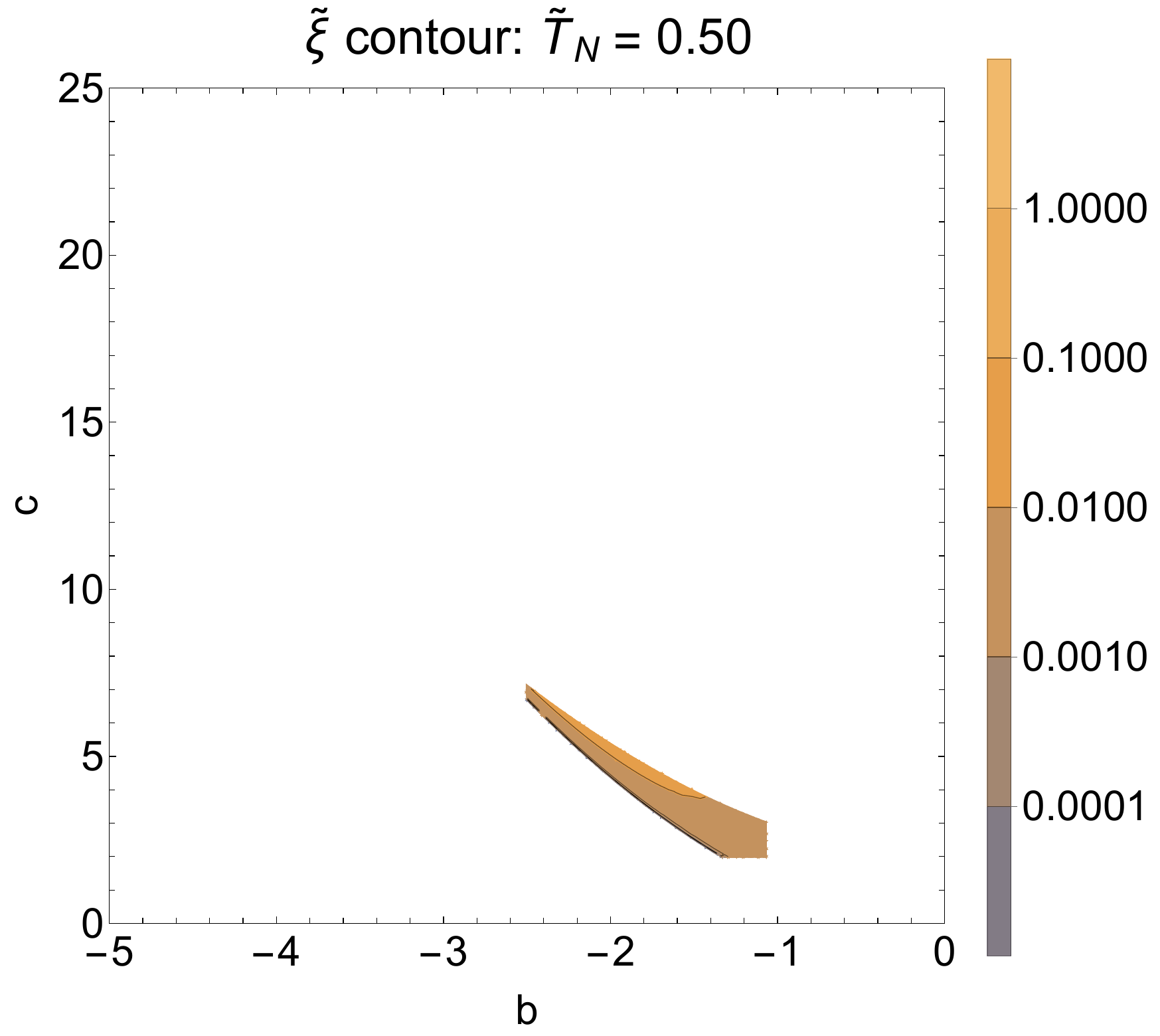}
\hfill
\includegraphics[width=.485\textwidth]{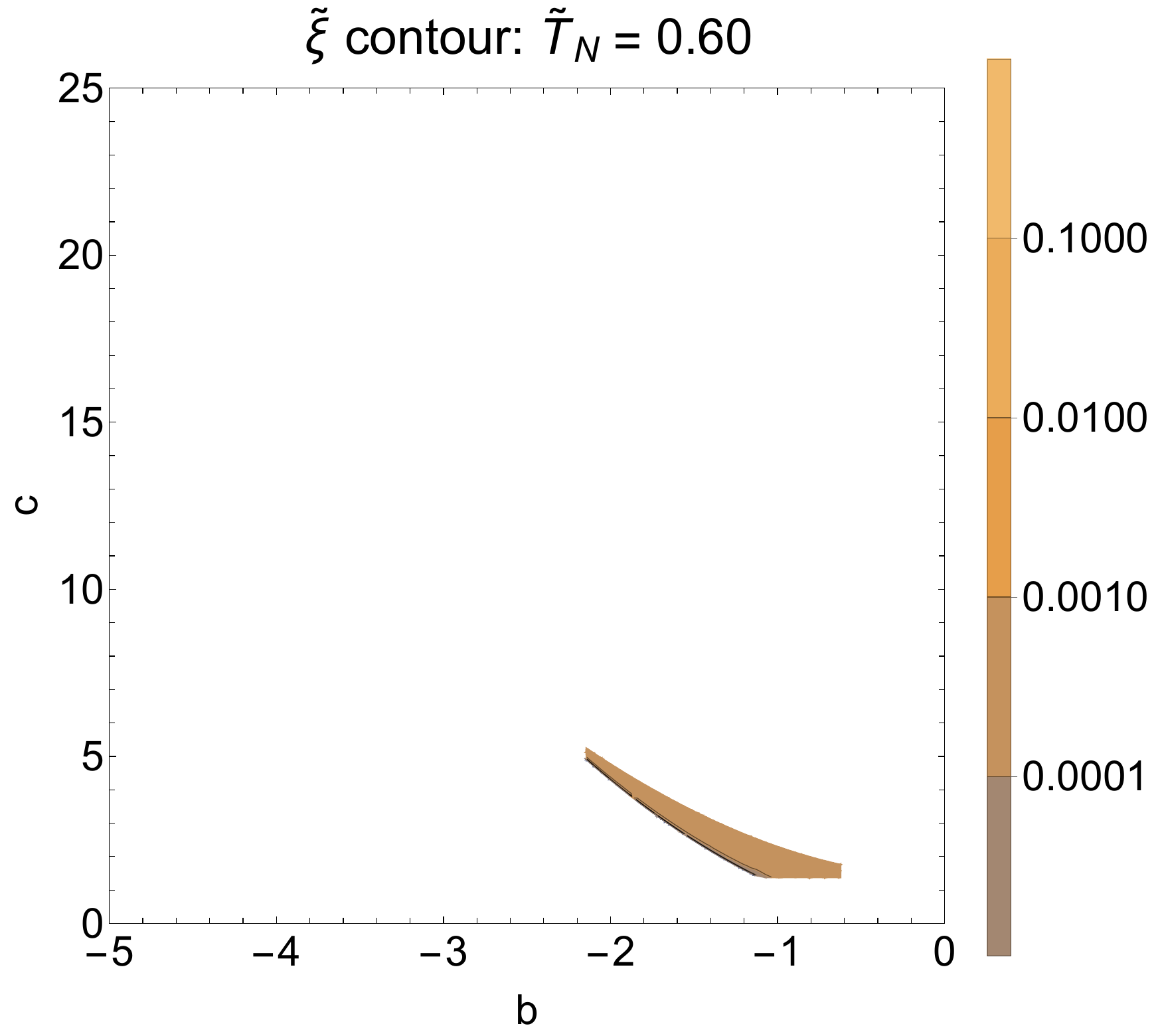}
\hfill
\includegraphics[width=.485\textwidth]{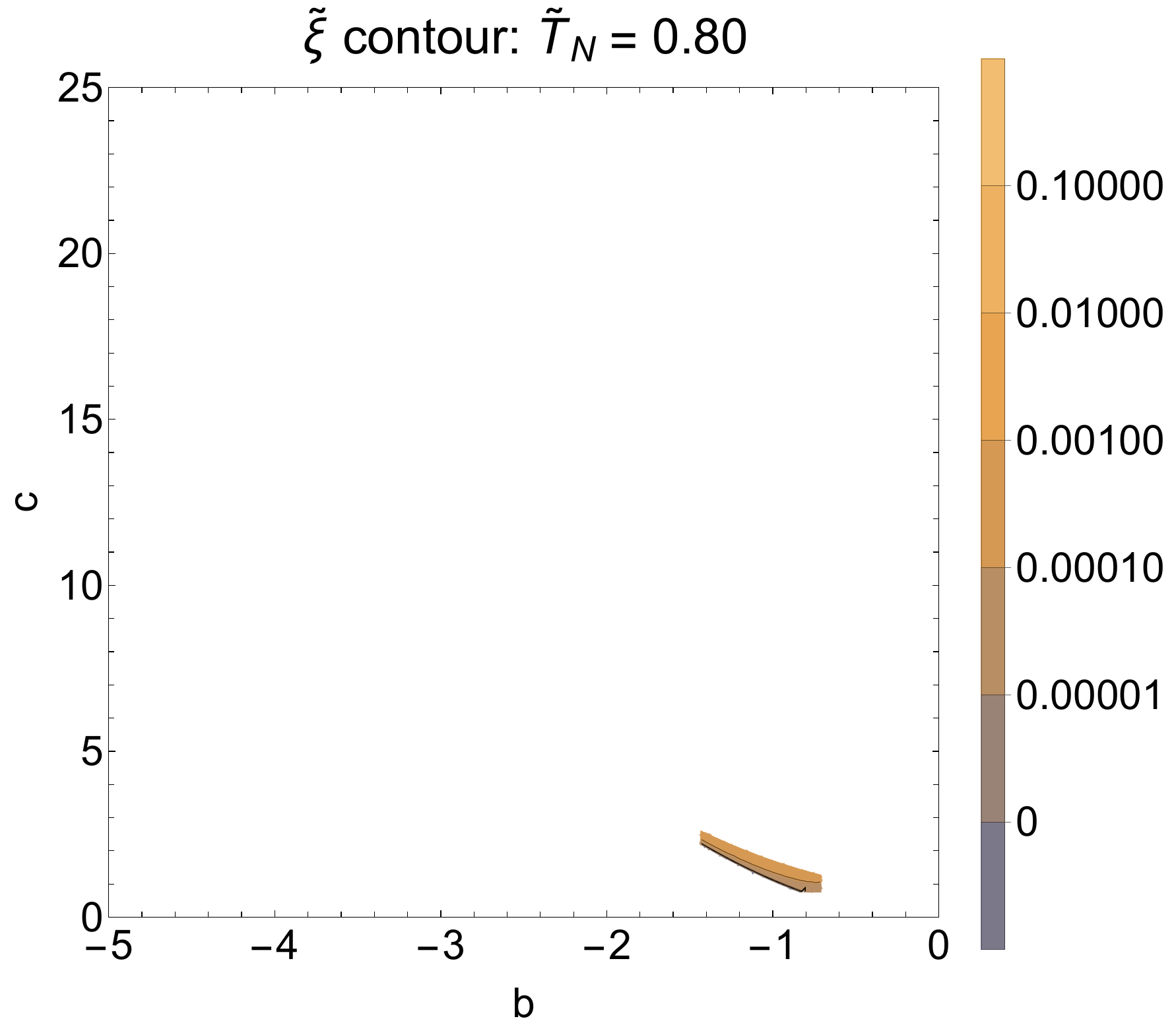}

\captionsetup{justification   = RaggedRight,
             labelfont = bf}
\caption{ Contours of $\tilde \xi$ 
are shown in the $(b,c)$ plane for the allowed parameter space  
for higher values of $\tilde{T}_N$: $\tilde T_N = 0.40$ (upper left), $0.50$ (upper right), $0.60$ (lower left), and $0.80$ 
(lower right).
The contours also satisfy the conditions: $c\ge b^2$, $c \tilde{T}_N \ge 1/2$, $\alpha \le 1 $, $0 \le \tilde{S}_E/140 \tilde{T}_N \le 1 $, $\tilde{\beta}/H\ge \tilde{S}_E/140 \tilde{T}_N$ and $\tilde{\xi} \ge 0 $. 
\label{fig:xicontourplothighT} } 
\end{figure}

Similarly, in Figures~\ref{fig:xicontourplotlowT} 
and~\ref{fig:xicontourplothighT}, we plot contours 
of $\tilde \xi$ in the $(b,c)$-plane for different 
choices of $\tilde T_N$.  Note that unless 
$\tilde T_N \ll 1$, we find $\tilde \xi \lesssim 
{\cal O}(10)$ throughout the entire parameter space.  
This implies that, unless the phase transition occurs 
well below the symmetry breaking scale, we will find
\bea
h^2 \tilde \Omega_{sw}^{max} &<& {\cal O}(0.1) , 
\nonumber\\
h^2  \Omega_{sw}^{max} &<& {\cal O}(0.1) 
\left(\frac{\Lambda}{v} \right)^{10+8n}
\label{eq:OmegaScaling}
\eea

The maximum sensitivity of LISA is roughly 
$h^2 \tilde \Omega_{sw} \sim {\cal O}(10^{-13})$, 
implying that LISA will be insensitive to any model 
for which $\Lambda / v \lesssim {\cal O}(10^{-1})$.  
Note that we can directly relate the parameters of 
the thermal effective potential to the mass and vev 
of the dark Higgs at zero temperature through the 
relation $(\Lambda / v)^2 = m / \sqrt{2} v$.  
We thus see that these simple scaling arguments imply 
a powerful connection between gravitational wave 
observations of a cosmological phase transition and 
laboratory probes of the hidden sector at zero 
temperature.  In particular, if the mass of the 
dark Higgs particle excitation is less than ${\cal O}(1\%)$ of the 
dark Higgs vev, then the gravitational wave signal 
arising from condensation of the dark Higgs in the early 
Universe would have an amplitude too small to 
be observed at LISA.  If the hidden sector is coupled to 
the Standard Model, then the dark Higgs mass and vev 
can be probed at fixed target or beam dump experiments, 
thus determining if a gravitational wave signal can be seen.  
Note that, although the maximum sensitivity of 
BBO will be 4-5 orders of magnitude better than LISA 
this leads to much less than an order of magnitude 
improvement in minimum value of $m/v$ for models 
which can be probed with gravitational waves.


\begin{figure}[t]
\begin{subfigure}[b]{0.49\textwidth}
\includegraphics[width=1.0\linewidth]{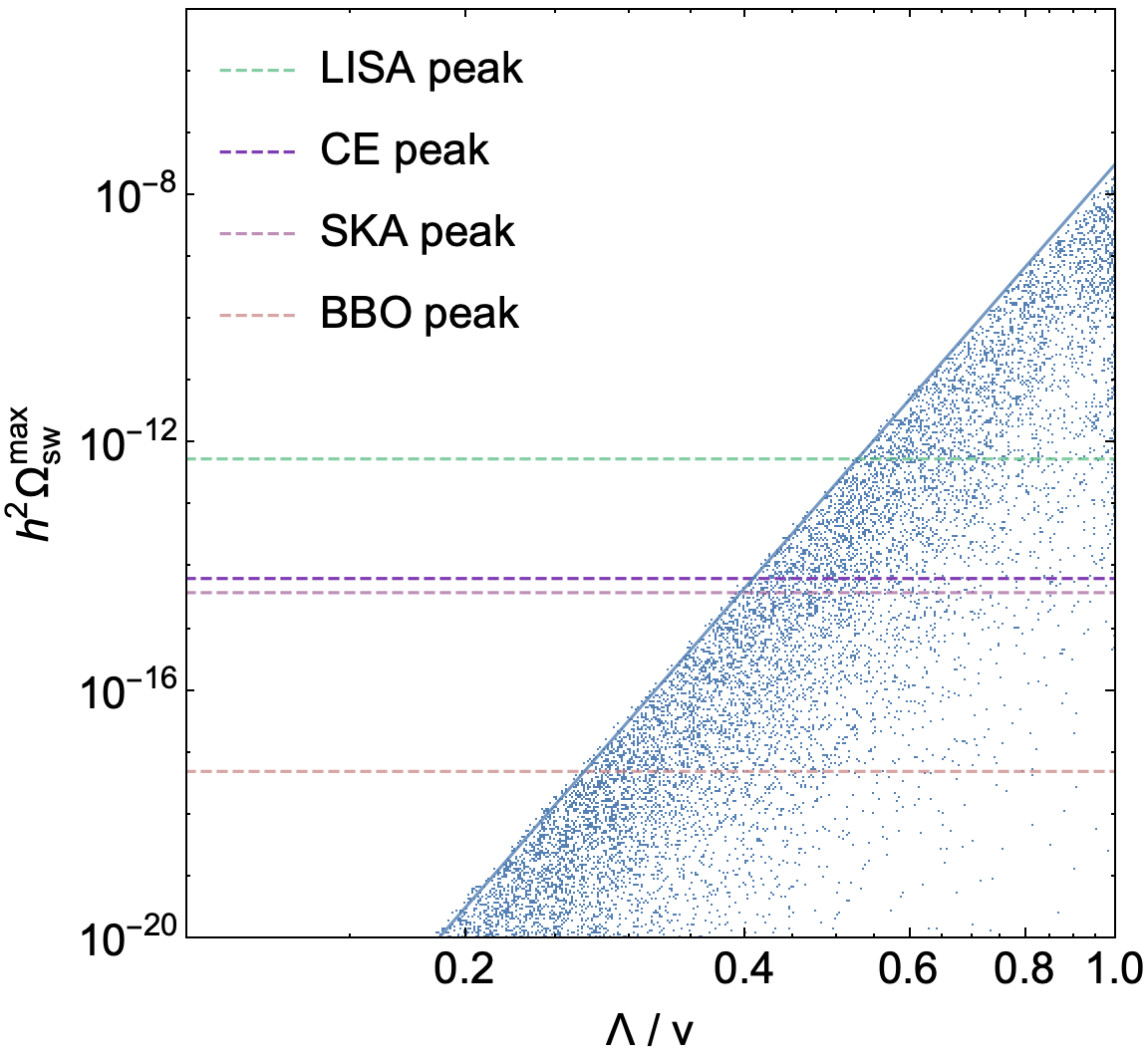}
\captionsetup{labelfont = bf}
\caption{\label{fig:Neff:hmaxvslambdaoverv}}
\end{subfigure}	
\hfill	
\begin{subfigure}[b]{0.49\textwidth}
\includegraphics[width=1.005\linewidth]{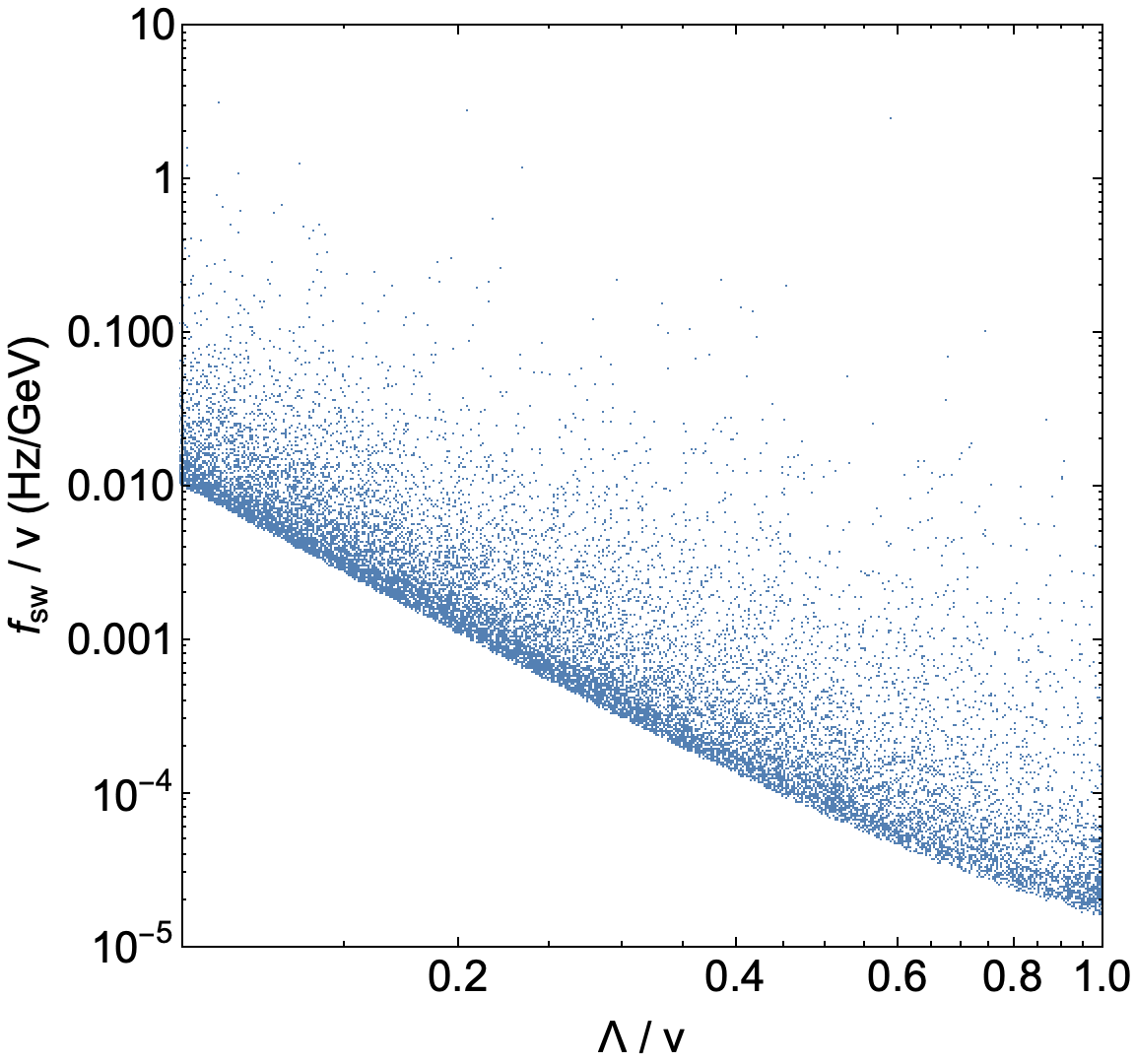}	
\captionsetup{labelfont = bf}
\caption{\label{fig:Neff:fswvslambdaoverv}}
\end{subfigure}	
\captionsetup{justification   = RaggedRight,
             labelfont = bf}
\caption{\label{fig:lambdaoverv} Plot of $h^2 \Omega_{sw}^{max}$ (left) and $f_{sw}/v$ (right) 
against $\Lambda/v$ for a set of models obtained by scanning of 
$0 \leq \Lambda/v \leq 1$, $-5 \leq b \leq 0$ and $0 \leq c \leq 25$.  The dashed lines in 
the left panel are the maximum sensitivities of 
LISA (green), BBO (beige), SKA (magenta) and CE (purple).  The solid blue line in the left 
panel presents the maximum value of $h^2 \Omega_{sw}^{max}$ obtained for a given 
$\Lambda / v$, for any $(b,c)$; this line has a slope of $\sim 17.2$.
} 
\end{figure}

We can provide an even tighter bound on the amplitude 
of the gravitational wave signal by accounting for the 
dependence of $h^2 \Omega_{sw}^{max}$ on $\xi$ and 
$\tilde \beta / H$.
In Figure~\ref{fig:Neff:hmaxvslambdaoverv}, we scan over 
parameters $(b,c,\Lambda / v)$, plotting each 
point on the $(h^2 \Omega_{sw}^{max}, \Lambda / v)$-plane.
We have assumed $v_w = 1$.  If we could ignore the 
effects of supercooling, then we would expect from 
Eq.~\ref{eq:BasicRelations} that the 
set of points with the largest values of 
$h^2  \Omega_{sw}^{max}$ correspond to the choice of 
$(b,c)$ for which $h^2  \tilde \Omega_{sw}^{max}$ is 
largest, and 
lie along the line 
$h^2  \Omega_{sw}^{max} \propto (\Lambda / v)^{18}$.  
In fact, the slope of this line (plotted as the solid 
blue line in Fig.~\ref{fig:Neff:hmaxvslambdaoverv}) is $\sim 17.2$, 
which is only 
slightly different 
from that given above, indicating that the 
dependence of 
$h^2  \tilde \Omega_{sw}^{max}$ on $\Lambda / v$ induced 
by the effects of supercooling is relatively minor.
Similarly, in Figure~\ref{fig:Neff:fswvslambdaoverv}, we scan over 
the parameters $(b,c,\Lambda/v)$, plotting each point in 
the $(f_{sw}, \Lambda / v)$-plane.  Again, we see that most 
of the points roughly follow the $f_{sw} /v \propto (\Lambda / v)^{-2}$ 
relation which we found earlier (ignoring the effects of supercooling), 
with a spread of roughly an order of magnitude in $f_{sw} /v$ for any 
$\Lambda / v$.

\begin{figure}[t]
\centering
\includegraphics[width=1.0\textwidth]{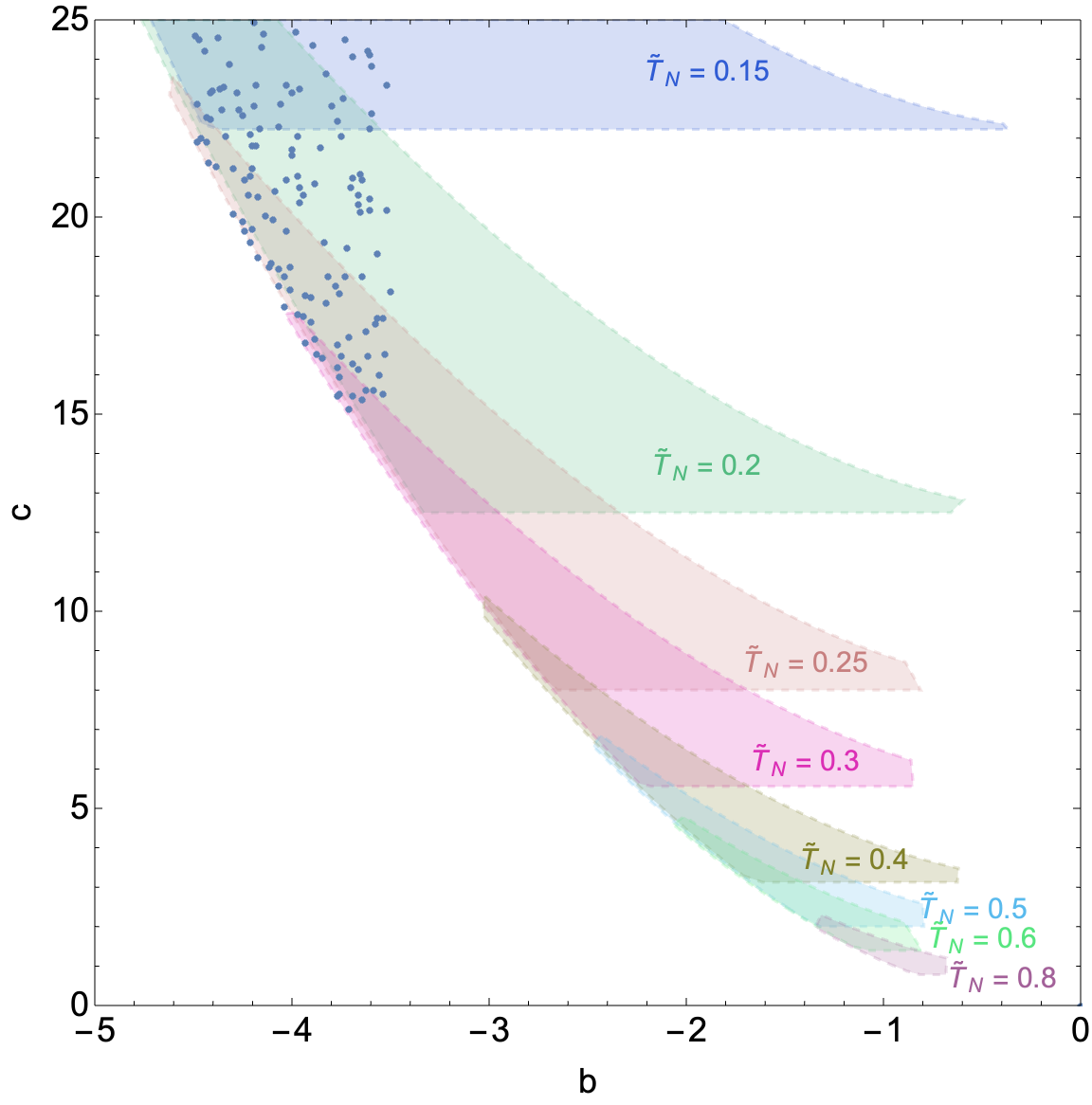}
\captionsetup{justification   = RaggedRight,
             labelfont = bf}
\caption{\label{fig:2} Plot in the $(b,c)$-plane of parameter points for 
which $h^2 \Omega_{sw}^{max}$ is within 2\% of 
solid blue line shown in Fig.~\ref{fig:Neff:hmaxvslambdaoverv}.  The shaded 
regions indicate regions in the $(b,c)$-plane for which various nucleation 
temperatures $\tilde T_N$ can be realized.
}
\end{figure}

In Figure~\ref{fig:2}, we plot the set of points in the 
$(b,c)$-plane for which $h^2 \Omega_{sw}^{max}$ is within $2\%$ of the maximum for any 
given value of $\Lambda / v$. These points correspond to the $(b,c)$ values of 
the points near the solid blue 
line in Figure~\ref{fig:Neff:hmaxvslambdaoverv}, discussed above. 
We see that scatter in these points is small, 
another indication that the effects of supercooling are relatively 
minor (if these effects were negligible, then all parameter points on this line 
would have the same values of $b$ and $c$, differing only in the parameter 
$\Lambda / v$).  We also plot shaded regions indicating values of $(b,c)$ for 
which a given nucleation temperature $\tilde T_N$ can be realized.  We see that 
these models with maximal signal amplitude can only be realized with 
$\tilde T_N < 0.4$.

\begin{figure}[t]
\centering
\includegraphics[width=1.0\textwidth]{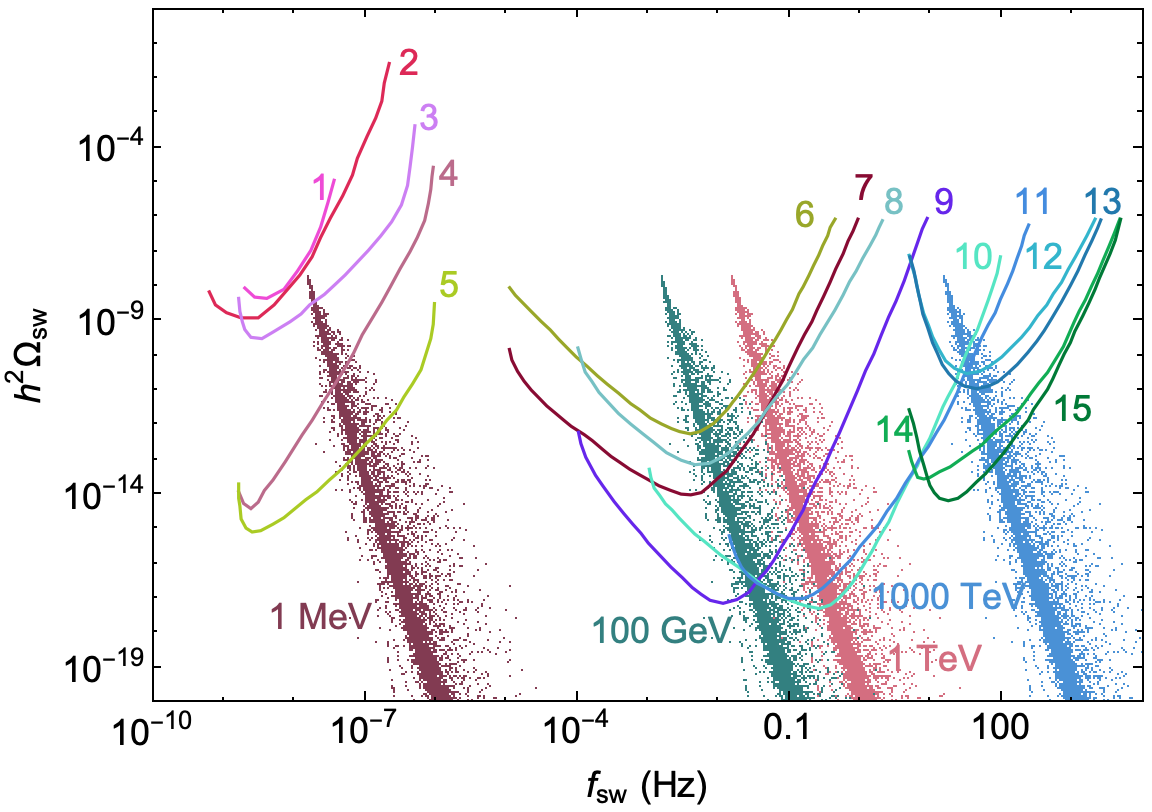}
\captionsetup{justification   = RaggedRight,
             labelfont = bf}
\caption{\label{fig:h2Omegafsw}  
Plot in the $(h^2 \Omega_{sw}, f_{sw})$-plane of the gravitational wave signal obtained from a scan of parameter points $(b, c, \Lambda /v)$.  The values of $v$ used are 1 MeV, 100 GeV, 1 TeV and 1000 TeV (from left to right). Also plotted are the sensitivities of the following current and upcoming experiments: 1. EPTA~\cite{vanHaasteren:2011ni}, 2. NANOGrav~\cite{NANOGrav:2020bcs, Lazio:2017fos}, 3. Gaia~\cite{Gaia:2018ydn}, 4. SKA~\cite{Janssen:2014dka}, 5. THEIA~\cite{10.3389/fspas.2018.00011}, 6. LISA~\cite{Caprini:2019egz, LISA:2017pwj, Robson:2018ifk}, 7. Taiji~\cite{Ruan:2018tsw}, 8.TianQin~\cite{TianQin:2015yph}, 9. ALIA~\cite{Gong:2014mca}, 10. BBO~\cite{Corbin:2005ny, Yagi:2011wg}, 11. DECIGO~\cite{Kawamura:2020pcg, Kawamura:2006up}, 12. aLIGO~\cite{Shoemaker:2019bqt, LIGOScientific:2014pky}, 13. A$+$~\cite{QuantumLIGO}, 14. ET~\cite{Punturo:2010zz}, 15. CE~\cite{LIGOScientific:2016wof}.}
\end{figure}

Finally, in Figure~\ref{fig:h2Omegafsw}, we scan over models, which are 
plotted in ths $(f_{sw}, h^2 \Omega_{sw}^{max})$-plane, taking $v_w =1$.  
In particular, we scan over 
$b$, $c$, and $\Lambda / v$, with $v=1~\mev$ (brown), $100~\gev$ (green), 
$1~\tev$ (pink), and $1000~\tev$ (blue).  We find $f_{sw} \propto v$, while for any choice of $v$, we roughly find 
$h^2 \Omega_{sw}^{max} \propto f_{sw}^{-9}$, as one would expect from 
eqn.~\ref{eq:BasicRelations}.  We also plot the sensitivity of a variety of 
current and upcoming gravitational wave observatories in Fig.~\ref{fig:h2Omegafsw}.

Note that the most sensitive upcoming experiments (BBO and ALIA) 
will probe models with symmetry-breaking scales in the range 
${\cal O}(100-1000)~\gev$.  For scenarios with new 
physics appearing at the MeV-scale, experiments such as THEIA 
would provide the leading sensitivity, while for scenarios with 
new physics appearing at the PeV-scale, experiments 
such as ET and CE would provide the leading sensitivity.  But 
the absence of sensitivity for models with 
$m / v < {\cal O}(0.01)$ is largely independent of scale; even 
at the scales for which future experiments will be most sensitive, 
this lower bound can be reduced by no more than a factor of a few.
These experiments are described in Table~\ref{tab:Experiments}.


Note from Figure~\ref{fig:Neff:hmaxvslambdaoverv} that, 
although the upper bound on $h^2 \Omega_{sw}^{max}$ follows 
fairly closely the scaling relations we have derived, and 
in particular, grows steeply with $\Lambda / v$, the actual 
amplitude may be much smaller than this maximum.  For example, 
the behavior of the SM electroweak phase transition is different, 
with the amplitude of a gravitational wave signal decreasing with 
$m_h / v$, and disappearing entirely for $m_h = 125~\gev$, at 
which point the phase transition is smooth.  This is because the 
thermal effective potential for the Standard Model Higgs sector 
corresponds to a choice of parameters which lie well below the 
upper bound on $h^2 \Omega_{sw}^{max}$.  For sufficiently large 
$m_h / v$, the parameter $c$ becomes small enough that the 
potential barrier disappears before reaching the nucleation 
temperature.  This example illustrates the point that, although 
this analysis provides an upper limit on the gravitational wave 
signal, it would be much more difficult to translate a measurement 
of $m / v$ into a prediction for the gravitational wave amplitude, 
or vice versa.

\section{Conclusions}
\label{sec:Conclusion}

We have considered gravitational wave signals from sound waves in the context of an effective field 
theory with a renormalizable thermal potential exhibiting a first-order phase 
transition in the early Universe.  We have considered a thermal effective potential 
of power law form, which would arise, for example, in the high temperature limit.  This 
approach allows us to consider the gravitational wave signals produced by phase 
transitions in a wide class of models.  

We find that for this class of models, there is a relatively clean relationship between 
the parameters of the gravitational wave signal and the parameters of the thermal effective 
potential.  The effective potential can be expressed 
in terms of one dimensionful parameter, the vacuum 
expectation value of the dark Higgs ($v$), which sets 
the frequency scale of the gravitational wave signal.
The amplitude of the gravitational wave signal is largely set by one dimensionless parameter, $\Lambda / v$, which 
sets the ratio of the potential energy scale to the 
dark Higgs vev.  Since the amplitude of the gravitational 
wave signal scales as $\Lambda / v$ to a very high 
power, the dependence on the other dimensionless 
parameters is subdominant.

This approach leads to a striking conclusion.  If the ratio of the dark Higgs 
mass to the dark Higgs vev is less than ${\cal O}(10^{-2})$, then the magnitude of the 
gravitational wave signal will be so small as to be undetectable by any upcoming 
observatories.  This bound on sensitivity is nearly independent of 
the scale of new physics.  
This result is also robust against future 
developments in our understanding of the amplitude of 
gravitational sound waves produced by a first-order cosmological 
phase transition.  Since the amplitude depends on $m_f / v$ raised 
to a high power, to modify our bound by an order of magnitude 
would require a new correction to the sound wave amplitude of 
several orders of magnitude.

This result leads to several intriguing possibilities.  For example, in scenarios of new MeV-scale physics, one may potentially be able 
to probe the symmetry-breaking 
scale with forward detectors at high-luminosity beam experiments, 
which could potentially determine the mass and coupling of the 
dark photon.  The detection of a gravitational wave signal 
from the corresponding first-order phase transition would 
then provide a lower bound on the Higgs mass.
Similarly, it is possible that future high-energy beam 
experiments could produce a heavy ($> {\cal O}(\tev)$) dark 
Higgs.  In this case, the amplitude of a gravitational wave 
signal from the phase transition would correspond to an upper 
bound on the symmetry-breaking scale.

\FloatBarrier
\begin{table}[tbp]
\Large
\centering
\begin{adjustbox}{width=1.0\textwidth,center=\textwidth}
\begin{tabular}{c|lll}
\hline
&&&\\
\makecell{Classifications \\ based on \\ frequency reach }& \makecell{ Dark sector \\ scale  that \\ can be probed  } & \makecell{ Type of \\ experiments}  &~~~~~~~~~~~~~~~ Experiments\\

&&&\\

\hline 

&&&\\

\makecell{Very low frequency  \\ $10^{-9}-10^{-8}$~Hz} & \makecell{ $\le 1$~MeV  }  & \makecell{ Ground-based  \\telescope using \\ pulsar timing  \\observation technique  } & \makecell{EPTA (European Pulsar Timing \\ Array)~\cite{vanHaasteren:2011ni} \\ \\ NANOGrav (North \\ American Nanohertz \\ Observatory for \\ Gravitational Wave)~\cite{NANOGrav:2020bcs, Lazio:2017fos}  \\ \\ SKA (Square Kilometer Array)~\cite{Janssen:2014dka}: \\ World’s largest radio telescope}\\
&&&\\
\cline{3-4}
&&&\\
&& \makecell{ Space-based  \\ telescope \\ using astrometry}& \makecell{Gaia~\cite{Gaia:2018ydn} 
\\ \\ THEIA (Telescope for \\ Habitable Exoplanets and \\ Intersteller/Intergalactic \\ Astronomy)~\cite{10.3389/fspas.2018.00011}: \\ Proposed upgrade of Gaia } \\
&&&\\
\hline
&&&\\

\makecell{$\mu$-frequency  \\ $10^{-6}-10^{-5}$~Hz} & \makecell{ $\sim$ 100 MeV  }  & \makecell{ Space-based\\ heliocentric constellation\\ of satellites  \\ using Laser  \\interferometry technique } & \makecell{$\mu$-Ares~\cite{Sesana:2019vho} }\\

&&&\\
\hline
&&&\\

\makecell{Low frequency  \\ $10^{-3}-1$~Hz} & \makecell{ 100 GeV-1 TeV  }  & \makecell{ Space-based  \\ interferometer \\ using Laser  \\interferometry technique } & \makecell{LISA (Laser Interferometer Space \\ Antenna)~\cite{Caprini:2019egz, LISA:2017pwj, Robson:2018ifk} \\ \\ALIA (Advanced Laser  Interferometer\\ Antenna)~\cite{Gong:2014mca} \\ \\  Taiji~\cite{Ruan:2018tsw} 
\\ \\ TianQin~\cite{TianQin:2015yph} 
\\ \\    BBO (Big Bang Observer)~\cite{Corbin:2005ny, Yagi:2011wg} \\ \\ DECIGO (Deci-Hertz Interferometer \\ Gravitational Wave Observatory)~\cite{Kawamura:2020pcg, Kawamura:2006up} }\\
&&&\\
\hline

\end{tabular} 
\end{adjustbox}

\end{table}
\FloatBarrier

\begin{table}[h]

\centering
\begin{adjustbox}{width=1.0\textwidth,center=\textwidth}
\begin{tabular}{c|lll}
\hline
&&&\\
\makecell{Classifications \\ based on \\ frequency reach }& \makecell{ Dark sector \\ scale  that \\ can be probed  } & \makecell{ Type of \\ experiments}  &~~~~~~~~~~~~~~~ Experiments\\

&&&\\

\hline 

&&&\\

\makecell{High frequency  \\ $10-100$~Hz} & \makecell{ $\sim$1000 TeV  }  & \makecell{Second generation \\ Ground-based  \\ interferometer\\ using Laser  \\interferometry technique } & \makecell{aLIGO (Advanced LIGO)~\cite{Shoemaker:2019bqt, LIGOScientific:2014pky} \\ \\ A$+$~\cite{QuantumLIGO} (Quantum LIGO)}\\
&&&\\
\cline{3-4}
&&&\\
&& \makecell{ Third generation \\ Ground-based  \\ interferometer\\ using Laser  \\interferometry technique}& \makecell{ET (Einstein Telescope)~\cite{Punturo:2010zz} \\ \\ CE (Cosmic Explorer)~\cite{LIGOScientific:2016wof}} \\
&&&\\
\hline
\end{tabular} 
\end{adjustbox}
\captionsetup{justification   = RaggedRight,
             labelfont = bf}
\caption{\label{tab:GWobservatory} Description of upcoming 
gravitational wave observatories, including the frequency 
range to which they would be sensitive, and the correspond 
symmetry-breaking scale. }
\label{tab:Experiments}
\end{table}

There are several avenues for further exploration.  
We have considered a power-law form for the thermal 
effective potential, which is expected to be a good 
approximation for renormalizable models, in the 
high-temperature limit.  But there are a variety 
of models in which the phase transition is 
supercooled (see, for example \cite{Kobakhidze:2017mru,Ellis:2019oqb,Ellis:2020nnr,Wang:2020jrd}), and nucleation temperature is relatively 
low.  In these cases, although the corrections to the 
form of the potential may be small, they need not be. 
There has been significant recent work regarding theoretical 
issues in finite temperature perturbation theory (see, 
for example,~\cite{Croon:2020cgk,Gould:2021oba}).

Similarly, we have assumed that the thermal effective 
potential is renormalizable.  There are interesting models 
of new physics in which the phase transition can only be 
seen with the inclusion of non-renormalizable terms.  
It would be interesting to consider more general 
forms of the thermal effective potential, to better determine 
the extent to which the lessons found here generalize.

\acknowledgments

We are grateful to Djuna Croon, Jeremy Sakstein, Tim Tait, and especially Graham White for useful discussions.
The work of JK is supported in part by DOE grant DE-SC0010504.
JR is supported by NSF grant AST-1934744. The work of BD and SG is supported in part by DOE grant DE-SC0010813. The work of SG is also supported in part by National Research Foundation of Korea (NRF)
Grant No. NRF-2019R1A2C3005009 (SG). JBD acknowledges support from the National Science Foundation under grants no. PHY-1820801 and PHY-1748958.

\bibliographystyle{JHEP.bst}
\bibliography{gw.bib}

\providecommand{\href}[2]{#2}\begingroup\raggedright\begin{thebibliography}{100}

\bibitem{LIGOScientific:2016aoc}
{\bf LIGO Scientific, Virgo} Collaboration, B.~P. Abbott et~al., {\it
  {Observation of Gravitational Waves from a Binary Black Hole Merger}},  {\em
  Phys. Rev. Lett.} {\bf 116} (2016), no.~6 061102,
  [\href{http://arxiv.org/abs/1602.03837}{{\tt arXiv:1602.03837}}].

\bibitem{Kosowsky:1992rz}
A.~Kosowsky, M.~S. Turner, and R.~Watkins, {\it {Gravitational waves from first
  order cosmological phase transitions}},  {\em Phys. Rev. Lett.} {\bf 69}
  (1992) 2026--2029.

\bibitem{Kosowsky:1991ua}
A.~Kosowsky, M.~S. Turner, and R.~Watkins, {\it {Gravitational radiation from
  colliding vacuum bubbles}},  {\em Phys. Rev. D} {\bf 45} (1992) 4514--4535.

\bibitem{Apreda:2001us}
R.~Apreda, M.~Maggiore, A.~Nicolis, and A.~Riotto, {\it {Gravitational waves
  from electroweak phase transitions}},  {\em Nucl. Phys. B} {\bf 631} (2002)
  342--368, [\href{http://arxiv.org/abs/gr-qc/0107033}{{\tt gr-qc/0107033}}].

\bibitem{Grojean:2006bp}
C.~Grojean and G.~Servant, {\it {Gravitational Waves from Phase Transitions at
  the Electroweak Scale and Beyond}},  {\em Phys. Rev. D} {\bf 75} (2007)
  043507, [\href{http://arxiv.org/abs/hep-ph/0607107}{{\tt hep-ph/0607107}}].

\bibitem{Huber:2008hg}
S.~J. Huber and T.~Konstandin, {\it {Gravitational Wave Production by
  Collisions: More Bubbles}},  {\em JCAP} {\bf 09} (2008) 022,
  [\href{http://arxiv.org/abs/0806.1828}{{\tt arXiv:0806.1828}}].

\bibitem{Cai:2017cbj}
R.-G. Cai, Z.~Cao, Z.-K. Guo, S.-J. Wang, and T.~Yang, {\it {The
  Gravitational-Wave Physics}},  {\em Natl. Sci. Rev.} {\bf 4} (2017), no.~5
  687--706, [\href{http://arxiv.org/abs/1703.00187}{{\tt arXiv:1703.00187}}].

\bibitem{Weir:2017wfa}
D.~J. Weir, {\it {Gravitational waves from a first order electroweak phase
  transition: a brief review}},  {\em Phil. Trans. Roy. Soc. Lond. A} {\bf 376}
  (2018), no.~2114 20170126, [\href{http://arxiv.org/abs/1705.01783}{{\tt
  arXiv:1705.01783}}].

\bibitem{Caprini:2015zlo}
C.~Caprini et~al., {\it {Science with the space-based interferometer eLISA. II:
  Gravitational waves from cosmological phase transitions}},  {\em JCAP} {\bf
  04} (2016) 001, [\href{http://arxiv.org/abs/1512.06239}{{\tt
  arXiv:1512.06239}}].

\bibitem{Mazumdar:2018dfl}
A.~Mazumdar and G.~White, {\it {Review of cosmic phase transitions: their
  significance and experimental signatures}},  {\em Rept. Prog. Phys.} {\bf 82}
  (2019), no.~7 076901, [\href{http://arxiv.org/abs/1811.01948}{{\tt
  arXiv:1811.01948}}].

\bibitem{Caprini:2018mtu}
C.~Caprini and D.~G. Figueroa, {\it {Cosmological Backgrounds of Gravitational
  Waves}},  {\em Class. Quant. Grav.} {\bf 35} (2018), no.~16 163001,
  [\href{http://arxiv.org/abs/1801.04268}{{\tt arXiv:1801.04268}}].

\bibitem{Caprini:2019egz}
C.~Caprini et~al., {\it {Detecting gravitational waves from cosmological phase
  transitions with LISA: an update}},  {\em JCAP} {\bf 03} (2020) 024,
  [\href{http://arxiv.org/abs/1910.13125}{{\tt arXiv:1910.13125}}].

\bibitem{Caldwell:2022qsj}
R.~Caldwell et~al., {\it {Detection of Early-Universe Gravitational Wave
  Signatures and Fundamental Physics}},  in {\em {2022 Snowmass Summer Study}},
  3, 2022.
\newblock \href{http://arxiv.org/abs/2203.07972}{{\tt arXiv:2203.07972}}.

\bibitem{Ashoorioon:2009nf}
A.~Ashoorioon and T.~Konstandin, {\it {Strong electroweak phase transitions
  without collider traces}},  {\em JHEP} {\bf 07} (2009) 086,
  [\href{http://arxiv.org/abs/0904.0353}{{\tt arXiv:0904.0353}}].

\bibitem{Alves:2018jsw}
A.~Alves, T.~Ghosh, H.-K. Guo, K.~Sinha, and D.~Vagie, {\it {Collider and
  Gravitational Wave Complementarity in Exploring the Singlet Extension of the
  Standard Model}},  {\em JHEP} {\bf 04} (2019) 052,
  [\href{http://arxiv.org/abs/1812.09333}{{\tt arXiv:1812.09333}}].

\bibitem{Balazs:2016tbi}
C.~Balazs, A.~Fowlie, A.~Mazumdar, and G.~White, {\it {Gravitational waves at
  aLIGO and vacuum stability with a scalar singlet extension of the Standard
  Model}},  {\em Phys. Rev. D} {\bf 95} (2017), no.~4 043505,
  [\href{http://arxiv.org/abs/1611.01617}{{\tt arXiv:1611.01617}}].

\bibitem{Kang:2017mkl}
Z.~Kang, P.~Ko, and T.~Matsui, {\it {Strong first order EWPT $\&$ strong
  gravitational waves in Z$_{3}$-symmetric singlet scalar extension}},  {\em
  JHEP} {\bf 02} (2018) 115, [\href{http://arxiv.org/abs/1706.09721}{{\tt
  arXiv:1706.09721}}].

\bibitem{Matsui:2017ggm}
T.~Matsui, {\it {Gravitational waves from the first order electroweak phase
  transition in the $Z_3$ symmetric singlet scalar model}},  {\em EPJ Web
  Conf.} {\bf 168} (2018) 05001, [\href{http://arxiv.org/abs/1709.05900}{{\tt
  arXiv:1709.05900}}].

\bibitem{Vaskonen:2016yiu}
V.~Vaskonen, {\it {Electroweak baryogenesis and gravitational waves from a real
  scalar singlet}},  {\em Phys. Rev. D} {\bf 95} (2017), no.~12 123515,
  [\href{http://arxiv.org/abs/1611.02073}{{\tt arXiv:1611.02073}}].

\bibitem{Lewicki:2021pgr}
M.~Lewicki, M.~Merchand, and M.~Zych, {\it {Electroweak bubble wall expansion:
  gravitational waves and baryogenesis in Standard Model-like thermal plasma}},
   {\em JHEP} {\bf 02} (2022) 017, [\href{http://arxiv.org/abs/2111.02393}{{\tt
  arXiv:2111.02393}}].

\bibitem{Croon:2018new}
D.~Croon and G.~White, {\it {Exotic Gravitational Wave Signatures from
  Simultaneous Phase Transitions}},  {\em JHEP} {\bf 05} (2018) 210,
  [\href{http://arxiv.org/abs/1803.05438}{{\tt arXiv:1803.05438}}].

\bibitem{Beniwal:2018hyi}
A.~Beniwal, M.~Lewicki, M.~White, and A.~G. Williams, {\it {Gravitational waves
  and electroweak baryogenesis in a global study of the extended scalar singlet
  model}},  {\em JHEP} {\bf 02} (2019) 183,
  [\href{http://arxiv.org/abs/1810.02380}{{\tt arXiv:1810.02380}}].

\bibitem{Hashino:2018wee}
K.~Hashino, R.~Jinno, M.~Kakizaki, S.~Kanemura, T.~Takahashi, and M.~Takimoto,
  {\it {Selecting models of first-order phase transitions using the synergy
  between collider and gravitational-wave experiments}},  {\em Phys. Rev. D}
  {\bf 99} (2019), no.~7 075011, [\href{http://arxiv.org/abs/1809.04994}{{\tt
  arXiv:1809.04994}}].

\bibitem{Ahriche:2018rao}
A.~Ahriche, K.~Hashino, S.~Kanemura, and S.~Nasri, {\it {Gravitational Waves
  from Phase Transitions in Models with Charged Singlets}},  {\em Phys. Lett.
  B} {\bf 789} (2019) 119--126, [\href{http://arxiv.org/abs/1809.09883}{{\tt
  arXiv:1809.09883}}].

\bibitem{Shajiee:2018jdq}
V.~R. Shajiee and A.~Tofighi, {\it {Electroweak Phase Transition, Gravitational
  Waves and Dark Matter in Two Scalar Singlet Extension of The Standard
  Model}},  {\em Eur. Phys. J. C} {\bf 79} (2019), no.~4 360,
  [\href{http://arxiv.org/abs/1811.09807}{{\tt arXiv:1811.09807}}].

\bibitem{Vieu:2018nfq}
T.~Vieu, A.~P. Morais, and R.~Pasechnik, {\it {Electroweak phase transitions in
  multi-Higgs models: the case of Trinification-inspired THDSM}},  {\em JCAP}
  {\bf 07} (2018) 014, [\href{http://arxiv.org/abs/1801.02670}{{\tt
  arXiv:1801.02670}}].

\bibitem{Alves:2018oct}
A.~Alves, T.~Ghosh, H.-K. Guo, and K.~Sinha, {\it {Resonant Di-Higgs Production
  at Gravitational Wave Benchmarks: A Collider Study using Machine Learning}},
  {\em JHEP} {\bf 12} (2018) 070, [\href{http://arxiv.org/abs/1808.08974}{{\tt
  arXiv:1808.08974}}].

\bibitem{Chala:2018opy}
M.~Chala, M.~Ramos, and M.~Spannowsky, {\it {Gravitational wave and collider
  probes of a triplet Higgs sector with a low cutoff}},  {\em Eur. Phys. J. C}
  {\bf 79} (2019), no.~2 156, [\href{http://arxiv.org/abs/1812.01901}{{\tt
  arXiv:1812.01901}}].

\bibitem{Fujikura:2018duw}
K.~Fujikura, K.~Kamada, Y.~Nakai, and M.~Yamaguchi, {\it {Phase Transitions in
  Twin Higgs Models}},  {\em JHEP} {\bf 12} (2018) 018,
  [\href{http://arxiv.org/abs/1810.00574}{{\tt arXiv:1810.00574}}].

\bibitem{Graf:2021xku}
L.~Gr\'af, S.~Jana, A.~Kaladharan, and S.~Saad, {\it {Gravitational Wave
  Imprints of Left-Right Symmetric Model with Minimal Higgs Sector}},
  \href{http://arxiv.org/abs/2112.12041}{{\tt arXiv:2112.12041}}.

\bibitem{Li:2020eun}
M.~Li, Q.-S. Yan, Y.~Zhang, and Z.~Zhao, {\it {Prospects of gravitational waves
  in the minimal left-right symmetric model}},  {\em JHEP} {\bf 03} (2021) 267,
  [\href{http://arxiv.org/abs/2012.13686}{{\tt arXiv:2012.13686}}].

\bibitem{Ghorbani:2017lyk}
P.~H. Ghorbani, {\it {Electroweak phase transition in the scale invariant
  standard model}},  {\em Phys. Rev. D} {\bf 98} (2018), no.~11 115016,
  [\href{http://arxiv.org/abs/1711.11541}{{\tt arXiv:1711.11541}}].

\bibitem{Marzo:2018nov}
C.~Marzo, L.~Marzola, and V.~Vaskonen, {\it {Phase transition and vacuum
  stability in the classically conformal B\textendash{}L model}},  {\em Eur.
  Phys. J. C} {\bf 79} (2019), no.~7 601,
  [\href{http://arxiv.org/abs/1811.11169}{{\tt arXiv:1811.11169}}].

\bibitem{Marzola:2017jzl}
L.~Marzola, A.~Racioppi, and V.~Vaskonen, {\it {Phase transition and
  gravitational wave phenomenology of scalar conformal extensions of the
  Standard Model}},  {\em Eur. Phys. J. C} {\bf 77} (2017), no.~7 484,
  [\href{http://arxiv.org/abs/1704.01034}{{\tt arXiv:1704.01034}}].

\bibitem{Verweij:2019xch}
D.~Verweij, {\it {On the construction and measurement of conformal extensions
  of the Standard Model}},  Master's thesis, Utrecht U. (main), 2019.

\bibitem{Prokopec:2018tnq}
T.~Prokopec, J.~Rezacek, and B.~\'Swie\.zewska, {\it {Gravitational waves from
  conformal symmetry breaking}},  {\em JCAP} {\bf 02} (2019) 009,
  [\href{http://arxiv.org/abs/1809.11129}{{\tt arXiv:1809.11129}}].

\bibitem{Zhou:2018zli}
R.~Zhou, W.~Cheng, X.~Deng, L.~Bian, and Y.~Wu, {\it {Electroweak phase
  transition and Higgs phenomenology in the Georgi-Machacek model}},  {\em
  JHEP} {\bf 01} (2019) 216, [\href{http://arxiv.org/abs/1812.06217}{{\tt
  arXiv:1812.06217}}].

\bibitem{Archer-Smith:2019gzq}
P.~Archer-Smith, D.~Linthorne, and D.~Stolarski, {\it {Gravitational Wave
  Signals from Multiple Hidden Sectors}},  {\em Phys. Rev. D} {\bf 101} (2020),
  no.~9 095016, [\href{http://arxiv.org/abs/1910.02083}{{\tt
  arXiv:1910.02083}}].

\bibitem{Cai:2022bcf}
R.-G. Cai, K.~Hashino, S.-J. Wang, and J.-H. Yu, {\it {Gravitational waves from
  patterns of electroweak symmetry breaking: an effective perspective}},
  \href{http://arxiv.org/abs/2202.08295}{{\tt arXiv:2202.08295}}.

\bibitem{Paul:2020wbz}
A.~Paul, U.~Mukhopadhyay, and D.~Majumdar, {\it {Gravitational Wave Signatures
  from Domain Wall and Strong First-Order Phase Transitions in a Two Complex
  Scalar extension of the Standard Model}},  {\em JHEP} {\bf 05} (2021) 223,
  [\href{http://arxiv.org/abs/2010.03439}{{\tt arXiv:2010.03439}}].

\bibitem{Chao:2017ilw}
W.~Chao, W.-F. Cui, H.-K. Guo, and J.~Shu, {\it {Gravitational wave imprint of
  new symmetry breaking}},  {\em Chin. Phys. C} {\bf 44} (2020), no.~12 123102,
  [\href{http://arxiv.org/abs/1707.09759}{{\tt arXiv:1707.09759}}].

\bibitem{Pimentel:2020rkw}
E.~M. Pimentel, {\em {Phenomenology of New Physics Models at Colliders and in
  Gravitational Waves}}.
\newblock PhD thesis, Mainz U., 2020.

\bibitem{Blinov:2015sna}
N.~Blinov, J.~Kozaczuk, D.~E. Morrissey, and C.~Tamarit, {\it {Electroweak
  Baryogenesis from Exotic Electroweak Symmetry Breaking}},  {\em Phys. Rev. D}
  {\bf 92} (2015), no.~3 035012, [\href{http://arxiv.org/abs/1504.05195}{{\tt
  arXiv:1504.05195}}].

\bibitem{Inoue:2015pza}
S.~Inoue, G.~Ovanesyan, and M.~J. Ramsey-Musolf, {\it {Two-Step Electroweak
  Baryogenesis}},  {\em Phys. Rev. D} {\bf 93} (2016) 015013,
  [\href{http://arxiv.org/abs/1508.05404}{{\tt arXiv:1508.05404}}].

\bibitem{Greljo:2019xan}
A.~Greljo, T.~Opferkuch, and B.~A. Stefanek, {\it {Gravitational Imprints of
  Flavor Hierarchies}},  {\em Phys. Rev. Lett.} {\bf 124} (2020), no.~17
  171802, [\href{http://arxiv.org/abs/1910.02014}{{\tt arXiv:1910.02014}}].

\bibitem{Croon:2018kqn}
D.~Croon, T.~E. Gonzalo, and G.~White, {\it {Gravitational Waves from a
  Pati-Salam Phase Transition}},  {\em JHEP} {\bf 02} (2019) 083,
  [\href{http://arxiv.org/abs/1812.02747}{{\tt arXiv:1812.02747}}].

\bibitem{Demidov:2017lzf}
S.~V. Demidov, D.~S. Gorbunov, and D.~V. Kirpichnikov, {\it {Gravitational
  waves from phase transition in split NMSSM}},  {\em Phys. Lett. B} {\bf 779}
  (2018) 191--194, [\href{http://arxiv.org/abs/1712.00087}{{\tt
  arXiv:1712.00087}}].

\bibitem{Bian:2017wfv}
L.~Bian, H.-K. Guo, and J.~Shu, {\it {Gravitational Waves, baryon asymmetry of
  the universe and electric dipole moment in the CP-violating NMSSM}},  {\em
  Chin. Phys. C} {\bf 42} (2018), no.~9 093106,
  [\href{http://arxiv.org/abs/1704.02488}{{\tt arXiv:1704.02488}}]. [Erratum:
  Chin.Phys.C 43, 129101 (2019)].

\bibitem{Eichhorn:2020upj}
A.~Eichhorn, J.~Lumma, J.~M. Pawlowski, M.~Reichert, and M.~Yamada, {\it
  {Universal gravitational-wave signatures from heavy new physics in the
  electroweak sector}},  {\em JCAP} {\bf 05} (2021) 006,
  [\href{http://arxiv.org/abs/2010.00017}{{\tt arXiv:2010.00017}}].

\bibitem{Miura:2018dsy}
K.~Miura, H.~Ohki, S.~Otani, and K.~Yamawaki, {\it {Gravitational Waves from
  Walking Technicolor}},  {\em JHEP} {\bf 10} (2019) 194,
  [\href{http://arxiv.org/abs/1811.05670}{{\tt arXiv:1811.05670}}].

\bibitem{Azatov:2019png}
A.~Azatov, D.~Barducci, and F.~Sgarlata, {\it {Gravitational traces of broken
  gauge symmetries}},  {\em JCAP} {\bf 07} (2020) 027,
  [\href{http://arxiv.org/abs/1910.01124}{{\tt arXiv:1910.01124}}].

\bibitem{Aoki:2017aws}
M.~Aoki, H.~Goto, and J.~Kubo, {\it {Gravitational Waves from Hidden QCD Phase
  Transition}},  {\em Phys. Rev. D} {\bf 96} (2017), no.~7 075045,
  [\href{http://arxiv.org/abs/1709.07572}{{\tt arXiv:1709.07572}}].

\bibitem{Chen:2017cyc}
Y.~Chen, M.~Huang, and Q.-S. Yan, {\it {Gravitation waves from QCD and
  electroweak phase transitions}},  {\em JHEP} {\bf 05} (2018) 178,
  [\href{http://arxiv.org/abs/1712.03470}{{\tt arXiv:1712.03470}}].

\bibitem{Croon:2019ugf}
D.~Croon, J.~N. Howard, S.~Ipek, and T.~M.~P. Tait, {\it {QCD baryogenesis}},
  {\em Phys. Rev. D} {\bf 101} (2020), no.~5 055042,
  [\href{http://arxiv.org/abs/1911.01432}{{\tt arXiv:1911.01432}}].

\bibitem{Croon:2019iuh}
D.~Croon, R.~Houtz, and V.~Sanz, {\it {Dynamical Axions and Gravitational
  Waves}},  {\em JHEP} {\bf 07} (2019) 146,
  [\href{http://arxiv.org/abs/1904.10967}{{\tt arXiv:1904.10967}}].

\bibitem{Costa:2022oaa}
F.~Costa, S.~Khan, and J.~Kim, {\it {A Two-Component Dark Matter Model and its
  Associated Gravitational Waves}},
  \href{http://arxiv.org/abs/2202.13126}{{\tt arXiv:2202.13126}}.

\bibitem{Chao:2017vrq}
W.~Chao, H.-K. Guo, and J.~Shu, {\it {Gravitational Wave Signals of Electroweak
  Phase Transition Triggered by Dark Matter}},  {\em JCAP} {\bf 09} (2017) 009,
  [\href{http://arxiv.org/abs/1702.02698}{{\tt arXiv:1702.02698}}].

\bibitem{Ghosh:2020ipy}
T.~Ghosh, H.-K. Guo, T.~Han, and H.~Liu, {\it {Electroweak phase transition
  with an SU(2) dark sector}},  {\em JHEP} {\bf 07} (2021) 045,
  [\href{http://arxiv.org/abs/2012.09758}{{\tt arXiv:2012.09758}}].

\bibitem{Chao:2021xqv}
W.~Chao, H.-K. Guo, and X.-F. Li, {\it {First Order Color Symmetry Breaking and
  Restoration Triggered by Electroweak Symmetry Non-restoration}},
  \href{http://arxiv.org/abs/2112.13580}{{\tt arXiv:2112.13580}}.

\bibitem{Romero:2021kby}
A.~Romero, K.~Martinovic, T.~A. Callister, H.-K. Guo, M.~Mart\'\i{}nez,
  M.~Sakellariadou, F.-W. Yang, and Y.~Zhao, {\it {Implications for First-Order
  Cosmological Phase Transitions from the Third LIGO-Virgo Observing Run}},
  {\em Phys. Rev. Lett.} {\bf 126} (2021), no.~15 151301,
  [\href{http://arxiv.org/abs/2102.01714}{{\tt arXiv:2102.01714}}].

\bibitem{Alves:2020bpi}
A.~Alves, D.~Gon\c{c}alves, T.~Ghosh, H.-K. Guo, and K.~Sinha, {\it {Di-Higgs
  Blind Spots in Gravitational Wave Signals}},  {\em Phys. Lett. B} {\bf 818}
  (2021) 136377, [\href{http://arxiv.org/abs/2007.15654}{{\tt
  arXiv:2007.15654}}].

\bibitem{Hall:2019ank}
E.~Hall, T.~Konstandin, R.~McGehee, H.~Murayama, and G.~Servant, {\it
  {Baryogenesis From a Dark First-Order Phase Transition}},  {\em JHEP} {\bf
  04} (2020) 042, [\href{http://arxiv.org/abs/1910.08068}{{\tt
  arXiv:1910.08068}}].

\bibitem{Hall:2019rld}
E.~Hall, T.~Konstandin, R.~McGehee, and H.~Murayama, {\it {Asymmetric Matters
  from a Dark First-Order Phase Transition}},
  \href{http://arxiv.org/abs/1911.12342}{{\tt arXiv:1911.12342}}.

\bibitem{Madge:2018gfl}
E.~Madge and P.~Schwaller, {\it {Leptophilic dark matter from gauged lepton
  number: Phenomenology and gravitational wave signatures}},  {\em JHEP} {\bf
  02} (2019) 048, [\href{http://arxiv.org/abs/1809.09110}{{\tt
  arXiv:1809.09110}}].

\bibitem{Croon:2018erz}
D.~Croon, V.~Sanz, and G.~White, {\it {Model Discrimination in Gravitational
  Wave spectra from Dark Phase Transitions}},  {\em JHEP} {\bf 08} (2018) 203,
  [\href{http://arxiv.org/abs/1806.02332}{{\tt arXiv:1806.02332}}].

\bibitem{Hashino:2018zsi}
K.~Hashino, M.~Kakizaki, S.~Kanemura, P.~Ko, and T.~Matsui, {\it {Gravitational
  waves from first order electroweak phase transition in models with the
  U(1)$_{X}$ gauge symmetry}},  {\em JHEP} {\bf 06} (2018) 088,
  [\href{http://arxiv.org/abs/1802.02947}{{\tt arXiv:1802.02947}}].

\bibitem{Schwaller:2015tja}
P.~Schwaller, {\it {Gravitational Waves from a Dark Phase Transition}},  {\em
  Phys. Rev. Lett.} {\bf 115} (2015), no.~18 181101,
  [\href{http://arxiv.org/abs/1504.07263}{{\tt arXiv:1504.07263}}].

\bibitem{Alanne:2014bra}
T.~Alanne, K.~Tuominen, and V.~Vaskonen, {\it {Strong phase transition, dark
  matter and vacuum stability from simple hidden sectors}},  {\em Nucl. Phys.
  B} {\bf 889} (2014) 692--711, [\href{http://arxiv.org/abs/1407.0688}{{\tt
  arXiv:1407.0688}}].

\bibitem{Addazi:2017gpt}
A.~Addazi and A.~Marciano, {\it {Gravitational waves from dark first order
  phase transitions and dark photons}},  {\em Chin. Phys. C} {\bf 42} (2018),
  no.~2 023107, [\href{http://arxiv.org/abs/1703.03248}{{\tt
  arXiv:1703.03248}}].

\bibitem{Fairbairn:2019xog}
M.~Fairbairn, E.~Hardy, and A.~Wickens, {\it {Hearing without seeing:
  gravitational waves from hot and cold hidden sectors}},  {\em JHEP} {\bf 07}
  (2019) 044, [\href{http://arxiv.org/abs/1901.11038}{{\tt arXiv:1901.11038}}].

\bibitem{Bertone:2019irm}
G.~Bertone et~al., {\it {Gravitational wave probes of dark matter: challenges
  and opportunities}},  {\em SciPost Phys. Core} {\bf 3} (2020) 007,
  [\href{http://arxiv.org/abs/1907.10610}{{\tt arXiv:1907.10610}}].

\bibitem{Huang:2021rrk}
F.~Huang, V.~Sanz, J.~Shu, and X.~Xue, {\it {LIGO as a probe of dark sectors}},
   {\em Phys. Rev. D} {\bf 104} (2021), no.~9 095001,
  [\href{http://arxiv.org/abs/2102.03155}{{\tt arXiv:2102.03155}}].

\bibitem{Halverson:2020xpg}
J.~Halverson, C.~Long, A.~Maiti, B.~Nelson, and G.~Salinas, {\it {Gravitational
  waves from dark Yang-Mills sectors}},  {\em JHEP} {\bf 05} (2021) 154,
  [\href{http://arxiv.org/abs/2012.04071}{{\tt arXiv:2012.04071}}].

\bibitem{Nakai:2020oit}
Y.~Nakai, M.~Suzuki, F.~Takahashi, and M.~Yamada, {\it {Gravitational Waves and
  Dark Radiation from Dark Phase Transition: Connecting NANOGrav Pulsar Timing
  Data and Hubble Tension}},  {\em Phys. Lett. B} {\bf 816} (2021) 136238,
  [\href{http://arxiv.org/abs/2009.09754}{{\tt arXiv:2009.09754}}].

\bibitem{Ratzinger:2020koh}
W.~Ratzinger and P.~Schwaller, {\it {Whispers from the dark side: Confronting
  light new physics with NANOGrav data}},  {\em SciPost Phys.} {\bf 10} (2021),
  no.~2 047, [\href{http://arxiv.org/abs/2009.11875}{{\tt arXiv:2009.11875}}].

\bibitem{Bhoonah:2020oov}
A.~Bhoonah, J.~Bramante, S.~Nerval, and N.~Song, {\it {Gravitational Waves From
  Dark Sectors, Oscillating Inflatons, and Mass Boosted Dark Matter}},  {\em
  JCAP} {\bf 04} (2021) 043, [\href{http://arxiv.org/abs/2008.12306}{{\tt
  arXiv:2008.12306}}].

\bibitem{Helmboldt:2019pan}
A.~J. Helmboldt, J.~Kubo, and S.~van~der Woude, {\it {Observational prospects
  for gravitational waves from hidden or dark chiral phase transitions}},  {\em
  Phys. Rev. D} {\bf 100} (2019), no.~5 055025,
  [\href{http://arxiv.org/abs/1904.07891}{{\tt arXiv:1904.07891}}].

\bibitem{Beniwal:2017eik}
A.~Beniwal, M.~Lewicki, J.~D. Wells, M.~White, and A.~G. Williams, {\it
  {Gravitational wave, collider and dark matter signals from a scalar singlet
  electroweak baryogenesis}},  {\em JHEP} {\bf 08} (2017) 108,
  [\href{http://arxiv.org/abs/1702.06124}{{\tt arXiv:1702.06124}}].

\bibitem{Baldes:2017ygu}
I.~Baldes, {\it {Generation of Asymmetric Dark Matter and Gravitational
  Waves}},  in {\em {29th Rencontres de Blois on Particle Physics and
  Cosmology}}, 11, 2017.
\newblock \href{http://arxiv.org/abs/1711.08251}{{\tt arXiv:1711.08251}}.

\bibitem{Baldes:2018emh}
I.~Baldes and C.~Garcia-Cely, {\it {Strong gravitational radiation from a
  simple dark matter model}},  {\em JHEP} {\bf 05} (2019) 190,
  [\href{http://arxiv.org/abs/1809.01198}{{\tt arXiv:1809.01198}}].

\bibitem{Huang:2017kzu}
F.~P. Huang and C.~S. Li, {\it {Probing the baryogenesis and dark matter
  relaxed in phase transition by gravitational waves and colliders}},  {\em
  Phys. Rev. D} {\bf 96} (2017), no.~9 095028,
  [\href{http://arxiv.org/abs/1709.09691}{{\tt arXiv:1709.09691}}].

\bibitem{Tsumura:2017knk}
K.~Tsumura, M.~Yamada, and Y.~Yamaguchi, {\it {Gravitational wave from dark
  sector with dark pion}},  {\em JCAP} {\bf 07} (2017) 044,
  [\href{http://arxiv.org/abs/1704.00219}{{\tt arXiv:1704.00219}}].

\bibitem{Croon:2019rqu}
D.~Croon, A.~Kusenko, A.~Mazumdar, and G.~White, {\it {Solitosynthesis and
  Gravitational Waves}},  {\em Phys. Rev. D} {\bf 101} (2020), no.~8 085010,
  [\href{http://arxiv.org/abs/1910.09562}{{\tt arXiv:1910.09562}}].

\bibitem{Addazi:2017nmg}
A.~Addazi, Y.-F. Cai, and A.~Marciano, {\it {Testing Dark Matter Models with
  Radio Telescopes in light of Gravitational Wave Astronomy}},  {\em Phys.
  Lett. B} {\bf 782} (2018) 732--736,
  [\href{http://arxiv.org/abs/1712.03798}{{\tt arXiv:1712.03798}}].

\bibitem{Imtiaz:2018dfn}
B.~Imtiaz, Y.-F. Cai, and Y.~Wan, {\it {Two-field cosmological phase
  transitions and gravitational waves in the singlet Majoron model}},  {\em
  Eur. Phys. J. C} {\bf 79} (2019), no.~1 25,
  [\href{http://arxiv.org/abs/1804.05835}{{\tt arXiv:1804.05835}}].

\bibitem{Addazi:2017oge}
A.~Addazi and A.~Marciano, {\it {Limiting majoron self-interactions from
  gravitational wave experiments}},  {\em Chin. Phys. C} {\bf 42} (2018), no.~2
  023105, [\href{http://arxiv.org/abs/1705.08346}{{\tt arXiv:1705.08346}}].

\bibitem{Bai:2018dxf}
Y.~Bai, A.~J. Long, and S.~Lu, {\it {Dark Quark Nuggets}},  {\em Phys. Rev. D}
  {\bf 99} (2019), no.~5 055047, [\href{http://arxiv.org/abs/1810.04360}{{\tt
  arXiv:1810.04360}}].

\bibitem{Bian:2018bxr}
L.~Bian and X.~Liu, {\it {Two-step strongly first-order electroweak phase
  transition modified FIMP dark matter, gravitational wave signals, and the
  neutrino mass}},  {\em Phys. Rev. D} {\bf 99} (2019), no.~5 055003,
  [\href{http://arxiv.org/abs/1811.03279}{{\tt arXiv:1811.03279}}].

\bibitem{Pandey:2020hoq}
M.~Pandey and A.~Paul, {\it {Gravitational Wave Emissions from First Order
  Phase Transitions with Two Component FIMP Dark Matter}},
  \href{http://arxiv.org/abs/2003.08828}{{\tt arXiv:2003.08828}}.

\bibitem{Davoudiasl:2021ijv}
H.~Davoudiasl, P.~B. Denton, and J.~Gehrlein, {\it {Supermassive Black Holes,
  Ultralight Dark Matter, and Gravitational Waves from a First Order Phase
  Transition}},  {\em Phys. Rev. Lett.} {\bf 128} (2022), no.~8 081101,
  [\href{http://arxiv.org/abs/2109.01678}{{\tt arXiv:2109.01678}}].

\bibitem{Huang:2017rzf}
F.~P. Huang and J.-H. Yu, {\it {Exploring inert dark matter blind spots with
  gravitational wave signatures}},  {\em Phys. Rev. D} {\bf 98} (2018), no.~9
  095022, [\href{http://arxiv.org/abs/1704.04201}{{\tt arXiv:1704.04201}}].

\bibitem{Paul:2019pgt}
A.~Paul, B.~Banerjee, and D.~Majumdar, {\it {Gravitational wave signatures from
  an extended inert doublet dark matter model}},  {\em JCAP} {\bf 10} (2019)
  062, [\href{http://arxiv.org/abs/1908.00829}{{\tt arXiv:1908.00829}}].

\bibitem{Hektor:2018esx}
A.~Hektor, K.~Kannike, and V.~Vaskonen, {\it {Modifying dark matter indirect
  detection signals by thermal effects at freeze-out}},  {\em Phys. Rev. D}
  {\bf 98} (2018), no.~1 015032, [\href{http://arxiv.org/abs/1801.06184}{{\tt
  arXiv:1801.06184}}].

\bibitem{Kannike:2019mzk}
K.~Kannike, K.~Loos, and M.~Raidal, {\it {Gravitational wave signals of
  pseudo-Goldstone dark matter in the $\mathbb{Z}_{3}$ complex singlet model}},
   {\em Phys. Rev. D} {\bf 101} (2020), no.~3 035001,
  [\href{http://arxiv.org/abs/1907.13136}{{\tt arXiv:1907.13136}}].

\bibitem{Kannike:2019wsn}
K.~Kannike and M.~Raidal, {\it {Phase Transitions and Gravitational Wave Tests
  of Pseudo-Goldstone Dark Matter in the Softly Broken U(1) Scalar Singlet
  Model}},  {\em Phys. Rev. D} {\bf 99} (2019), no.~11 115010,
  [\href{http://arxiv.org/abs/1901.03333}{{\tt arXiv:1901.03333}}].

\bibitem{Mohamadnejad:2019vzg}
A.~Mohamadnejad, {\it {Gravitational waves from scale-invariant vector dark
  matter model: Probing below the neutrino-floor}},  {\em Eur. Phys. J. C} {\bf
  80} (2020), no.~3 197, [\href{http://arxiv.org/abs/1907.08899}{{\tt
  arXiv:1907.08899}}].

\bibitem{Breitbach:2018ddu}
M.~Breitbach, J.~Kopp, E.~Madge, T.~Opferkuch, and P.~Schwaller, {\it {Dark,
  Cold, and Noisy: Constraining Secluded Hidden Sectors with Gravitational
  Waves}},  {\em JCAP} {\bf 07} (2019) 007,
  [\href{http://arxiv.org/abs/1811.11175}{{\tt arXiv:1811.11175}}].

\bibitem{Arakawa:2021wgz}
J.~Arakawa, A.~Rajaraman, and T.~M.~P. Tait, {\it {Annihilogenesis}},
  \href{http://arxiv.org/abs/2109.13941}{{\tt arXiv:2109.13941}}.

\bibitem{Azatov:2021ifm}
A.~Azatov, M.~Vanvlasselaer, and W.~Yin, {\it {Dark Matter production from
  relativistic bubble walls}},  {\em JHEP} {\bf 03} (2021) 288,
  [\href{http://arxiv.org/abs/2101.05721}{{\tt arXiv:2101.05721}}].

\bibitem{Azatov:2020ufh}
A.~Azatov and M.~Vanvlasselaer, {\it {Bubble wall velocity: heavy physics
  effects}},  {\em JCAP} {\bf 01} (2021) 058,
  [\href{http://arxiv.org/abs/2010.02590}{{\tt arXiv:2010.02590}}].

\bibitem{Chung:2012vg}
D.~J.~H. Chung, A.~J. Long, and L.-T. Wang, {\it {125 GeV Higgs boson and
  electroweak phase transition model classes}},  {\em Phys. Rev. D} {\bf 87}
  (2013), no.~2 023509, [\href{http://arxiv.org/abs/1209.1819}{{\tt
  arXiv:1209.1819}}].

\bibitem{Dev:2016feu}
P.~S.~B. Dev and A.~Mazumdar, {\it {Probing the Scale of New Physics by
  Advanced LIGO/VIRGO}},  {\em Phys. Rev. D} {\bf 93} (2016), no.~10 104001,
  [\href{http://arxiv.org/abs/1602.04203}{{\tt arXiv:1602.04203}}].

\bibitem{Jinno:2017ixd}
R.~Jinno, S.~Lee, H.~Seong, and M.~Takimoto, {\it {Gravitational waves from
  first-order phase transitions: Towards model separation by bubble nucleation
  rate}},  {\em JCAP} {\bf 11} (2017) 050,
  [\href{http://arxiv.org/abs/1708.01253}{{\tt arXiv:1708.01253}}].

\bibitem{Chala:2019rfk}
M.~Chala, V.~V. Khoze, M.~Spannowsky, and P.~Waite, {\it {Mapping the shape of
  the scalar potential with gravitational waves}},  {\em Int. J. Mod. Phys. A}
  {\bf 34} (2019), no.~33 1950223, [\href{http://arxiv.org/abs/1905.00911}{{\tt
  arXiv:1905.00911}}].

\bibitem{Croon:2019kpe}
D.~Croon, T.~E. Gonzalo, L.~Graf, N.~Ko\v{s}nik, and G.~White, {\it {GUT
  Physics in the era of the LHC}},  {\em Front. in Phys.} {\bf 7} (2019) 76,
  [\href{http://arxiv.org/abs/1903.04977}{{\tt arXiv:1903.04977}}].

\bibitem{Ellis:2019flb}
S.~A.~R. Ellis, S.~Ipek, and G.~White, {\it {Electroweak Baryogenesis from
  Temperature-Varying Couplings}},  {\em JHEP} {\bf 08} (2019) 002,
  [\href{http://arxiv.org/abs/1905.11994}{{\tt arXiv:1905.11994}}].

\bibitem{Ellis:2020awk}
J.~Ellis, M.~Lewicki, and J.~M. No, {\it {Gravitational waves from first-order
  cosmological phase transitions: lifetime of the sound wave source}},  {\em
  JCAP} {\bf 07} (2020) 050, [\href{http://arxiv.org/abs/2003.07360}{{\tt
  arXiv:2003.07360}}].

\bibitem{Megevand:2021llq}
A.~Megevand and F.~A. Membiela, {\it {Model-independent features of
  gravitational waves from bubble collisions}},  {\em Phys. Rev. D} {\bf 104}
  (2021), no.~12 123532, [\href{http://arxiv.org/abs/2108.07034}{{\tt
  arXiv:2108.07034}}].

\bibitem{Schmitz:2020syl}
K.~Schmitz, {\it {New Sensitivity Curves for Gravitational-Wave Signals from
  Cosmological Phase Transitions}},  {\em JHEP} {\bf 01} (2021) 097,
  [\href{http://arxiv.org/abs/2002.04615}{{\tt arXiv:2002.04615}}].

\bibitem{Alanne:2019bsm}
T.~Alanne, T.~Hugle, M.~Platscher, and K.~Schmitz, {\it {A fresh look at the
  gravitational-wave signal from cosmological phase transitions}},  {\em JHEP}
  {\bf 03} (2020) 004, [\href{http://arxiv.org/abs/1909.11356}{{\tt
  arXiv:1909.11356}}].

\bibitem{Bai:2021ibt}
Y.~Bai and M.~Korwar, {\it {Cosmological Constraints on First-Order Phase
  Transitions}},  \href{http://arxiv.org/abs/2109.14765}{{\tt
  arXiv:2109.14765}}.

\bibitem{Kumar:2021ffi}
S.~Kumar, R.~Sundrum, and Y.~Tsai, {\it {Non-Gaussian stochastic gravitational
  waves from phase transitions}},  {\em JHEP} {\bf 11} (2021) 107,
  [\href{http://arxiv.org/abs/2102.05665}{{\tt arXiv:2102.05665}}].

\bibitem{Ellis:2020uid}
J.~Ellis, M.~Fairbairn, M.~Lewicki, V.~Vaskonen, and A.~Wickens, {\it
  {Detecting circular polarisation in the stochastic gravitational-wave
  background from a first-order cosmological phase transition}},  {\em JCAP}
  {\bf 10} (2020) 032, [\href{http://arxiv.org/abs/2005.05278}{{\tt
  arXiv:2005.05278}}].

\bibitem{Espinosa:2010hh}
J.~R. Espinosa, T.~Konstandin, J.~M. No, and G.~Servant, {\it {Energy Budget of
  Cosmological First-order Phase Transitions}},  {\em JCAP} {\bf 06} (2010)
  028, [\href{http://arxiv.org/abs/1004.4187}{{\tt arXiv:1004.4187}}].

\bibitem{Giese:2020rtr}
F.~Giese, T.~Konstandin, and J.~van~de Vis, {\it {Model-independent energy
  budget of cosmological first-order phase transitions\textemdash{}A sound
  argument to go beyond the bag model}},  {\em JCAP} {\bf 07} (2020), no.~07
  057, [\href{http://arxiv.org/abs/2004.06995}{{\tt arXiv:2004.06995}}].

\bibitem{Giese:2020znk}
F.~Giese, T.~Konstandin, K.~Schmitz, and J.~van~de Vis, {\it {Model-independent
  energy budget for LISA}},  {\em JCAP} {\bf 01} (2021) 072,
  [\href{http://arxiv.org/abs/2010.09744}{{\tt arXiv:2010.09744}}].

\bibitem{Bodeker:2009qy}
D.~Bodeker and G.~D. Moore, {\it {Can electroweak bubble walls run away?}},
  {\em JCAP} {\bf 05} (2009) 009, [\href{http://arxiv.org/abs/0903.4099}{{\tt
  arXiv:0903.4099}}].

\bibitem{Bodeker:2017cim}
D.~Bodeker and G.~D. Moore, {\it {Electroweak Bubble Wall Speed Limit}},  {\em
  JCAP} {\bf 05} (2017) 025, [\href{http://arxiv.org/abs/1703.08215}{{\tt
  arXiv:1703.08215}}].

\bibitem{Hoche:2020ysm}
S.~H\"oche, J.~Kozaczuk, A.~J. Long, J.~Turner, and Y.~Wang, {\it {Towards an
  all-orders calculation of the electroweak bubble wall velocity}},  {\em JCAP}
  {\bf 03} (2021) 009, [\href{http://arxiv.org/abs/2007.10343}{{\tt
  arXiv:2007.10343}}].

\bibitem{Schmitz:2020rag}
K.~Schmitz, {\it {LISA Sensitivity to Gravitational Waves from Sound Waves}},
  {\em Symmetry} {\bf 12} (2020), no.~9 1477,
  [\href{http://arxiv.org/abs/2005.10789}{{\tt arXiv:2005.10789}}].

\bibitem{Hindmarsh:2013xza}
M.~Hindmarsh, S.~J. Huber, K.~Rummukainen, and D.~J. Weir, {\it {Gravitational
  waves from the sound of a first order phase transition}},  {\em Phys. Rev.
  Lett.} {\bf 112} (2014) 041301, [\href{http://arxiv.org/abs/1304.2433}{{\tt
  arXiv:1304.2433}}].

\bibitem{Hindmarsh:2015qta}
M.~Hindmarsh, S.~J. Huber, K.~Rummukainen, and D.~J. Weir, {\it {Numerical
  simulations of acoustically generated gravitational waves at a first order
  phase transition}},  {\em Phys. Rev. D} {\bf 92} (2015), no.~12 123009,
  [\href{http://arxiv.org/abs/1504.03291}{{\tt arXiv:1504.03291}}].

\bibitem{Hindmarsh:2017gnf}
M.~Hindmarsh, S.~J. Huber, K.~Rummukainen, and D.~J. Weir, {\it {Shape of the
  acoustic gravitational wave power spectrum from a first order phase
  transition}},  {\em Phys. Rev. D} {\bf 96} (2017), no.~10 103520,
  [\href{http://arxiv.org/abs/1704.05871}{{\tt arXiv:1704.05871}}]. [Erratum:
  Phys.Rev.D 101, 089902 (2020)].

\bibitem{Jinno:2017fby}
R.~Jinno and M.~Takimoto, {\it {Gravitational waves from bubble dynamics:
  Beyond the Envelope}},  {\em JCAP} {\bf 01} (2019) 060,
  [\href{http://arxiv.org/abs/1707.03111}{{\tt arXiv:1707.03111}}].

\bibitem{Cutting:2018tjt}
D.~Cutting, M.~Hindmarsh, and D.~J. Weir, {\it {Gravitational waves from vacuum
  first-order phase transitions: from the envelope to the lattice}},  {\em
  Phys. Rev. D} {\bf 97} (2018), no.~12 123513,
  [\href{http://arxiv.org/abs/1802.05712}{{\tt arXiv:1802.05712}}].

\bibitem{Cutting:2020nla}
D.~Cutting, E.~G. Escartin, M.~Hindmarsh, and D.~J. Weir, {\it {Gravitational
  waves from vacuum first order phase transitions II: from thin to thick
  walls}},  {\em Phys. Rev. D} {\bf 103} (2021), no.~2 023531,
  [\href{http://arxiv.org/abs/2005.13537}{{\tt arXiv:2005.13537}}].

\bibitem{Jinno:2020eqg}
R.~Jinno, T.~Konstandin, and H.~Rubira, {\it {A hybrid simulation of
  gravitational wave production in first-order phase transitions}},  {\em JCAP}
  {\bf 04} (2021) 014, [\href{http://arxiv.org/abs/2010.00971}{{\tt
  arXiv:2010.00971}}].

\bibitem{Dahl:2021wyk}
J.~Dahl, M.~Hindmarsh, K.~Rummukainen, and D.~Weir, {\it {Decay of acoustic
  turbulence in two dimensions and implications for cosmological gravitational
  waves}},  \href{http://arxiv.org/abs/2112.12013}{{\tt arXiv:2112.12013}}.

\bibitem{Auclair:2022jod}
P.~Auclair, C.~Caprini, D.~Cutting, M.~Hindmarsh, K.~Rummukainen, D.~A. Steer,
  and D.~J. Weir, {\it {Generation of gravitational waves from freely decaying
  turbulence}},  \href{http://arxiv.org/abs/2205.02588}{{\tt
  arXiv:2205.02588}}.

\bibitem{Quiros:1999jp}
M.~Quiros, {\it {Finite temperature field theory and phase transitions}},  in
  {\em {ICTP Summer School in High-Energy Physics and Cosmology}},
  pp.~187--259, 1, 1999.
\newblock \href{http://arxiv.org/abs/hep-ph/9901312}{{\tt hep-ph/9901312}}.

\bibitem{wainwright2012:cosmotransitions}
C.~L. Wainwright, {\it Cosmotransitions: Computing cosmological phase
  transition temperatures and bubble profiles with multiple fields},  {\em
  Computer Physics Communications} {\bf 183} (Sep, 2012) 2006–2013.

\bibitem{Ramsey-Musolf:2019lsf}
M.~J. Ramsey-Musolf, {\it {The electroweak phase transition: a collider
  target}},  {\em JHEP} {\bf 09} (2020) 179,
  [\href{http://arxiv.org/abs/1912.07189}{{\tt arXiv:1912.07189}}].

\bibitem{Akula:2016gpl}
S.~Akula, C.~Bal\'azs, and G.~A. White, {\it {Semi-analytic techniques for
  calculating bubble wall profiles}},  {\em Eur. Phys. J. C} {\bf 76} (2016),
  no.~12 681, [\href{http://arxiv.org/abs/1608.00008}{{\tt arXiv:1608.00008}}].

\bibitem{Dine:1992wr}
M.~Dine, R.~G. Leigh, P.~Y. Huet, A.~D. Linde, and D.~A. Linde, {\it {Towards
  the theory of the electroweak phase transition}},  {\em Phys. Rev. D} {\bf
  46} (1992) 550--571, [\href{http://arxiv.org/abs/hep-ph/9203203}{{\tt
  hep-ph/9203203}}].

\bibitem{Guo:2020grp}
H.-K. Guo, K.~Sinha, D.~Vagie, and G.~White, {\it {Phase Transitions in an
  Expanding Universe: Stochastic Gravitational Waves in Standard and
  Non-Standard Histories}},  {\em JCAP} {\bf 01} (2021) 001,
  [\href{http://arxiv.org/abs/2007.08537}{{\tt arXiv:2007.08537}}].

\bibitem{Kosowsky:1992vn}
A.~Kosowsky and M.~S. Turner, {\it {Gravitational radiation from colliding
  vacuum bubbles: envelope approximation to many bubble collisions}},  {\em
  Phys. Rev. D} {\bf 47} (1993) 4372--4391,
  [\href{http://arxiv.org/abs/astro-ph/9211004}{{\tt astro-ph/9211004}}].

\bibitem{Lewicki:2019gmv}
M.~Lewicki and V.~Vaskonen, {\it {On bubble collisions in strongly supercooled
  phase transitions}},  {\em Phys. Dark Univ.} {\bf 30} (2020) 100672,
  [\href{http://arxiv.org/abs/1912.00997}{{\tt arXiv:1912.00997}}].

\bibitem{Hindmarsh:2016lnk}
M.~Hindmarsh, {\it {Sound shell model for acoustic gravitational wave
  production at a first-order phase transition in the early Universe}},  {\em
  Phys. Rev. Lett.} {\bf 120} (2018), no.~7 071301,
  [\href{http://arxiv.org/abs/1608.04735}{{\tt arXiv:1608.04735}}].

\bibitem{Guo:2021qcq}
H.-K. Guo, K.~Sinha, D.~Vagie, and G.~White, {\it {The benefits of diligence:
  how precise are predicted gravitational wave spectra in models with phase
  transitions?}},  {\em JHEP} {\bf 06} (2021) 164,
  [\href{http://arxiv.org/abs/2103.06933}{{\tt arXiv:2103.06933}}].

\bibitem{Chiang:2017zbz}
C.-W. Chiang and E.~Senaha, {\it {On gauge dependence of gravitational waves
  from a first-order phase transition in classical scale-invariant $U(1)'$
  models}},  {\em Phys. Lett. B} {\bf 774} (2017) 489--493,
  [\href{http://arxiv.org/abs/1707.06765}{{\tt arXiv:1707.06765}}].

\bibitem{Niemi:2021qvp}
L.~Niemi, P.~Schicho, and T.~V.~I. Tenkanen, {\it {Singlet-assisted electroweak
  phase transition at two loops}},  {\em Phys. Rev. D} {\bf 103} (2021), no.~11
  115035, [\href{http://arxiv.org/abs/2103.07467}{{\tt arXiv:2103.07467}}].

\bibitem{Cai:2017tmh}
R.-G. Cai, M.~Sasaki, and S.-J. Wang, {\it {The gravitational waves from the
  first-order phase transition with a dimension-six operator}},  {\em JCAP}
  {\bf 08} (2017) 004, [\href{http://arxiv.org/abs/1707.03001}{{\tt
  arXiv:1707.03001}}].

\bibitem{Chala:2018ari}
M.~Chala, C.~Krause, and G.~Nardini, {\it {Signals of the electroweak phase
  transition at colliders and gravitational wave observatories}},  {\em JHEP}
  {\bf 07} (2018) 062, [\href{http://arxiv.org/abs/1802.02168}{{\tt
  arXiv:1802.02168}}].

\bibitem{Gould:2021ccf}
O.~Gould and J.~Hirvonen, {\it {Effective field theory approach to thermal
  bubble nucleation}},  {\em Phys. Rev. D} {\bf 104} (2021), no.~9 096015,
  [\href{http://arxiv.org/abs/2108.04377}{{\tt arXiv:2108.04377}}].

\bibitem{Gould:2021oba}
O.~Gould and T.~V.~I. Tenkanen, {\it {On the perturbative expansion at high
  temperature and implications for cosmological phase transitions}},  {\em
  JHEP} {\bf 06} (2021) 069, [\href{http://arxiv.org/abs/2104.04399}{{\tt
  arXiv:2104.04399}}].

\bibitem{Schicho:2021gca}
P.~M. Schicho, T.~V.~I. Tenkanen, and J.~\"Osterman, {\it {Robust approach to
  thermal resummation: Standard Model meets a singlet}},  {\em JHEP} {\bf 06}
  (2021) 130, [\href{http://arxiv.org/abs/2102.11145}{{\tt arXiv:2102.11145}}].

\bibitem{Gould:2021dzl}
O.~Gould, {\it {Real scalar phase transitions: a nonperturbative analysis}},
  {\em JHEP} {\bf 04} (2021) 057, [\href{http://arxiv.org/abs/2101.05528}{{\tt
  arXiv:2101.05528}}].

\bibitem{Postma:2020toi}
M.~Postma and G.~White, {\it {Cosmological phase transitions: is effective
  field theory just a toy?}},  {\em JHEP} {\bf 03} (2021) 280,
  [\href{http://arxiv.org/abs/2012.03953}{{\tt arXiv:2012.03953}}].

\bibitem{Cutting:2019zws}
D.~Cutting, M.~Hindmarsh, and D.~J. Weir, {\it {Vorticity, kinetic energy, and
  suppressed gravitational wave production in strong first order phase
  transitions}},  {\em Phys. Rev. Lett.} {\bf 125} (2020), no.~2 021302,
  [\href{http://arxiv.org/abs/1906.00480}{{\tt arXiv:1906.00480}}].

\bibitem{RoperPol:2019wvy}
A.~Roper~Pol, S.~Mandal, A.~Brandenburg, T.~Kahniashvili, and A.~Kosowsky, {\it
  {Numerical simulations of gravitational waves from early-universe
  turbulence}},  {\em Phys. Rev. D} {\bf 102} (2020), no.~8 083512,
  [\href{http://arxiv.org/abs/1903.08585}{{\tt arXiv:1903.08585}}].

\bibitem{LatticeStrongDynamics:2020jwi}
{\bf Lattice Strong Dynamics} Collaboration, R.~C. Brower et~al., {\it {Stealth
  dark matter confinement transition and gravitational waves}},  {\em Phys.
  Rev. D} {\bf 103} (2021), no.~1 014505,
  [\href{http://arxiv.org/abs/2006.16429}{{\tt arXiv:2006.16429}}].

\bibitem{Lewicki:2020azd}
M.~Lewicki and V.~Vaskonen, {\it {Gravitational waves from colliding vacuum
  bubbles in gauge theories}},  {\em Eur. Phys. J. C} {\bf 81} (2021), no.~5
  437, [\href{http://arxiv.org/abs/2012.07826}{{\tt arXiv:2012.07826}}].

\bibitem{Cutting:2021tqt}
D.~Cutting, {\em {Simulations of early universe phase transitions and
  gravitational waves}}.
\newblock PhD thesis, University of Sussex, Sussex U., 2021.

\bibitem{Jinno:2021ury}
R.~Jinno, T.~Konstandin, H.~Rubira, and J.~van~de Vis, {\it {Effect of density
  fluctuations on gravitational wave production in first-order phase
  transitions}},  {\em JCAP} {\bf 12} (2021), no.~12 019,
  [\href{http://arxiv.org/abs/2108.11947}{{\tt arXiv:2108.11947}}].

\bibitem{Schicho:2022wty}
P.~Schicho, T.~V.~I. Tenkanen, and G.~White, {\it {Combining thermal
  resummation and gauge invariance for electroweak phase transition}},
  \href{http://arxiv.org/abs/2203.04284}{{\tt arXiv:2203.04284}}.

\bibitem{vanHaasteren:2011ni}
R.~van Haasteren et~al., {\it {Placing limits on the stochastic
  gravitational-wave background using European Pulsar Timing Array data}},
  {\em Mon. Not. Roy. Astron. Soc.} {\bf 414} (2011), no.~4 3117--3128,
  [\href{http://arxiv.org/abs/1103.0576}{{\tt arXiv:1103.0576}}]. [Erratum:
  Mon.Not.Roy.Astron.Soc. 425, 1597 (2012)].

\bibitem{NANOGrav:2020bcs}
{\bf NANOGrav} Collaboration, Z.~Arzoumanian et~al., {\it {The NANOGrav 12.5 yr
  Data Set: Search for an Isotropic Stochastic Gravitational-wave Background}},
   {\em Astrophys. J. Lett.} {\bf 905} (2020), no.~2 L34,
  [\href{http://arxiv.org/abs/2009.04496}{{\tt arXiv:2009.04496}}].

\bibitem{Lazio:2017fos}
T.~J.~W. Lazio, S.~Bhaskaran, C.~Cutler, W.~M. Folkner, R.~S. Park, J.~A.
  Ellis, T.~Ely, S.~R. Taylor, and M.~Vallisneri, {\it {Solar System
  Ephemerides, Pulsar Timing, Gravitational Waves, \textbackslash{}\&
  Navigation}},  {\em IAU Symp.} {\bf 337} (2017) 150--153,
  [\href{http://arxiv.org/abs/1801.02898}{{\tt arXiv:1801.02898}}].

\bibitem{Gaia:2018ydn}
{\bf Gaia} Collaboration, A.~G.~A. Brown et~al., {\it {Gaia Data Release 2}:
  {Summary of the contents and survey properties}},  {\em Astron. Astrophys.}
  {\bf 616} (2018) A1, [\href{http://arxiv.org/abs/1804.09365}{{\tt
  arXiv:1804.09365}}].

\bibitem{Janssen:2014dka}
G.~Janssen et~al., {\it {Gravitational wave astronomy with the SKA}},  {\em
  PoS} {\bf AASKA14} (2015) 037, [\href{http://arxiv.org/abs/1501.00127}{{\tt
  arXiv:1501.00127}}].

\bibitem{10.3389/fspas.2018.00011}
A.~Vallenari, {\it The future of astrometry in space},  {\em Frontiers in
  Astronomy and Space Sciences} {\bf 5} (2018).

\bibitem{LISA:2017pwj}
{\bf LISA} Collaboration, P.~Amaro-Seoane et~al., {\it {Laser Interferometer
  Space Antenna}},  \href{http://arxiv.org/abs/1702.00786}{{\tt
  arXiv:1702.00786}}.

\bibitem{Robson:2018ifk}
T.~Robson, N.~J. Cornish, and C.~Liu, {\it {The construction and use of LISA
  sensitivity curves}},  {\em Class. Quant. Grav.} {\bf 36} (2019), no.~10
  105011, [\href{http://arxiv.org/abs/1803.01944}{{\tt arXiv:1803.01944}}].

\bibitem{Ruan:2018tsw}
W.-H. Ruan, Z.-K. Guo, R.-G. Cai, and Y.-Z. Zhang, {\it {Taiji program:
  Gravitational-wave sources}},  {\em Int. J. Mod. Phys. A} {\bf 35} (2020),
  no.~17 2050075, [\href{http://arxiv.org/abs/1807.09495}{{\tt
  arXiv:1807.09495}}].

\bibitem{TianQin:2015yph}
{\bf TianQin} Collaboration, J.~Luo et~al., {\it {TianQin: a space-borne
  gravitational wave detector}},  {\em Class. Quant. Grav.} {\bf 33} (2016),
  no.~3 035010, [\href{http://arxiv.org/abs/1512.02076}{{\tt
  arXiv:1512.02076}}].

\bibitem{Gong:2014mca}
X.~Gong et~al., {\it {Descope of the ALIA mission}},  {\em J. Phys. Conf. Ser.}
  {\bf 610} (2015), no.~1 012011, [\href{http://arxiv.org/abs/1410.7296}{{\tt
  arXiv:1410.7296}}].

\bibitem{Corbin:2005ny}
V.~Corbin and N.~J. Cornish, {\it {Detecting the cosmic gravitational wave
  background with the big bang observer}},  {\em Class. Quant. Grav.} {\bf 23}
  (2006) 2435--2446, [\href{http://arxiv.org/abs/gr-qc/0512039}{{\tt
  gr-qc/0512039}}].

\bibitem{Yagi:2011wg}
K.~Yagi and N.~Seto, {\it {Detector configuration of DECIGO/BBO and
  identification of cosmological neutron-star binaries}},  {\em Phys. Rev. D}
  {\bf 83} (2011) 044011, [\href{http://arxiv.org/abs/1101.3940}{{\tt
  arXiv:1101.3940}}]. [Erratum: Phys.Rev.D 95, 109901 (2017)].

\bibitem{Kawamura:2020pcg}
S.~Kawamura et~al., {\it {Current status of space gravitational wave antenna
  DECIGO and B-DECIGO}},  {\em PTEP} {\bf 2021} (2021), no.~5 05A105,
  [\href{http://arxiv.org/abs/2006.13545}{{\tt arXiv:2006.13545}}].

\bibitem{Kawamura:2006up}
S.~Kawamura et~al., {\it {The Japanese space gravitational wave antenna
  DECIGO}},  {\em Class. Quant. Grav.} {\bf 23} (2006) S125--S132.

\bibitem{Shoemaker:2019bqt}
{\bf LIGO Scientific} Collaboration, D.~Shoemaker, {\it {Gravitational wave
  astronomy with LIGO and similar detectors in the next decade}},
  \href{http://arxiv.org/abs/1904.03187}{{\tt arXiv:1904.03187}}.

\bibitem{LIGOScientific:2014pky}
{\bf LIGO Scientific} Collaboration, J.~Aasi et~al., {\it {Advanced LIGO}},
  {\em Class. Quant. Grav.} {\bf 32} (2015) 074001,
  [\href{http://arxiv.org/abs/1411.4547}{{\tt arXiv:1411.4547}}].

\bibitem{QuantumLIGO}
T.~L.~S. Collaboration, {\em {LIGO DCC-T1400316}}, 2014.

\bibitem{Punturo:2010zz}
M.~Punturo et~al., {\it {The Einstein Telescope: A third-generation
  gravitational wave observatory}},  {\em Class. Quant. Grav.} {\bf 27} (2010)
  194002.

\bibitem{LIGOScientific:2016wof}
{\bf LIGO Scientific} Collaboration, B.~P. Abbott et~al., {\it {Exploring the
  Sensitivity of Next Generation Gravitational Wave Detectors}},  {\em Class.
  Quant. Grav.} {\bf 34} (2017), no.~4 044001,
  [\href{http://arxiv.org/abs/1607.08697}{{\tt arXiv:1607.08697}}].

\bibitem{Sesana:2019vho}
A.~Sesana et~al., {\it {Unveiling the gravitational universe at $\mu$-Hz
  frequencies}},  {\em Exper. Astron.} {\bf 51} (2021), no.~3 1333--1383,
  [\href{http://arxiv.org/abs/1908.11391}{{\tt arXiv:1908.11391}}].

\bibitem{Kobakhidze:2017mru}
A.~Kobakhidze, C.~Lagger, A.~Manning, and J.~Yue, {\it {Gravitational waves
  from a supercooled electroweak phase transition and their detection with
  pulsar timing arrays}},  {\em Eur. Phys. J. C} {\bf 77} (2017), no.~8 570,
  [\href{http://arxiv.org/abs/1703.06552}{{\tt arXiv:1703.06552}}].

\bibitem{Ellis:2019oqb}
J.~Ellis, M.~Lewicki, J.~M. No, and V.~Vaskonen, {\it {Gravitational wave
  energy budget in strongly supercooled phase transitions}},  {\em JCAP} {\bf
  06} (2019) 024, [\href{http://arxiv.org/abs/1903.09642}{{\tt
  arXiv:1903.09642}}].

\bibitem{Ellis:2020nnr}
J.~Ellis, M.~Lewicki, and V.~Vaskonen, {\it {Updated predictions for
  gravitational waves produced in a strongly supercooled phase transition}},
  {\em JCAP} {\bf 11} (2020) 020, [\href{http://arxiv.org/abs/2007.15586}{{\tt
  arXiv:2007.15586}}].

\bibitem{Wang:2020jrd}
X.~Wang, F.~P. Huang, and X.~Zhang, {\it {Phase transition dynamics and
  gravitational wave spectra of strong first-order phase transition in
  supercooled universe}},  {\em JCAP} {\bf 05} (2020) 045,
  [\href{http://arxiv.org/abs/2003.08892}{{\tt arXiv:2003.08892}}].

\bibitem{Croon:2020cgk}
D.~Croon, O.~Gould, P.~Schicho, T.~V.~I. Tenkanen, and G.~White, {\it
  {Theoretical uncertainties for cosmological first-order phase transitions}},
  {\em JHEP} {\bf 04} (2021) 055, [\href{http://arxiv.org/abs/2009.10080}{{\tt
  arXiv:2009.10080}}].

\end{thebibliography}\endgroup

\end{document}